\documentclass[preprint,3p,times,twocolumn,authoryear]{elsarticle}

\usepackage[switch]{lineno}
\usepackage[utf8]{inputenc}
\usepackage[T1]{fontenc}
\usepackage{amssymb}
\usepackage{amsmath}
\usepackage{latexsym}
\usepackage{bm} 

\usepackage{graphicx}  
\usepackage{adjustbox}
\usepackage{pdflscape}
\usepackage{booktabs}   
\usepackage{siunitx}  
\usepackage{nicematrix}
\usepackage{caption}
\usepackage{subfig}
\usepackage{framed,multirow}
\usepackage{placeins}
\usepackage{float}
\usepackage{rotating}
\usepackage{xcolor}
\usepackage{url}
\usepackage{hyperref}
\hypersetup{colorlinks=true,linkcolor={blue},citecolor={blue},urlcolor={red}}
\usepackage{pdfpages}

\usepackage{natbib}

\def\orcid#1{}

\providecommand{\DOIprefix}{doi: }
\providecommand{\URLprefix}{URL: }
\providecommand{\doi}[1]{\href{https://doi.org/#1}{#1}}
\providecommand{\bibinfo}[2]{#2}

\journal{Advances in Space Research}

\providecommand{\aap}{Astronomy \& Astrophysics}
\providecommand{\aj}{The Astronomical Journal}
\providecommand{\apj}{The Astrophysical Journal}

\providecommand{\mnras}{Monthly Notices of the Royal Astronomical Society}

\providecommand{\apss}{Astrophysics and Space Science}

\providecommand{\pasp}{Publications of the Astronomical Society of the Pacific}
\providecommand{\araa}{Annual Review of Astronomy and Astrophysics}
\providecommand{\apjl}{The Astrophysical Journal Letters}
\providecommand{\physscr}{Physica Scripta}
\providecommand{\cjaa}{Chinese Journal of Astronomy and Astrophysics}
\providecommand{\pasa}{Publications of the Astronomical Society of Australia}

\DeclareUnicodeCharacter{03C7}{\chi}

\begin{document}
\begin{frontmatter}

\title{Kings of the Milky Way: A Homogeneous Gaia DR3 Analysis of King Open Clusters and the Galactic Disc Metallicity Gradient} 

\author[label1]{Talar Yontan\corref{cor1}\orcid{0000-0002-5657-6194}}
\ead{talar.yontan@istanbul.edu.tr}
\author[label1]{Selçuk Bilir\orcid{0000-0003-3510-1509}}
\author[label2]{Hikmet Çakmak\orcid{0000-0002-1959-6049}}
\author[label1]{Olcay Plevne\orcid{0000-0002-0435-4493}}
\author[label3,label4]{Esin Soydugan\orcid{0000-0003-0321-3469}}
\author[label5]{Seval Taşdemir\orcid{0000-0003-1339-9148}}
\author[label6]{Remziye Canbay\orcid{0000-0003-2575-9892}}
\author[label5]{Deniz Cennet Çınar\orcid{0000-0001-7940-3731}}
\author[label7]{Seliz Koç\orcid{0000-0001-7420-0994}}
\author[label5]{Hülya Karagöz\orcid{0009-0004-5726-3749}}
\author[label1]{Burçin Tanık Öztürk\orcid{0000-0002-6372-2372}}
\author[label5]{Furkan Akbaba\orcid{0009-0002-9993-7244}}
\author[label5]{Özcan Çalışkan\orcid{0009-0003-5839-8007}}
\author[label8,label9]{Timothy Banks\orcid{0000-0001-9445-4588}}

\affiliation[label1]{
    organization={Istanbul University, Faculty of Science, Department of Astronomy and Space Sciences}, addressline={Beyaz\i t}, 
    city={Istanbul}, 
    postcode={34119}, 
    country={Turkey}}

\affiliation[label2]{
    organization={Istanbul University, Faculty of Science, Department of Computer Sciences}, addressline={Vezneciler}, 
    city={Istanbul}, 
    postcode={34134}, 
    country={Turkey}}

\affiliation[label3]{
    organization={\c{C}anakkale Onsekiz Mart University, Faculty of  Sciences, Department of Physics},
    city={\c{C}anakkale}, 
    postcode={17100}, 
    country={Turkey}}

\affiliation[label4]{
    organization={\c{C}anakkale Onsekiz Mart University, Astrophysics Research Center and Ulup{\i}nar Observatory}, 
    city={\c{C}anakkale}, 
    postcode={17100}, 
    country={Turkey}}

\affiliation[label5]{
    organization={Istanbul University, Institute of Graduate Studies in Science, Programme of Astronomy and Space Sciences}, 
    addressline={Beyaz\i t}, 
    city={Istanbul}, 
    postcode={34116}, 
    country={Turkey}}

\affiliation[label6]{
    organization={Akdeniz University, Department of Space Sciences and Technologies }, 
    addressline={Konyaalt\i}, 
    city={Antalya}, 
    postcode={07058}, 
    country={Turkey}}

\affiliation[label7]{
    organization={Istanbul Galata University, Faculty of Engineering and Natural Sciences, Department of Data Science and Analytics},
    city={Beyo\u{g}lu},
    state={Istanbul},
    postcode={34440},
    country={T\"{u}rkiye}}

\affiliation[label8]{
    organization={Komatsu},
    addressline={8770 W Bryn Mawr Ave, Suite 100},
    city={Chicago},
    state={IL},
    postcode={60631},
    country={USA}}
\affiliation[label9]{
    organization={Harper College, Department of Physical Science and Engineering},
    addressline={1200 W Algonquin Rd},
    city={Palatine},
    state={IL},
    postcode={60067},
    country={USA}}

\cortext[cor1]{Corresponding author}


\begin{abstract}

This study aims to establish a internally homogeneous set of structural, astrophysical, and kinematical parameters for King open clusters and to assess their role as tracers of the Galactic thin disc. We present a homogeneous structural, astrophysical, and kinematical analysis of 27 King open clusters using high-precision \textit{Gaia} DR3 astrometry, photometry, and spectroscopy. Probable members are selected through astrometric criteria, from which mean trigonometric parallaxes and proper motions are derived, while structural parameters are obtained by fitting King surface-density profiles. Cluster reddenings, distances, metallicities, and ages are determined via a homogeneous Markov Chain Monte Carlo isochrone-fitting approach applied to \textit{Gaia} colour-magnitude diagrams. The derived parameters span $E(G_{\rm BP}-G_{\rm RP}) = 0.113$-$1.933$~mag, metallicities of $\mathrm{[Fe/H]} = -0.40$ to $+0.28$~dex, ages from 17 to 6166~Myr, and heliocentric distances between 739 and 6272~pc. A comparison between astrometric and isochrone-based distances yields a median parallax difference of $-13~\mu$as, indicating a good level of internal consistency between the photometric and astrometric distance scales within the adopted homogeneous framework. Using the resulting homogeneous parameter set, we then investigate the spatial distribution and orbital properties of the King clusters within the Galactic disc.The radial metallicity behaviour of the King clusters is also examined using three complementary Galactic distance definitions and compared with the metallicity-radius relations obtained independently from 164 spectroscopically analysed open clusters from the literature. The King-cluster gradients are close to $-0.060$~dex~kpc$^{-1}$ and are broadly consistent with those of the external reference sample, indicating that the homogeneously analysed King clusters follow the negative metallicity-radius trend traced by the broader thin-disc open-cluster population. Flatter gradients are found for young clusters, suggesting more homogeneous chemical enrichment at recent epochs. When dynamical constraints are applied to reduce the effects of radial migration and orbital blurring, the inferred gradient remains close to $-0.059$~dex~kpc$^{-1}$. The use of a strictly homogeneous analysis framework reduces inter-study systematics and provides a consistent astrophysical, chemical, kinematical, and orbital reference dataset for the King open-cluster sample.

\end{abstract}

\end{frontmatter}
\clearpage




\section{Introduction} \label{sec:introduction}

Most stars in the Milky Way Galaxy are believed to have formed in open star clusters, which constitute young, physically associated Population I groupings within the Galaxy \citep{Lada_2003, Moraux_2016}. Open clusters (OCs) are gravitationally bound systems formed by the collapse of molecular clouds, leading to stellar populations with nearly common distances, ages, and chemical compositions. These properties make OCs fundamental tracers of the Galactic disc, providing key constraints on its structure, kinematics, and evolutionary history \citep{Gilmore_2012, Moraux_2016}. Their fundamental parameters are typically derived through isochrone fitting to colour–magnitude diagrams (CMDs), enabling reliable determination of distances, reddening, and ages.

Large catalogues compiling OC parameters \citep{Dias_2002, Kharchenko_2005, Kharchenko_2013, Cantat-Gaudin_2018, Liu_2019, Dias_2021, Hunt_2024} have enabled population-level studies of the Galactic disc. However, combining results from heterogeneous analyses introduces systematic differences that can limit the precision of large-scale Galactic investigations \citep{Bukowiecki_2011, Kharchenko_2013, Tadross_2014, Loktin_2017, Reddy_2020, Joshi_2023}. In this context, the availability of homogeneous, high-precision data from the \textit{Gaia} mission \citep{Gaia_DR1, Gaia_DR2, Gaia_EDR3, Gaia_DR3} represents a major step forward, significantly improving membership determination and the reliability of derived cluster parameters \citep{Soubiran_2019, Tarricq_2021, Hunt_2024}.

A key strength of \textit{Gaia} is the availability of trigonometric parallaxes, providing a direct distance scale across the Galactic disc. Nevertheless, these parallaxes are affected by small but non-negligible systematic errors, commonly expressed as a parallax zero-point offset, which depends on magnitude, colour, and sky position \citep[e.g.,][]{Lindegren_2021a, Akbulut_2021}. Since accurate distances underpin all derived cluster parameters and subsequent Galactic interpretations, ensuring an internally consistent distance scale is essential for any homogeneous OC analysis.

OCs are also among the most powerful tracers of the Galactic radial metallicity gradient, one of the key observables constraining chemical evolution models. Observations consistently indicate a negative gradient in the inner disc, typically between $-0.05$ and $-0.12$~dex~kpc$^{-1}$, with possible flattening in the outer regions \citep[e.g.,][]{Donor_2020, Netopil_2016, Jacobson_2016}. However, the observed dispersion reflects not only measurement uncertainties but also the effects of age differences, sampling, and dynamical processes such as radial migration. In particular, churning and blurring can displace clusters from their birth radii, complicating the interpretation of present-day gradients \citep{Sellwood_2002, Schonrich2009, Minchev_2013}. For this reason, analyses that incorporate guiding radii or reconstructed birth locations provide a more physically meaningful characterization of the Galactic disc.

The King clusters represent a distinct subset of Galactic OCs that remain relatively underexplored within a internally homogeneous \textit{Gaia}-based framework. Rather than analysing these systems individually, treating them as a uniform sample enables a systematic investigation that minimizes inter-study systematics and allows direct comparison with other OC populations. This approach also places the King clusters within a broader, coherent framework for studying Galactic structure in the \textit{Gaia} era. In this context, the present study may also be regarded as a modest tribute to Ivan R. King (1925–2021), whose empirical formulation of cluster density profiles continues to underpin modern analyses of stellar systems.

The primary aim of this study is threefold. First, we establish a strictly homogeneous set of structural, astrophysical, and kinematical parameters for all 27 catalogued King OCs using \textit{Gaia} DR3 astrometry and photometry, applying a unified MCMC isochrone-fitting framework that enables direct inter-cluster comparisons free of methodological systematics. Second, we assess the internal consistency of the \textit{Gaia} DR3 parallax scale for Galactic-plane clusters by comparing trigonometric parallaxes with independent MCMC-based photometric distances. Third, we use the resulting homogeneous King-cluster parameter set to examine the radial metallicity behaviour of this OC subgroup within the broader chemical structure of the Galactic thin disc. The King sample is not intended to independently redefine the global Galactic radial metallicity gradient, because its size is limited compared with recent large OC catalogues. Instead, we derive metallicity-radius relations for the King clusters and compare them with those obtained from the 164 spectroscopically analysed OCs of \citet{Otto_2026}, which are used as an external reference sample. This approach allows us to assess whether the homogeneously derived King-cluster metallicities follow the chemical pattern traced by recent large OC studies.
To account for possible dynamical effects associated with radial migration, the analysis is performed using three complementary definitions of Galactocentric distance: the present-day Galactocentric radius ($R_{\rm gc}$), the guiding radius ($R_{\rm gui}$), and the model-based early orbital/birth-radius ($R_{\rm teo}$). In this framework, the King sample provides a homogeneous benchmark for evaluating the chemical and dynamical consistency of this poorly consolidated OC subgroup within the Galactic thin disc, while the \citet{Otto_2026} sample serves as an independent reference against which the King-cluster trends are assessed.

The analysis pipeline consists of (i) membership determination using \textit{Gaia} DR3 astrometry, (ii) structural parameter estimation via radial density profiles, (iii) MCMC-based derivation of astrophysical parameters from CMDs, (iv) kinematic and orbital analysis, and (v) investigation of the Galactic metallicity gradient using the resulting homogeneous parameter set.

\section{Data}
\label{sec:data}

\begin{figure}
\centering
\includegraphics[width=0.95\linewidth]{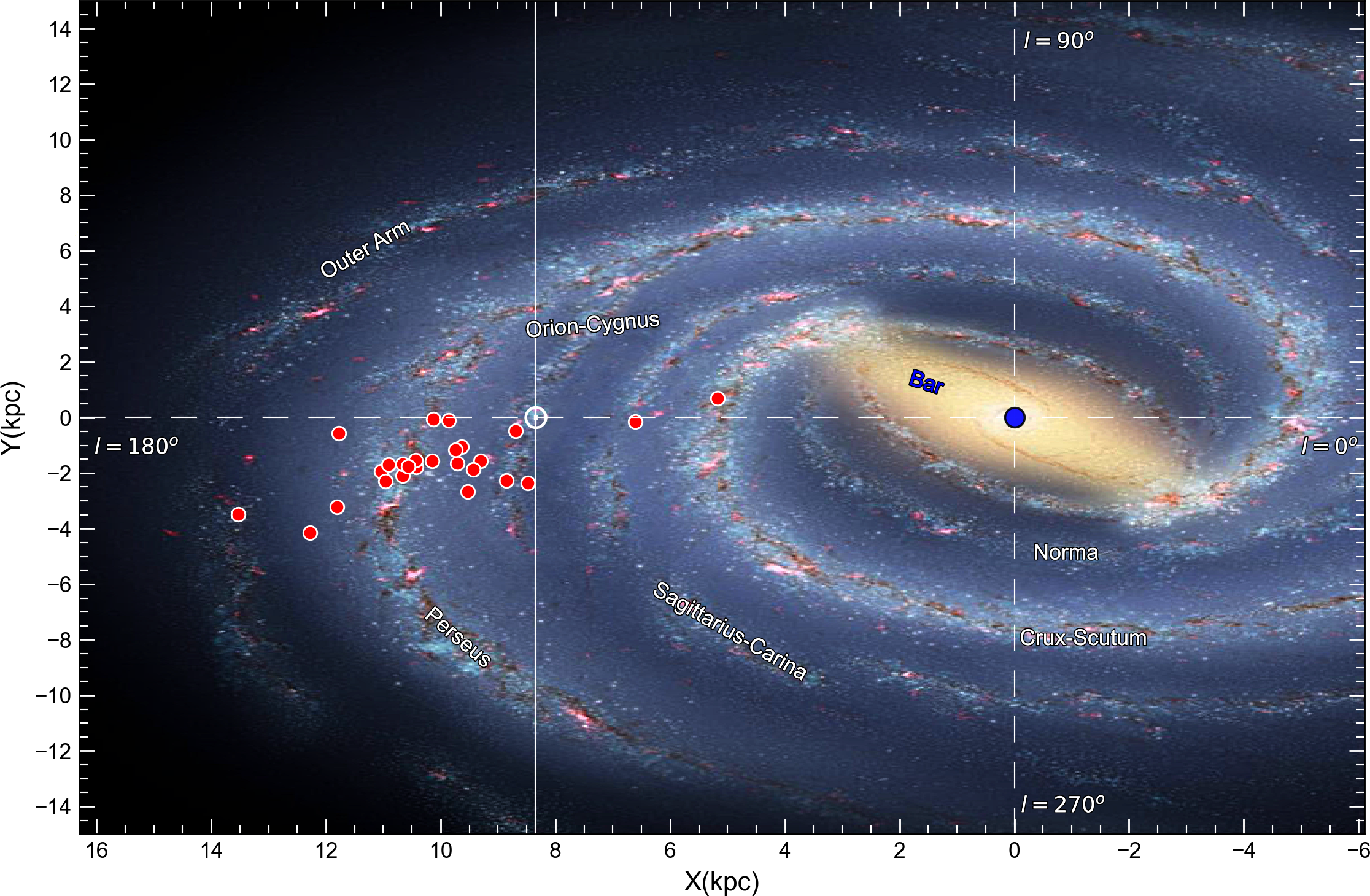} \vspace*{-5pt}
\caption{Positions of 27 King OCs on the Milky Way Galaxy.} 
\label{fig:positions}
\end {figure}

In this study, we selected the 27 King OCs from recently published catalogues \citep{Cantat-Gaudin_2020, Hunt_2024}. Their positions in the Milky Way Galaxy are shown in Figure~\ref{fig:positions}. For each King OC, we selected stars from the {\it Gaia} DR3 \citep{Gaia_DR3} database using the central coordinates reported by \citet{Cantat-Gaudin_2020}. Following this, catalogues of these clusters were constructed for $40 \times 40$ arcmin cluster fields by using astrometric, photometric, and spectroscopic data of {\it Gaia} DR3. However, the analysis was not restricted to stars that simultaneously have complete astrometric, photometric, and spectroscopic data. Astrometric and photometric parameters were primarily used for membership determination and CMD analyses, while spectroscopic data (when available) were incorporated only in the relevant parts of the analysis. Therefore, stars lacking spectroscopic measurements were still included in the astrometric and photometric analyses. 

\begin{figure*}
\centering
\includegraphics[width=0.85\linewidth]{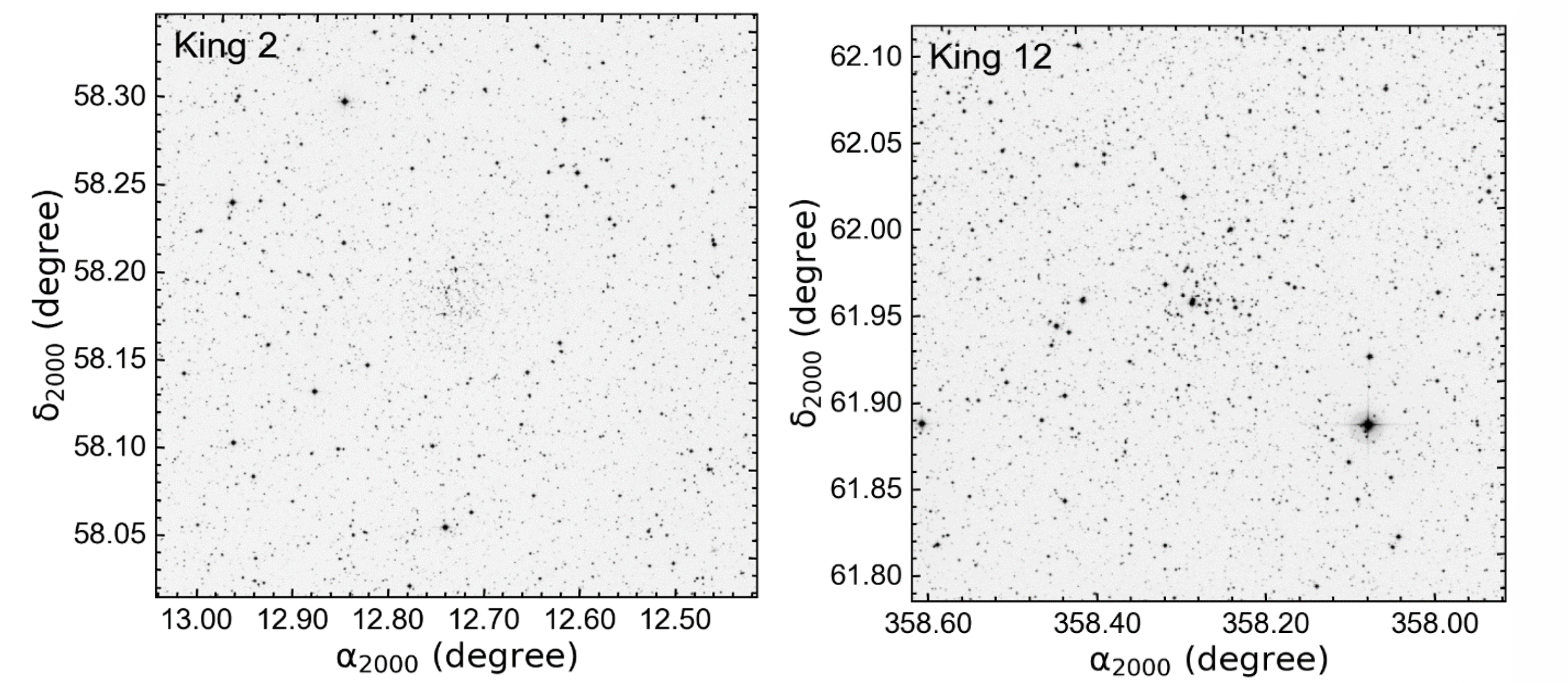}
\caption{Identification maps for King 2 (left panel) and King 12 (right panel). The field of view of the charts is $40' \times 40'$. North is towards up, and East is leftwards.} 
\label{fig:charts}
\end {figure*}

This uniform field size ensures a homogeneous selection function across the sample and provides a sufficiently wide surrounding region for field-star characterization. For all clusters, the derived limiting radii lie well within the adopted fields, indicating that the structural fits are not affected by field boundaries. Any extremely sparse outlying members in the nearest systems do not influence the derived structural or fundamental cluster parameters. To illustrate the analyses performed in this paper, we focus on King 2 and King 12 as examples, representing the oldest and youngest clusters in our sample, respectively. These two systems highlight the broad age range of the 27 King OCs studied. The number of stars within the applied cluster areas spans from 5,905 to 211,958, with 91,523 stars detected for King 2 and 12,861 for King 12. Their identification charts are shown in Figure~\ref{fig:charts}.

\begin{figure}[!t]
\centering
\includegraphics[width=0.98\columnwidth]{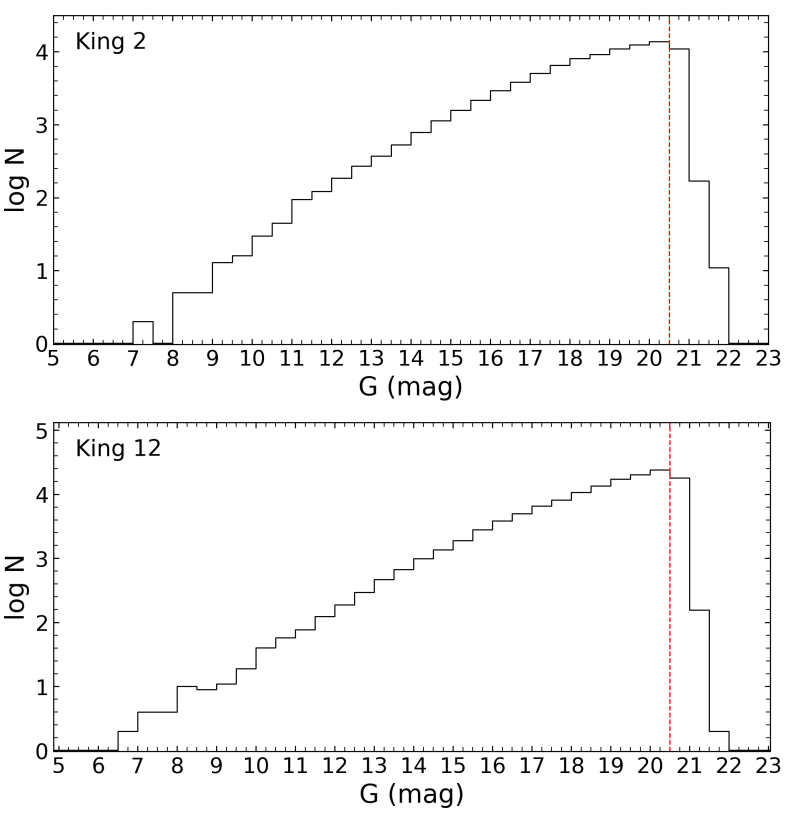}
\caption{Stellar counts of King 2 and King 12 for per magnitude bin in $G$ bands. The vertical red dashed lines show the adopted faint limiting apparent magnitudes $G$ bands.} 
\label{fig:histograms}
\end {figure}

The catalogues for the King OCs contain the stellar equatorial coordinates ($\alpha, \delta$), apparent $G$ magnitude and $G_{\rm BP}-G_{\rm RP}$ colour index, proper-motion components ($\mu_{\alpha}\cos\delta, \mu_{\delta}$), trigonometric parallax ($\varpi$) along with the radial velocity ($V_{\rm R}$) and membership probability ($P$) as calculated in the study. Before starting the analyses, as described in the following sections, we limited the photometric data to stars with reliable magnitudes and colour indices by applying a faint $G$-band limiting magnitude based on the completeness of the data. This selection was achieved by estimating the faint $G$ limiting magnitudes for each cluster. To do this, we constructed the number of stars as a function of $G$ magnitudes, as shown for King 2 and King 12 in Figure~\ref{fig:histograms}. In the figure, the number of stars increases toward the fainter magnitudes and drops after reaching a maximum. This point of ``turn over'' or maximum was defined as the faint $G$ limiting magnitude for a given cluster. These limits are $G = 20.5$ mag for all the King OCs in this study, and the uncertainties for faint limit magnitude in the $G$ magnitudes and the colour indices $G_{\rm BP}-G_{\rm RP}$ reach 0.005 and 0.102 mag when we consider the data of all the King OCs.

\section{Structural Parameters}
\label{sec:structural}

\begin{figure*}
\centering
\includegraphics[width=0.85\linewidth]{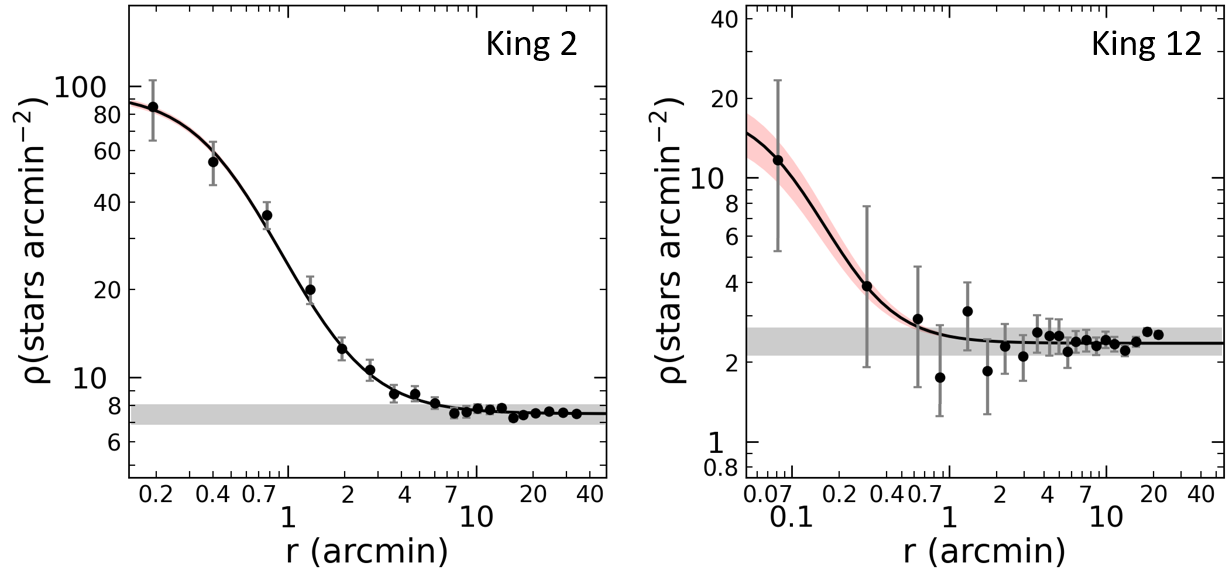}
\caption{The stellar density distribution of King 2 and King 12. The fitted black curve shows the empirical RDP profile} of \citet{King_1962}, whereas the horizontal gray band represents the background stellar density. The $1\sigma$ King fit uncertainty is pictured by the red-shaded area. 
\label{fig:rdps}
\end{figure*}

To determine the limiting radii of the King OCs and obtain their structural parameters, we utilized Radial Density Profile (RDP) analyses using {\it Gaia} DR3 data within 40 arcmin of the cluster centers given by \citet{Cantat-Gaudin_2020}. Considering the stars brighter than the $G$ limiting magnitude, we divided the field of view into concentric rings and calculated the stellar number density ($\rho(r)$) for each $\it i$th ring as $R_{i}=N_{i}/A_{i}$, where $N_{i}$ is the number of stars in that ring and $A_{i}$ is the ring area. We used Poisson statistics ($1/\sqrt N$, where $N$ indicates the number of stars) to calculate the uncertainties in the stellar number densities. To visualize the radial density profile, we fitted the empirical RDP of \citet{King_1962} to the calculated stellar number densities. This expression is an approximation to the empirical King surface-density profile and is mainly applicable to the central region of the cluster, sufficiently far from the tidal radius.

\begin{equation}
\rho(r)=f_{\rm bg}+\frac{f_0}{1+(r/r_{\rm c})^2}
\end{equation}
where $r$ represents the radius, $f_{\rm bg}$ the background stellar density, $f_0$ the central stellar density, and $r_{\rm c}$ the core radius of the OC. Finding the best fitting models involved $\chi^{2}$ minimization. RDPs of the King 2 and King 12 OCs are shown in Figure~\ref{fig:rdps}, where a smooth, continuous line is the best-fit solution of the empirical King profile. The correlation coefficients ($R^2$) of the best-fit solutions are higher than 0.90 for all of the King OCs. The estimated structural parameters for each cluster are listed in Table~\ref{app:fundamental}.

We determined limiting radii $r_{\rm lim}^{\rm obs}$ visually according to the RDPs of each OC. We considered the radius at which the RDP flattens and begins to merge with the background stellar density as being the extent of an OC. The estimated limiting radii for all the studied King OCs were in the range of $8 \leq r_{\rm lim}^{\rm obs} \leq 25$ arcmin. For King 2 and King 12, these values were determined as 15 and 10 arcmin, respectively. Observed radii for all King OCs are also listed in Table~\ref{app:fundamental}. 

The concentration parameter $C=\log(r_{\rm lim}/{r_{\rm c}})$ is a key indicator of the structural properties of OCs, providing insights into their internal stellar distributions. Here, $r_{\rm lim}$ is the limiting radius, adopted as $r_{\rm lim}^{\rm obs}$ in this study. The parameter $C$ is also useful for distinguishing OCs from the background field stars, as it quantifies the degree of central concentration of the cluster \citep{Richstone_1986}. We calculated the $C$ concentration parameter for all King OCs, finding the range of these values to be within $0.21\leq C \lesssim 2.01$. For King 2 and King 12, the $C$ parameter was calculated to be 1.49 and 2.01, respectively. The $C$ values calculated for the King OCs are listed in Table~\ref{app:fundamental}. When the distribution of clusters across the Galactic quadrants on the Galactic plane is examined, it is found that three OCs are located in the first quadrant, with a median $C$ value of 0.77. The second quadrant contains 22 OCs, for which the median $C$ value is calculated to be 1.035. In the third quadrant, only two OCs are identified, yielding a median $C$ value of 1.11. 

An evaluation based on Galactic quadrants reveals that clusters located in regions farther from the Galactic center exhibit systematically higher $C$ values compared to those situated in the inner Galactic regions. This trend is consistent with previous studies demonstrating that dynamical processes such as enhanced tidal forces, interactions with molecular clouds, and disc shocking become increasingly significant toward the inner Galaxy, strongly influencing the structural evolution of OCs \citep[e.g.,][]{Spitzer_1958, Gieles_2006, Lamers_2006}.


\section{Membership analyses and Astrometric Parameters}
\label{sec:membership}

The presence of background/foreground stars across a studied OC affects the reliable estimation of astrophysical parameters, being contaminants. Proper-motion components of stars are one useful tool to separate OC members from field stars. Thanks to the precise astrometric data of {\it Gaia} DR3, members of the OC can be identified reliably. We used the Photometric Membership Assignment in stellar Cluster \citep[{\sc upmask};][]{Krone-Martins_2014} method on {\it Gaia} DR3 astrometric data to calculate membership probabilities of individual stars located in the direction of studied clusters. This method was performed previously by our group and various researchers for OCs \citep[e.g.,][]{Cantat-Gaudin_2018, Cantat-Gaudin_2020, Castro-Ginard_2018, Wang_2022, Yontan_2021, Yontan_2022, Yontan-Canbay_2023, Cinar_2024, Cakmak_2024, TasdemirCinar_2025}. 

\begin{figure}[!t]
\centering
\includegraphics[width=0.8\linewidth]{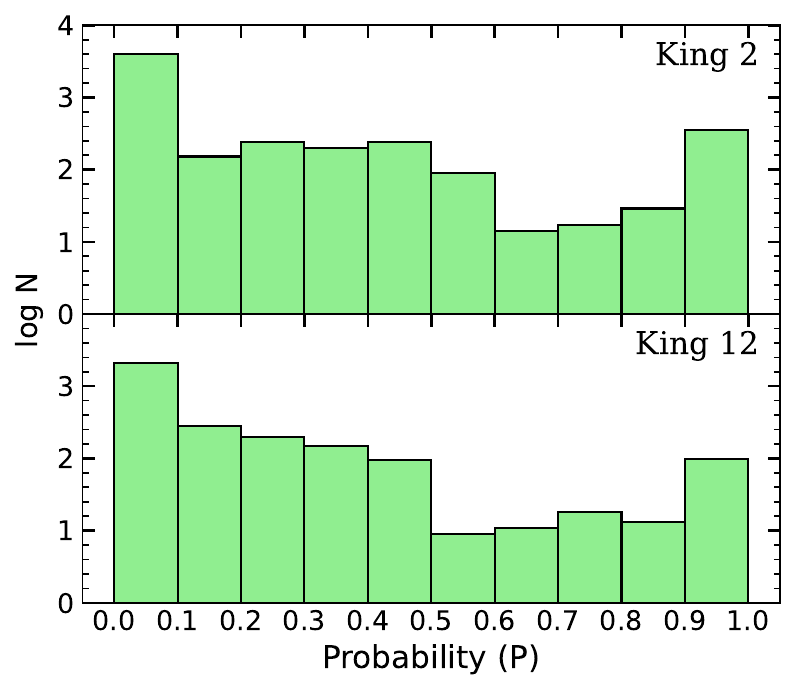}
\caption{Histograms of the {\it Gaia}-based membership probabilities for sources within the observational limiting radii ($r_{\rm lim}^{\rm obs}$) of the two clusters. The upper panel shows King~2, while the lower panel shows King~12.}
\label{fig:probability_hist}
\end{figure}

{\sc upmask} is based on the $k$-means clustering method. This technique accounts for measurements of proper-motion components, together with trigonometric parallaxes and their uncertainties, and detects similar groups of stars, providing statistical membership probabilities for these groups. Detailed background on {\sc upmask} can be found in \citet{Krone-Martins_2014}. We calculated the membership probabilities ($P$) of each detected star by using their astrometric parameters and uncertainties ($\alpha$, $\delta$, $\mu_{\alpha}\cos \delta$, $\mu_{\delta}$, $\varpi$) along with 100 iterations of {\sc upmask}. Then, we selected the stars with probabilities $P \geq 0.5$ as possible cluster members. In addition to the statistical method, we selected member stars based on photometric criteria. This took into account the possible binary star contamination in the main sequence, an effect which widens the main sequence of OCs. We considered the distribution of OC stars with membership probabilities $P \geq 0.5$ on {\it Gaia} photometry-based CMDs and identified the lower bound of the cluster's main sequence by fitting {\sc PARSEC} models \citep{Bressan_2012}. After this, we shifted the lower bound by $+0.75$ mag through the brighter stars and determined the upper bound for the main sequence of the clusters that cover the possible binary star contamination. In addition to this criterion, stars with membership probabilities $P \geq 0.5$, brighter than the $G$ limiting magnitude, and located within the observationally determined limiting radius ($r_{\rm lim}^{\rm obs}$) were classified as being probable cluster stars and used in further analyses in this study. With these limitations described above, we identified 474 and 141 most probable member stars for King 2 and King 12, respectively (for the number of most probable member stars of the 27 King OCs, see Table~\ref{app:fundamental}).
Figure~\ref{fig:probability_hist} shows the histograms of the computed membership 
probabilities for stars brighter than the $G$-band completeness limit and located within the $r_{\rm lim}^{\rm obs}$ radius of King~2 and King~12. Thus, the histograms refer to the same spatial and photometric 
sample used for the final membership selection. Within these samples, stars 
satisfying $P \geq 0.5$ constitute approximately 9\% and 5\% of all stars with 
$0 < P \leq 1$ for King~2 and King~12, respectively. The relatively small fractions 
of high-probability members indicate that both cluster regions are strongly affected 
by field-star contamination.

To picture the proper-motion (PM) distribution and movement vectors of the most probable cluster members versus the field stars, we plotted vector point diagrams (VPDs) and profiles of PM vectors and calculated mean PM components for each studied cluster. We concluded that the PM components of the most probable member stars are found to be within a radius of at most 0.5 mas yr$^{-1}$ from the mean PM components of the King OCs in which they are located. 

\begin{figure*}[!t]
\centering
\includegraphics[width=0.8\textwidth]{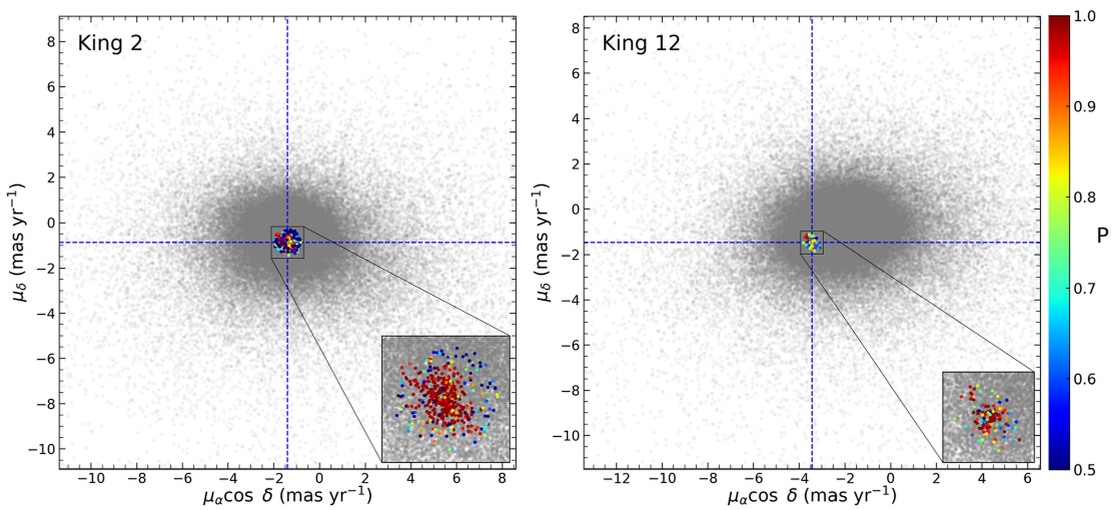}
\caption{Left and right panels show the VPDs for the King 2 and King 12, respectively. The most probable cluster members ($P\geq0.5$) are presented by colour-scaled circles. The zoomed-in boxes indicate the distribution of the OCs according to field stars (gray circles). The intersection of blue dashed lines shows the point of the mean PM values of the OCs.}
\label{fig:VPD_all} 
\end{figure*}

The VPDs of the King 2 and King 12 OCs are shown in the left and right panels of Figure~\ref{fig:VPD_all}, respectively. It can be seen from the panels that both OCs are embedded in the field stars (gray circles); however, the concentration of most probable member stars ($P\geq0.5$) (the colour-scaled circles) indicates that the clusters are distinct from the field stars. The mean PM values, which are presented by the intersection of the blue dashed lines in Figure~\ref{fig:VPD_all}, were obtained from the most probable OC members, and the results are listed in Table~\ref{app:fundamental} for each King OC. We derived mean PM values for King 2 as $\mu_{\alpha}\cos \delta = -1.398 \pm 0.005$ and  $\mu_{\delta} = -0.876 \pm 0.008$ mas yr$^{-1}$ and for King 12 as $\mu_{\alpha}\cos \delta = -3.437 \pm 0.003 $ and $\mu_{\delta} = -1.471 \pm 0.003$ mas yr$^{-1}$. These results match well with the values of \citet{Cantat-Gaudin_2020}.

\begin{figure}[!t]
\centering
\includegraphics[width=0.7\linewidth]{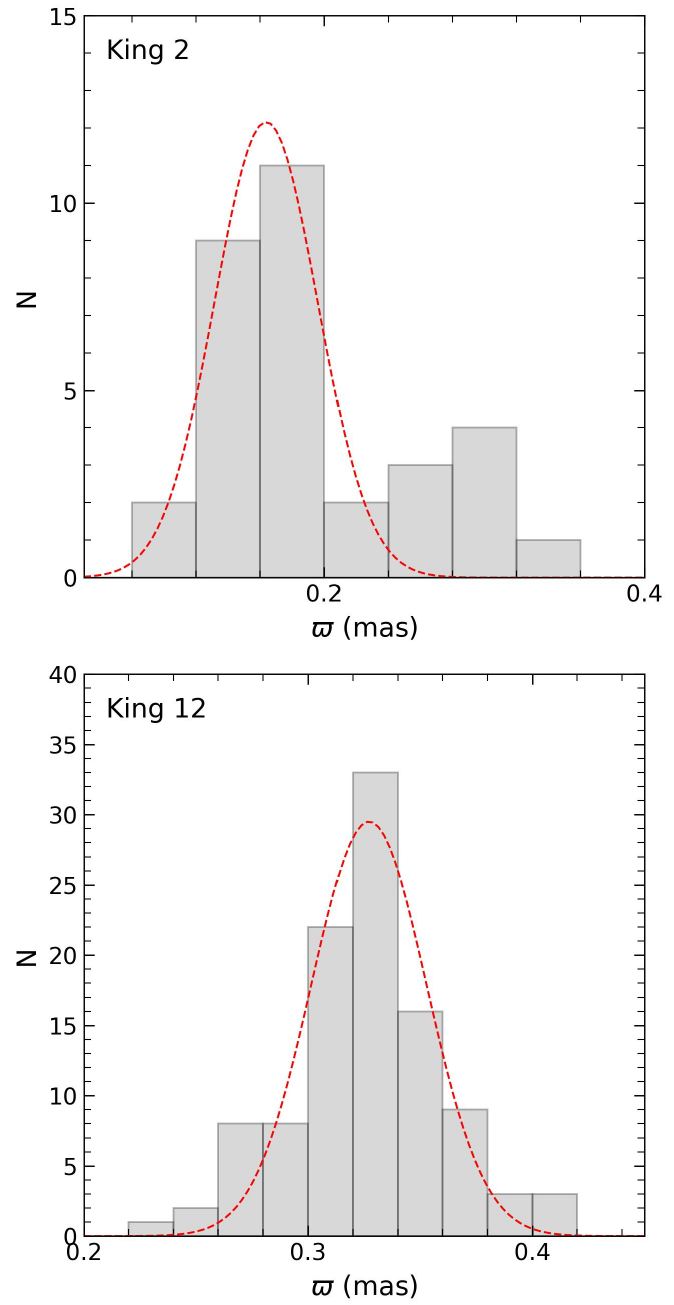}
\caption{Trigonometric parallax histograms for King 2 and King 12. The plots are based on cluster stars with membership probabilities $P \geq 0.5$, located within the clusters' limiting radii and with $\sigma_{\varpi} / \varpi \leq $ 0.25. The red dashed curves represent Gaussian fits. $N$ is the number of stars.
\label{fig:plx_hist} }
\end{figure}

\begin{figure*}
\centering
\includegraphics[width=\linewidth]{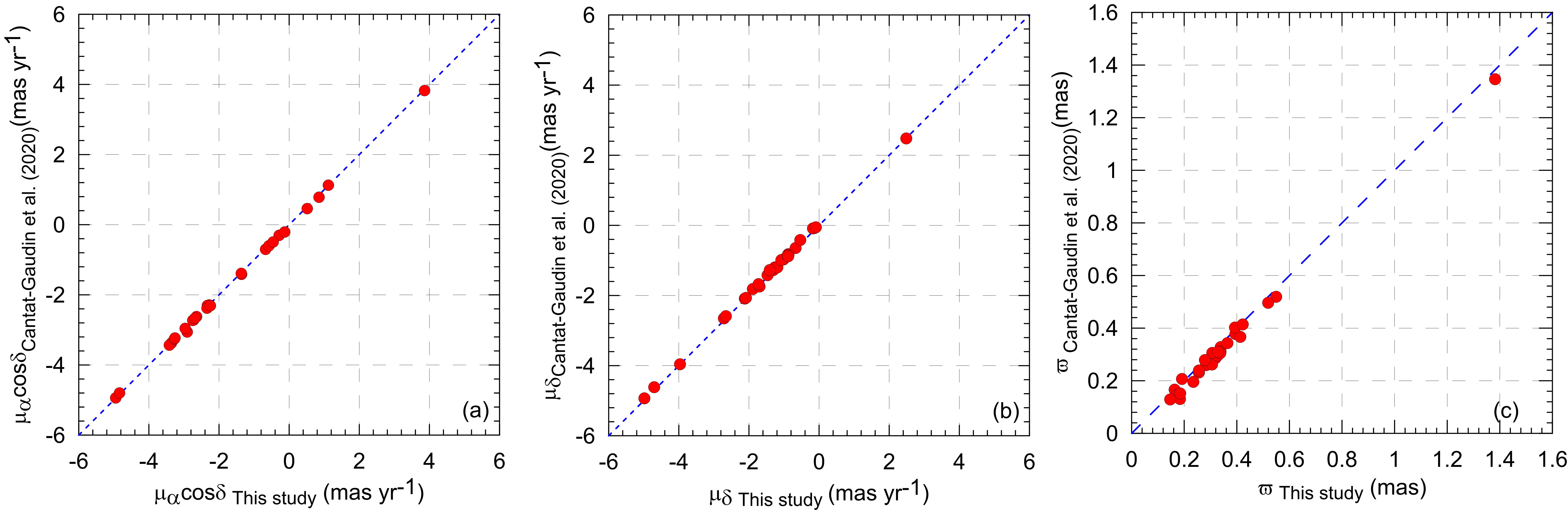}\\
\caption{Comparison of the mean PM components in RA (panel a) and DEC (panel b) directions as well as trigonometric parallax (panel c) of 27 King OCs, calculated using {\it Gaia} DR3 astrometric data, with those derived from {\it Gaia} DR2 data by \citet{Cantat-Gaudin_2020}. The dashed blue line represents the one-to-one line.}
\label{fig:ozhrkt_plx} 
\end{figure*}

The mean trigonometric parallax of each OC was determined from the most probable member stars ($P \geq 0.5$) with relative parallax errors ($\sigma_{\varpi}/\varpi $) smaller than 0.25. For each OC, we constructed a histogram of trigonometric parallaxes and fitted a Gaussian function to this distribution, as shown in Figure~\ref{fig:plx_hist}. For King 2, the trigonometric parallax histogram reveals the presence of a second population beyond approximately 0.3 mas. The analyses indicate that this structure is characterized by relative parallax errors larger than 0.15, and therefore does not statistically affect the mean trigonometric parallax value of King 2. We also applied the linear equation of $d({\rm pc})=1000/\varpi$ (mas) to the estimated mean trigonometric parallaxes and calculated the astrometric distances ($d_{\varpi}$) for each King OC. The mean trigonometric parallaxes for King 2 and King 12 were derived as $\varpi=0.164 \pm 0.032 $ and $\varpi=0.327 \pm 0.025$ mas, respectively. The corresponding astrometric distances are $d_{\varpi} = 6098\pm 1190$ and $d_{\varpi} = 3058\pm 234$ pc. The histograms of the trigonometric parallaxes are shown in Figure~\ref{fig:plx_hist} for King 2 and King 12.

We compared our astrometric results with those of \citet{Cantat-Gaudin_2020} to assess the consistency of our measurements. Figure~\ref{fig:ozhrkt_plx} presents the outcomes of this comparison, where the abscissa corresponds to the mean PM components, in RA and DEC directions, and trigonometric parallaxes derived in this study, while the ordinate represents the values reported by \citet{Cantat-Gaudin_2020}. The mean differences, calculated as our results minus those of  \citet{Cantat-Gaudin_2020}, are $-0.020$ mas yr$^{-1}$, $-0.047$ mas yr$^{-1}$ and $24~\mu$as for $\Delta \mu_{\alpha}\cos \delta$ (panel a), $\Delta \mu_{\delta}$ (panel b) and $\Delta \varpi$ (panel c), respectively. The corresponding standard deviations are $\sigma_{\mu_{\alpha}\cos \delta}=0.034$ mas yr$^{-1}$, $\sigma_{\mu_{\delta}} =0.038$ mas yr$^{-1}$ and $\sigma_{\varpi} =16~\mu$as. The astrometric data obtained from {\it Gaia} DR3 exhibit greater consistency and reduced systematic errors compared to {\it Gaia} DR2. The extended 34-month observational period of {\it Gaia} DR3, in contrast to the 22 months covered by {\it Gaia} DR2, contributes to improved precision and accuracy. In our comparison with \citet{Cantat-Gaudin_2020}, the observed median parallax difference ($\Delta \varpi = 24 \pm 16~ \mu$as) between {\it Gaia} DR3 and {\it Gaia} DR2, while small, appears to be systematic. This may indicate the presence of a small-scale zero-point offset in {\it Gaia} DR3, indicating that source-dependent corrections are more appropriate than a single, fixed correction. For precise astrometric studies, such as those involving OCs, additional analyses using {\it Gaia} DR3’s parallax zero-point correction models should be conducted, and the error analyses should be reassessed accordingly.

\section{MCMC-based Astrophysical Parameter Estimation}
\label{sec:parameter}

Accurate determination of fundamental astrophysical parameters such as age, distance, reddening, metallicity, and mass is essential for constraining stellar-evolution models and improving our understanding of the chemical and dynamical evolution of the Galaxy. The analysis of photometric and spectroscopic data using isochrone-fitting techniques and probabilistic methods enables robust estimation of these parameters. Recent advances in wide-field sky surveys and high-precision astrometric measurements have significantly enhanced the accuracy with which the physical properties of OCs can be derived \citep[e.g.,][]{Speagle_2019, Spitoni2020, Tasdemir_2023, Karagoz_2025, Tasdemir_2025, Tasdemir_2026}. Consequently, the application of homogeneous and consistent methodologies in parameter determination has become a critical requirement for reliable large-scale Galactic studies. A detailed description of the adopted methodology is provided in our previous studies \citep[e.g.,][]{Bilir_2006a, Bilir_2010, Bilir_2016, Bilir_2026, Yontan_2015, Yontan_2019, Bostanci_2015, Bostanci_2018, Ak_2016, Ak_2024, Canbay_2026}.

To obtain a homogeneous set of astrophysical parameters for the 27 King OCs, we apply a Markov Chain Monte Carlo (MCMC) isochrone–fitting procedure to their {\it Gaia} DR3 CMDs. 
This approach enables the distance ($d$), extinction in the $G$ band ($A_{\rm G}$), age ($\tau$), and metallicity ($Z$) of each cluster to be constrained simultaneously while accounting for parameter covariances intrinsic to stellar-evolution models. By deriving all parameters within a single methodological scheme, this analysis provides a uniform set of cluster properties that avoids the heterogeneity introduced by combining results from different literature sources. The following subsections describe the construction of the isochrone grid, the likelihood formulation, and the sampling strategy adopted to obtain the final posterior distributions of the astrophysical parameters.

The method is built upon a pre-computed grid of {\sc parsec} stellar isochrones created through the CMD~3.9 interface\footnote{\url{http://stev.oapd.inaf.it/cgi-bin/cmd}} \citep{Bressan_2012}. The grid covers a broad range of evolutionary stages, with $\log \tau$ sampled from 6 to 10.13 in intervals of 0.02 dex, and heavy-element mass fraction $Z$ spanning 0 to 0.03 with steps of 0.0005. This resolution provides adequate coverage of the parameter space relevant to the King OCs, which contains young to intermediate-age systems with varying chemical compositions. As the MCMC sampler explores a continuous parameter space, interpolated isochrones are required for values lying between the tabulated grid points. We perform this interpolation using Delaunay triangulation implemented in the \texttt{Scipy} library \citep{Scipy}. This approach ensures that the isochrone grid is smoothly sampled and minimizes artifacts that can arise when using coarse or irregular grids. The free parameters of the model are ($d, A_{\rm G}, \tau, Z$) sampled within broad priors designed to encompass all astrophysically plausible solutions $0 \leq d~\rm (kpc) \leq 30$ and $0 \leq A_{\rm G} \rm~(mag) \leq 10$ with $\log\tau$ and $Z$ restricted to the isochrone grid. These broad limits guarantee that physically plausible solutions for all clusters remain within the explored domain, while the posterior distribution naturally concentrates around the parameter combinations favored by the observations.

For each trial parameter set, synthetic magnitudes are computed in the {\it Gaia} $G$, $G_{\rm BP}$, and $G_{\rm RP}$ passbands via:
\begin{equation}
    m_{X} = M_{X}(\tau, Z) + 5 \log d - 5 + A_{X},
\end{equation}
where $M_{X}(\tau, Z)$ is obtained from the interpolated isochrones, and the extinction in each band is expressed as  $A_{X} = k_{X} A_{\rm G}$ using extinction coefficients $k_{X}$ derived from the reddening laws of \citet{Cardelli89} and \citet{ODonnell94} for $R_{\rm V}=3.1$. This forward-modeling step ensures that the observed CMD is compared directly with synthetic photometry that reflects the full set of input parameters.

Assuming Gaussian uncertainties in the {\it Gaia} photometry, the likelihood of observing magnitudes $m_{X,i}$ for star $i$ is \citep[see also,][]{Tanik2025, Tasdemir_2026}:
\begin{equation}
    P(m_{X,i} | \theta) = 
    \frac{1}{\sqrt{2\pi \sigma_{X,i}^2}}
    \exp\left[-\frac{\left(m_{X,i}-\hat{m}_{X,i}\right)^2}{2\sigma_{X,i}^2}\right],
\end{equation}
where $\hat{m}_{X,i}$ is the model-predicted magnitude. The full likelihood for all stars and bands is:

\begin{equation}
    \mathcal{L}(\theta) = 
    \prod_{i=1}^{N_{\rm stars}}
    \prod_{X \in \{G, G_{\rm BP}, G_{\rm RP}\}}
    P(m_{X,i}|\theta),
\end{equation}
and the log-likelihood used for numerical optimization is:

\begin{equation}
    \ln \mathcal{L}(\theta) = 
    -\sum_{i=1}^{N_{\rm stars}} 
    \sum_{X}
    \left[
        \frac{\left(m_{X,i}-\hat{m}_{X,i}\right)^2}{2\sigma_{X,i}^2} +
        \ln(\sqrt{2\pi}\sigma_{X,i})
    \right].
\end{equation}

This formulation naturally gives less weight to low-S/N photometry and reduces the impact of outliers. Posterior distributions are explored using the affine-invariant ensemble sampler {\sc emcee} \citep{Foreman-Mackey2013}. For each cluster, an ensemble of walkers is initialized near reasonable astrophysical starting points and evolved for 5000 steps following a burn-in phase. Convergence is assessed through the stabilization of autocorrelation times and visual inspection of chain evolution. Final parameter estimates are adopted as the median of the marginalized posterior distributions, with uncertainties given by the 16th and 84th percentiles.

\begin{figure*}[!t]
\centering
\includegraphics[width=0.9\textwidth]{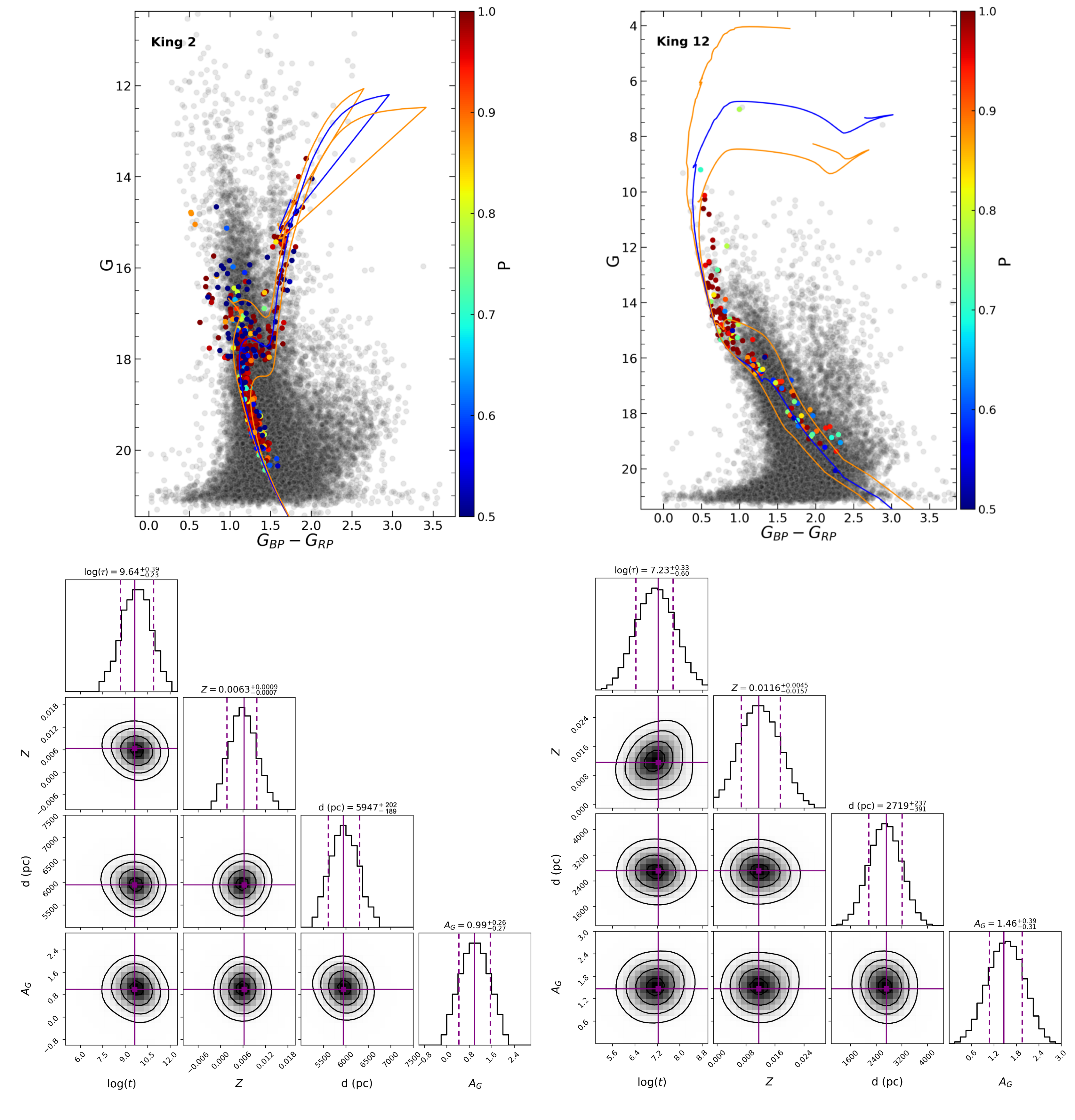}
\caption{CMDs of the King 2 and King 12 OCs showing the member stars. Stars with $P\geq 0.5$ are shown in different colours, while field stars are displayed as grey circles. The best-fitting PARSEC isochrones are shown as blue curves, and the age uncertainties are indicated by orange curves (upper panels). The corner diagram illustrating the posterior distributions of the inferred astrophysical parameters for King 2 and King 12 derived from the \texttt{MCMC} analysis. The solid purple lines indicate the median values of the parameters, while the dashed purple lines represent the corresponding 16th and 84th percentile confidence intervals (lower panels).
\label{fig:age_cmds} }
\end{figure*}

The upper panels of Figure~\ref{fig:age_cmds} illustrate the MCMC-derived isochrone fits for King~2 and King~12. Membership probabilities are indicated by symbol colour, while low-probability stars (\(P<0.5\)) and field stars (\(P=0\)) are shown in grey. The blue curves denote the best-fitting isochrones, and the orange curves represent the \(1\sigma\) uncertainty envelopes. The two-dimensional posterior projections display contour levels corresponding to approximately \(68\%\), \(90\%\), and \(95\%\) of the marginalized probability. In the one-dimensional marginalized distributions, the 16th, 50th, and 84th percentiles are indicated by vertical lines, marking the median and the associated \(1\sigma\) confidence intervals, as shown in the lower panels of Figure~\ref{fig:age_cmds}. The corresponding CMDs and MCMC-derived best-fitting isochrones for the remaining King OCs, excluding King~2 and King~12, are presented in Figure~\ref{app:all_cmds}.

The MCMC returns metallicity in terms of heavy-element mass fraction $Z$, which we convert to logarithmic [Fe/H] using the {\sc parsec}-consistent formulation implemented by Bovy's equations\footnote{\url{https://github.com/jobovy/isodist/blob/master/isodist/Isochrone.py}}. The conversion uses an intermediate variable $Z_{\rm x}$ together with the solar heavy–element abundance $Z_{\odot}$ and is given by \citep[see also][]{Gokmen_2023, Yontan_2023a, Yontan_2023b}:

\begin{equation}
Z_{\rm x} = \frac{{\rm Z}}{0.7515 - 2.78 \times {\rm Z}}
\end{equation}
followed by the calculation of the OC iron abundances:
\begin{equation}
{\rm [Fe/H]} = \log \left({\rm Z_{\rm x}} \right) - \log \left( \frac{Z_{\odot}}{1 - 0.248 - 2.78 \times Z_{\odot}} \right)
\end{equation}
We adopt $Z_{\odot} = 0.0152$ for the solar metallicity \citep{Bressan_2012}. These relations allow the MCMC-derived $Z$ values to be mapped onto the [Fe/H] scale used in Galactic chemical evolution studies.

\begin{figure*}[t]
\centering
\includegraphics[width=\linewidth]{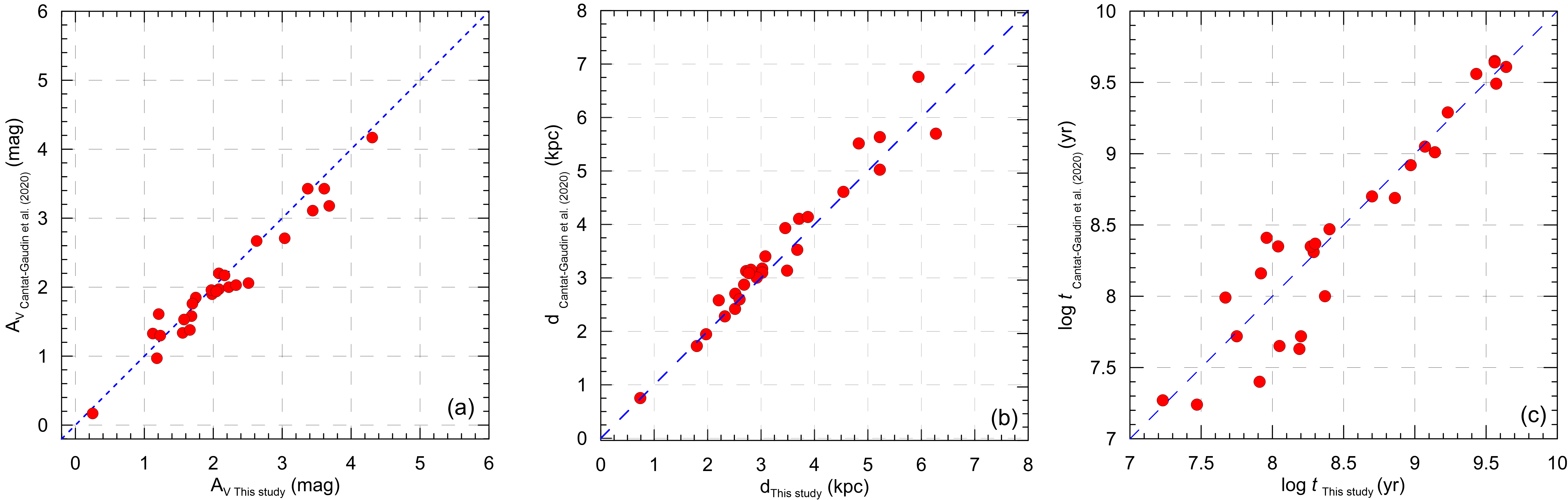}\\
\caption{Comparison of the calculated $V$-band extinction (a), distance (b), and age (c) values of the 27 King OCs with the results derived from {\it Gaia} DR2 data by \citet{Cantat-Gaudin_2020}. The blue dashed line represents the one-to-one line. The $V$-band extinction calculated using equation of $A_{\rm V}$=$A_{\rm G}$/0.83627.}
\label{fig:logt-FeH-d-Av} 
\end{figure*}

To convert the extinction and distance parameters obtained from the MCMC analysis into the colour excess $E(G_{\rm BP}-G_{\rm RP})$ and the distance modulus $\mu_{\rm G}$, we use the relations $A_{\rm G} = 1.8626 \times E(G_{\rm BP}-G_{\rm RP})$ \citep{Canbay_2023} and $\mu_{\rm G} = 5\log d - 5 + A_{\rm G}$. These transformations ensure consistency between the extinction-corrected MCMC distances and the photometric CMD fits. Using these relations, the inferred mass fractions for King~2 and King~12 correspond to metallicities consistent with the ranges obtained from the posterior distributions. The Galacto\-cen\-tric $(X, Y, Z)$ coordinates (where $X, Y$ and $Z$ are the distances towards the Galactic center, toward the Galactic rotation, and toward the Galactic north pole, respectively) of the OCs were also calculated \citep[see also,][]{Iyisan_2025, Canbay_2025, Caliskan_2025}. We used isochrone-based distances and the Galactic coordinates ($l, b$) of all clusters to perform the calculations. The parameter estimates for all clusters under study are listed in Table~\ref{app:fundamental}.

We compared astrophysical parameters $A_{V}$, $d$, and $\log (\tau)$ estimated in this study with the results of \citet{Cantat-Gaudin_2020}, which Figure~\ref{fig:logt-FeH-d-Av} presents in panels a, b, and c, respectively. The mean differences and standard deviations, calculated as our results minus those of \citet{Cantat-Gaudin_2020}, are $\langle \Delta A_{\rm V} \rangle =0.098 \pm 0.199$ mag, $\langle \Delta d \rangle =-151 \pm 299$ pc, and $\langle \Delta \log (\tau) \rangle =0.041 \pm 0.249$ yr. Astrophysical parameter comparisons show that the mean offsets are generally small and lie within the range expected from methodological differences between studies. The relatively large standard deviations, particularly in distance and extinction, are likely caused by differences in membership selection, extinction laws, and isochrone sets adopted in the compared catalogs. Regarding age estimates, clusters with $\log (\tau) > 8.5$ show better agreement between the two studies, whereas clusters with $\log (\tau) < 8.5$ exhibit a significantly larger scatter. This increased dispersion is most likely due to the higher sensitivity of young cluster parameter determinations to membership selection, since differences in the inclusion or exclusion of faint and kinematically marginal stars between the two studies can significantly affect isochrone fitting results. In contrast, older clusters show more consistent agreement, reflecting more stable and well-defined member distributions, which leads to more reliable parameter estimates.
Overall, this comparison provides an internal consistency check and supports the reliability of the derived astrophysical parameters within a homogeneous analysis framework.


\section{Internal Consistency of the Gaia DR3 Parallax Scale}

Small but non-negligible zero-point offsets are known to affect {\it Gaia} EDR3 and DR3 trigonometric parallaxes at the level of a few tens of $\mu$as, with amplitudes depending on tracer type and sky position (e.g., \citealt{Lindegren_2021a, Riess_2021, Ding_2024}). Rather than attempting to derive an independent correction for this effect, our goal here is to assess whether the parallaxes of the King OCs are consistent with expectations within a homogeneous analysis framework. 

In this study, we transformed the MCMC-based distances ($d_{\rm MCMC}$) of the 27 King OCs into trigonometric parallaxes using the relation $\varpi_{\rm MCMC} ({\rm mas}) = 1000/d_{\rm MCMC}$ (pc), and compared these values with the corresponding {\it Gaia} DR3 parallaxes $\varpi_{\rm Gaia DR3}$. The parallax difference for each cluster was defined as $\Delta\varpi = \varpi_{\rm Gaia DR3} - \varpi_{\rm MCMC}$.

Although the MCMC distances are not directly derived from \textit{Gaia} astrometric parallaxes, this comparison provides a meaningful internal consistency check of the trigonometric parallax scale. We emphasize that this analysis should not be interpreted as an independent external calibration of the \textit{Gaia} parallax zero-point, as the MCMC-based distances rely on stellar evolution models and are therefore subject to systematic uncertainties related to isochrone physics, extinction, and parameter degeneracies. Consequently, the results presented here should be regarded as an internal consistency test within a homogeneous framework rather than an absolute calibration of the \textit{Gaia} parallax scale.

For illustration, the inferred MCMC parallaxes for King~2 and King~12 are $\varpi_{\rm MCMC} = 0.168^{+0.005}_{-0.006}$ mas and $0.368^{+0.030}_{-0.033}$ mas, yielding $\Delta \varpi = -4^{+11}_{-12}$ $\mu$as and $-41^{+33}_{-36}$ $\mu$as, respectively.

To quantify the behaviour of the parallax differences across the full sample, we computed $\Delta\varpi$ for all 27 King OCs (Table~\ref{tab:distances}). The resulting values span the range $-41 \leq \Delta\varpi~(\mu{\rm as}) \leq 29$. The sample yields a median offset of $\widetilde{\Delta\varpi} = -13~\mu$as, a mean value of $-14.1~\mu$as, and a standard deviation of $15.3~\mu$as. In addition to the classical mean and standard deviation, we characterise the dispersion using the 16th–84th percentile interval as a non-parametric measure of spread. This interval is $[-28.4,-5.2]~\mu$as, corresponding to a half-width of $\sim 11.6~\mu$as. Furthermore, no statistically significant correlation is found between $\Delta\varpi$ and Galactic latitude (Spearman $\rho=-0.07$, $p=0.73$), indicating that the observed offset does not exhibit a measurable dependence on Galactic latitude within the $|b| \lesssim 6^\circ$ range covered by the sample.

Overall, the King OC ensemble indicates a median parallax difference of approximately $-13~\mu$as with a dispersion of $\sim 12$–$15~\mu$as, consistent with recent determinations for Galactic-plane samples \citep{Ding_2024}. The agreement of the derived offset with literature values suggests that no additional large systematic bias is introduced by the adopted MCMC-based distance scale. These results provide an internally consistent characterization of the \textit{Gaia} DR3 parallax scale for the King OCs and support the reliability of the adopted distance scale within the context of our homogeneous analysis.

\begin{table}[ht]
\centering
\setlength{\tabcolsep}{4pt}
\renewcommand{\arraystretch}{1}
\scriptsize
\caption{Comparison of trigonometric parallaxes ($\varpi_{\rm MCMC}$) estimated from MCMC analyses ($d_{\rm MCMC}$) and {\it Gaia} DR3 ($\varpi_{\rm Gaia DR3}$). $\Delta \varpi$ is the difference between $ \varpi_{\rm Gaia DR3}$ and $\varpi_{\rm MCMC}$, given in $\mu$as.}
\label{tab:distances}
\begin{tabular}{lrccc}
\hline
Cluster & $d_{\rm MCMC}$ & $\varpi_{\rm MCMC}$ & $\varpi_{\rm Gaia DR3}$ & $\Delta \varpi$ \\
        & (pc)           & (mas)               & (mas)                   & ($\mu$as)       \\
\hline
King 01	& $1795^{+160}_{-188}$  & $0.557^{+0.045}_{-0.065}$ & $0.550\pm0.002$ & $-07^{+47}_{-67}$ \\
King 02	& $5947^{+202}_{-189}$  & $0.168^{+0.005}_{-0.006}$ & $0.164\pm0.006$ & $-04^{+11}_{-12}$ \\
King 03	& $5222^{+177}_{-201}$  & $0.191^{+0.006}_{-0.008}$ & $0.170\pm0.003$ & $-21^{+09}_{-11}$ \\
King 04	& $2517^{+258}_{-270}$  & $0.397^{+0.037}_{-0.048}$ & $0.395\pm0.005$ & $-02^{+42}_{-53}$ \\
King 05	& $2208^{+280}_{-291}$  & $0.453^{+0.051}_{-0.069}$ & $0.413\pm0.003$ & $-40^{+54}_{-72}$ \\
King 06	& $739^{+11}_{-11}$     & $1.353^{+0.020}_{-0.021}$ & $1.382\pm0.003$ & $+29^{+23}_{-24}$ \\
King 07	& $3020^{+103}_{-91}$   & $0.331^{+0.011}_{-0.010}$ & $0.318\pm0.005$ & $-13^{+16}_{-15}$ \\
King 08	& $4828^{+269}_{-289}$  & $0.207^{+0.011}_{-0.013}$ & $0.184\pm0.009$ & $-23^{+20}_{-22}$ \\
King 09	& $6272^{+133}_{-147}$  & $0.159^{+0.003}_{-0.004}$ & $0.147\pm0.011$ & $-12^{+14}_{-15}$ \\
King 10	& $3452^{+115}_{-79}$   & $0.290^{+0.010}_{-0.006}$ & $0.283\pm0.003$ & $-07^{+13}_{-09}$ \\
King 11	& $3023^{+278}_{-311}$  & $0.331^{+0.028}_{-0.038}$ & $0.306\pm0.003$ & $-25^{+31}_{-41}$ \\
King 12	& $2719^{+237}_{-223}$  & $0.368^{+0.030}_{-0.033}$ & $0.327\pm0.003$ & $-41^{+33}_{-36}$ \\
King 13	& $3678^{+244}_{-251}$  & $0.272^{+0.017}_{-0.020}$ & $0.256\pm0.002$ & $-16^{+19}_{-22}$ \\
King 14	& $2515^{+356}_{-347}$  & $0.398^{+0.050}_{-0.063}$ & $0.393\pm0.004$ & $-05^{+54}_{-67}$ \\
King 15	& $3080^{+51}_{-59}$    & $0.325^{+0.006}_{-0.006}$ & $0.307\pm0.006$ & $-18^{+12}_{-12}$ \\
King 16	& $2685^{+226}_{-243}$  & $0.372^{+0.028}_{-0.038}$ & $0.339\pm0.003$ & $-33^{+31}_{-41}$ \\
King 17	& $3713^{+464}_{-471}$  & $0.269^{+0.030}_{-0.039}$ & $0.256\pm0.005$ & $-13^{+35}_{-44}$ \\
King 18	& $2811^{+209}_{-217}$  & $0.356^{+0.025}_{-0.030}$ & $0.338\pm0.003$ & $-18^{+28}_{-33}$ \\
King 19	& $2599^{+252}_{-261}$  & $0.385^{+0.034}_{-0.043}$ & $0.364\pm0.004$ & $-21^{+38}_{-47}$ \\
King 20	& $1973^{+133}_{-124}$  & $0.507^{+0.032}_{-0.034}$ & $0.519\pm0.004$ & $+12^{+36}_{-38}$ \\
King 21	& $2919^{+198}_{-206}$  & $0.343^{+0.022}_{-0.026}$ & $0.337\pm0.005$ & $-06^{+27}_{-31}$ \\
King 22	& $5222^{+510}_{-480}$  & $0.191^{+0.017}_{-0.020}$ & $0.185\pm0.003$ & $-06^{+20}_{-23}$ \\
King 23	& $3487^{+144}_{-160}$	& $0.287^{+0.012}_{-0.014}$ & $0.279\pm0.004$ & $-08^{+16}_{-18}$ \\
King 24	& $3876^{+89}_{-139}$	& $0.258^{+0.006}_{-0.010}$ & $0.235\pm0.003$ & $-23^{+09}_{-13}$ \\
King 25	& $2772^{+202}_{-240}$	& $0.361^{+0.025}_{-0.034}$ & $0.329\pm0.006$ & $-32^{+31}_{-40}$ \\
King 26	& $2321^{+154}_{-163}$	& $0.431^{+0.027}_{-0.032}$ & $0.423\pm0.005$ & $-08^{+32}_{-37}$ \\
King 27	& $4542^{+253}_{-264}$	& $0.220^{+0.011}_{-0.014}$ & $0.191\pm0.005$ & $-29^{+16}_{-19}$ \\

\hline
\end{tabular}
\end{table}


\section{Kinematics and the Galactic Orbital Parameters}
\label{sec:kinematics}
\subsection{Radial Velocities}

Accurate determination of the space velocities and Galactic orbital parameters of the King OCs requires robust estimates of their mean radial velocities ($V_{\rm R}$). In this study, the analysis was not limited to radial velocity measurements from the {\it Gaia} DR3 catalogue \citep{Gaia_DR3}. Instead, additional radial velocity data were incorporated from the compilation of \citet{Tsantaki_2022}, which includes measurements from large-scale spectroscopic surveys such as LAMOST and APOGEE. Candidate stars were selected through cross-matching the radial velocity catalogues with the OC membership lists, followed by the application of stringent quality criteria. In particular, only stars with membership probabilities of $P \geq 0.5$ and reliable astrometric solutions, characterized by $\texttt{RUWE} \leq 1.4$, were retained for further analysis.

As a result of this procedure, 370 OC member stars with available radial velocity measurements were identified in the {\it Gaia} DR3 catalogue. In addition, the \citet{Tsantaki_2022} catalogue contributed radial velocity data for 33 stars, including 19 from APOGEE and 14 from LAMOST. For King 20, no radial velocity measurements were available for any of the identified member stars; consequently, mean radial velocities could not be determined for King 20. Therefore, a total of 26 King OCs with measurable radial velocities were included in the kinematic and dynamical orbital analyses. Among these OCs, five (King 4, 14, 16, 17, and 26) have radial velocity estimates based on a single member star, while King 11 contains the largest number of member stars with radial velocity measurements, totalling 69. Using the criteria mentioned above, we detected 11 and 2 member stars for the mean radial velocity calculations in King 2 and King 12, respectively. The mean $V_{\rm R}$ with uncertainties of clusters was computed according to expressions given by \citet{Carrera_2022} that are based on the weighted mean and standard deviation. Hence, we determined the mean radial velocities as $V_{\rm R}= -110.36 \pm 9.87$ and $V_{\rm R}= -41.26\pm 0.17$ km s$^{-1}$, for King 2 and King 12, respectively. For clusters containing only one member star with radial velocity measurements, we adopted the corresponding radial velocity as the OC's radial velocity.

Recently, \citet{Hunt_2024} presented $V_{\rm R}$ values estimated using one star for the King 21 OC. Since this star is located outside the 1 mas yr$^{-1}$ radius of the cluster's mean PM components and the trigonometric parallax values significantly differ from the mean values estimated in this study, we did not consider this as an OC member. Additionally, because $V_{\rm R}$ measurements for King 20 in the literature are lacking, we excluded this cluster from the kinematic analyses.

\begin{figure}[!t]
\centering
\includegraphics[width=0.9\linewidth]{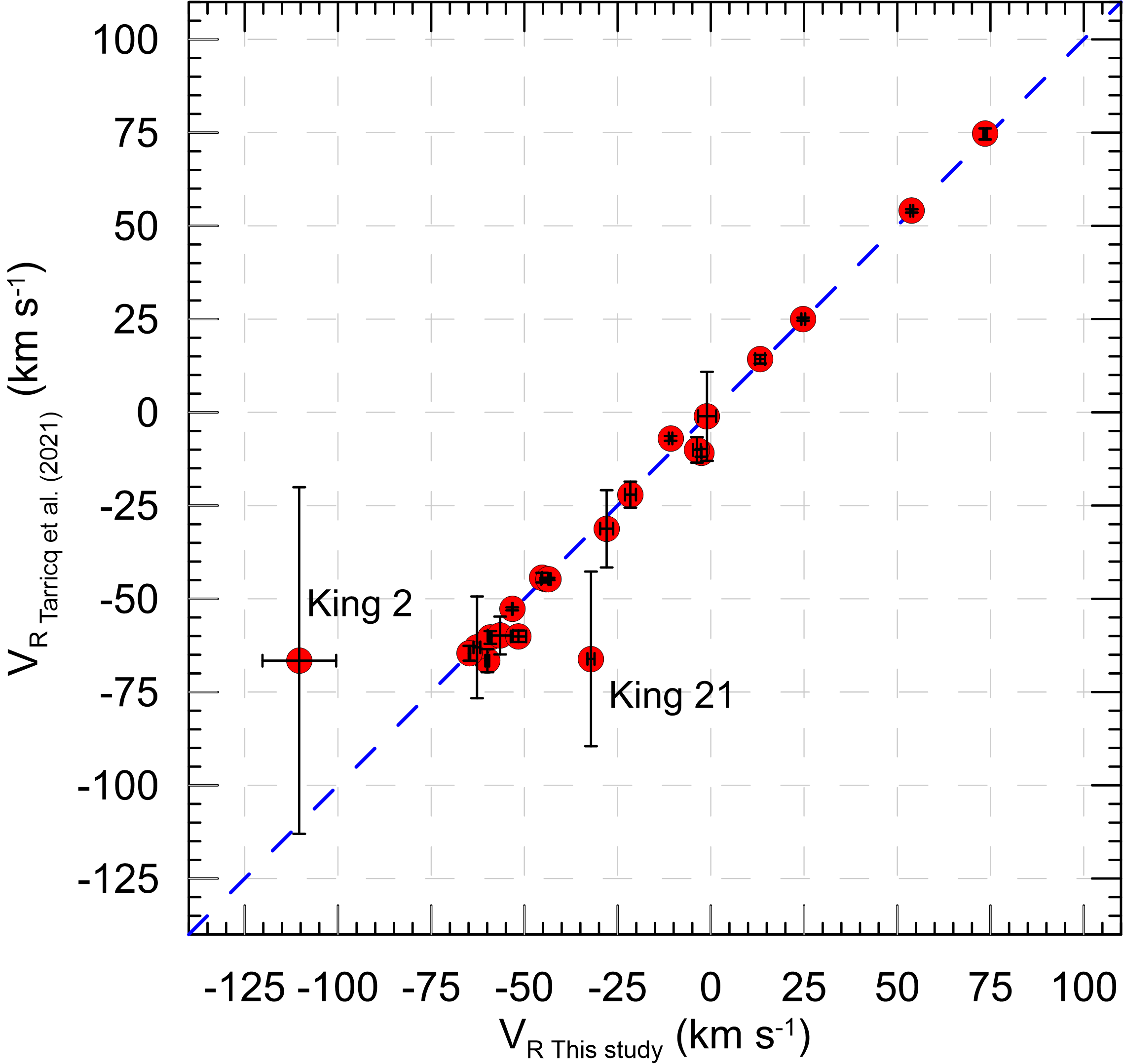}
\caption{The comparison of $V_{\rm R}$ for 22 King OCs calculated using {\it Gaia} DR3 data in this study with those derived by \citet{Tarricq_2021} from {\it Gaia} DR2 data.}
\label{fig:Rv} 
\end{figure}

In this study, the mean $V_{\rm R}$ was calculated for only 26 of the analyzed King OCs, using a total of 403 OC member stars (see also Table~\ref{app:fundamental}). The estimated $V_{\rm R}$ values for these clusters were compared with the $V_{\rm R}$ of member stars compiled from the {\it Gaia} DR2 catalog by \citet{Tarricq_2021}, and this comparison is presented in Figure~\ref{fig:Rv}. In their study, \citet{Tarricq_2021} estimated the mean radial velocities of 22 King OCs using 132 cluster member stars. In the catalog of \citet{Tarricq_2021}, data were available for only 22 of the 26 King OCs analyzed in this study, and the comparison of $V_{\rm R}$ from both studies is shown in Figure~\ref{fig:Rv}.

Overall, the mean radial velocity values obtained in both studies show a good agreement with the one-to-one relation. However, two OCs, King 2 (2 member stars) and King 21 (6 member stars), exhibit significant deviations from this relation. This discrepancy may be attributed to the limited number of cluster member stars used by \citet{Tarricq_2021} in determining the radial velocities of both OCs. The use of a small sample of member stars may lead to increased uncertainties and possible systematic offsets. The mean difference between the radial velocity values derived from the two {\it Gaia} catalogs and the corresponding standard deviation is found to be 1.02 and 12.57 km s$^{-1}$, respectively. These relatively small offsets and their dispersion indicate that the overall agreement between the two datasets is satisfactory. Consequently, the radial velocity values calculated from the mean velocities of cluster member stars, together with the corresponding number of stars used in each calculation, are provided in the electronically available table.

\subsection{Space Velocity Components and Galactic Orbit Parameters}

The metallicity gradient of the Milky Way is usually investigated from the present-day galacto\-cen\-tric distances of the celestial objects under study \citep[e.g.,][]{Luck_2011, Coskunoglu_2012, Bilir_2012, Recio-Blanco_2014, Huang_2015}. Consideration of the Galactic orbital properties of the studied OCs is important for the precise interpretation of the metallicity gradient. For example, the birth radius of an OC may be different from its current distance from the Galaxy center. An OC or star may suffer radial migration due to the resonance effects caused by the Galactic bar and spiral arms, or it can move inward or outward from its origin as it follows an eccentric orbit \citep{Wu_2009, Onal_Tas_2018, Doner_2023, Viscasillas_2023, Iles_2024, Wang_2024,  Cinar_2025}.

Orbit integration necessitates knowledge of equatorial coordinates ($\alpha, \delta$), helio\-cen\-tric distance ($d$), PM components ($\mu_{\alpha}\cos\delta, \mu_{\delta}$) and radial velocity ($V_{\rm R}$) parameters. With these parameters and on the assumption of a certain gravitational potential, the position of a celestial object in the Milky Way at a future time or a past time can be determined by integrating the orbit forward or backward. The kinematic analyses of the King OCs were performed by {\sc galpy} (the galactic dynamics library) package presented by \citet{Bovy_2015}\footnote{See also, {\href{https://galpy.readthedocs.io/}{https://galpy.readthedocs.io/}}} including the {\sc MWPotential2014} potential model for the Milky Way. This model aims to provide a detailed and accurate representation of the gravitational potential of the Milky Way, taking into account the distribution of mass and its effects on the motion of celestial objects within the Galaxy. {\sc MWPotential2014} is implemented as the combination of bulge, disc, and dark matter halo, which contribute to the gravitational potential that governs the dynamics of stars and gas in the Milky Way. The bulge component is explained as a spherical power-law density profile by \citet{Bovy_2015}, which illustrates the mass distribution of the Milky Way. The disc component is defined as the gravitational potential of a disc-like structure based on Galactic dynamics described by \citet{Miyamoto_1975}. The dark matter halo potential is represented by the Navarro-Frenk-White profile \citep{Navarro_1996}, considering the mass of this component.

\begin{figure*}
\centering
\includegraphics[width=\textwidth]{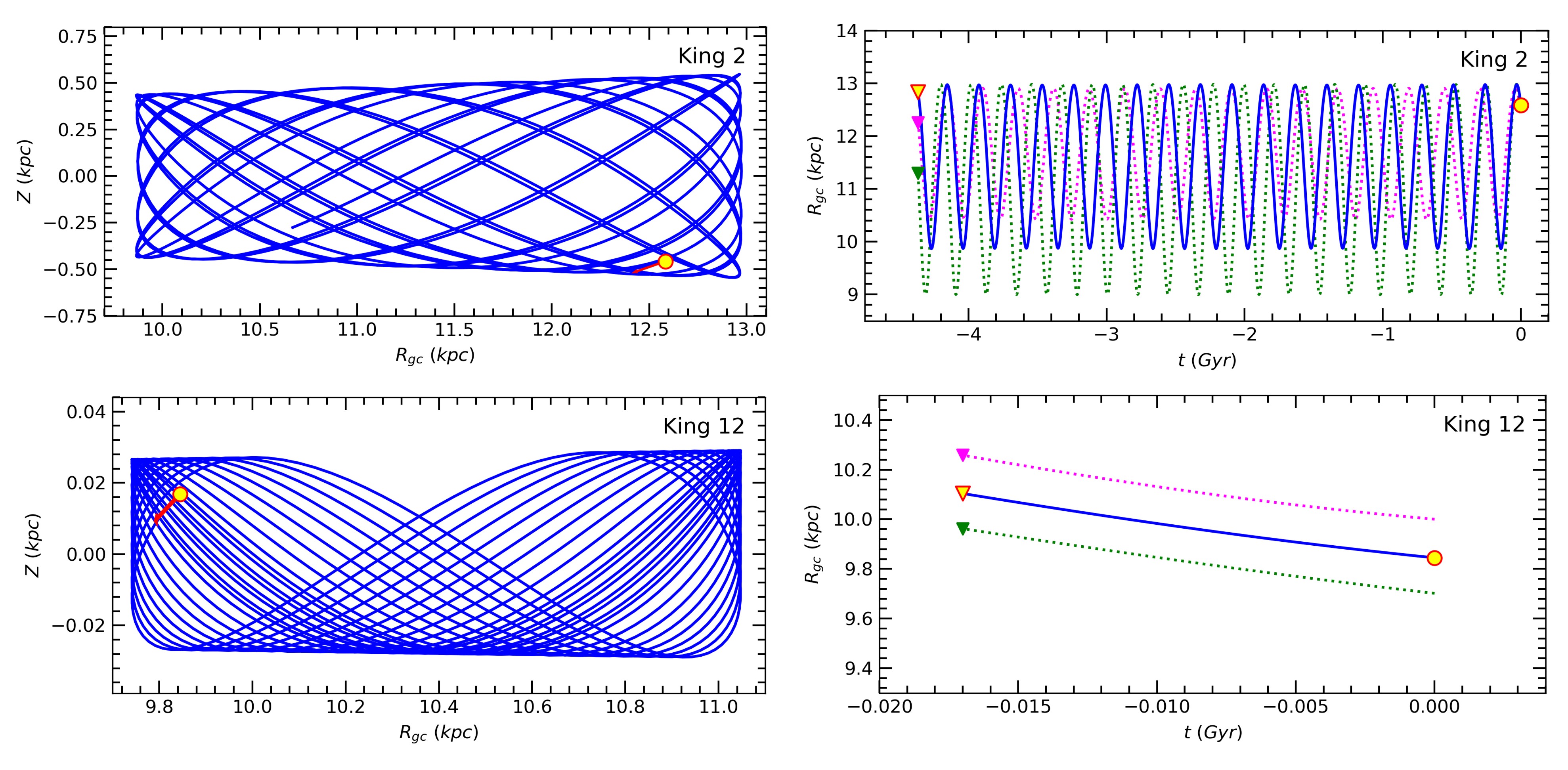}
\caption{\label{fig:galactic_orbits}
Orbits of King 2 and King 12 in the $Z \times R_{\rm gc}$ (left) and $R_{\rm gc} \times t$ (right) planes. The filled yellow triangles and circles represent birthplace and current locations, respectively. The upper and lower input parameters of orbit and birthplace are shown by pink and green filled triangles and relevant dashed lines. Also, red arrows illustrate the motion vectors.} 
\end{figure*}

Before performing orbit integration, we calculated the current galacto\-cen\-tric distance of the 27 King OCs by using the equation of: $R_{\rm gc}=\sqrt{R_{\rm 0}^2 +d^2\cos^2b - 2R_{\rm 0}d\cos l\cos b}$ \citep{Tuncel-Guctekin_2019}, where $R_{\rm 0}$ is the galacto\-cen\-tric distance of the Sun \citep{Bovy_2015, Bovy_2012}, $l$ and $b$ are the Galactic longitude and latitude of the OCs, respectively, and $d$ is the isochrone distance determined above for the studied clusters. According to the results, the nearest and most distant clusters from the Galactic center are King 25 and King 22, with the distances of $R_{\rm gc}=6.93$ and  $R_{\rm gc}=13.28$ kpc, respectively. The $R_{\rm gc}$ values for King 2 and King 12, which are analyzed in detail in our study, are calculated to be 12.50 and 9.80 kpc, respectively. 

To perform kinematic analyses for the King OCs we used their central equatorial coordinates ($\alpha, \delta$) taken from \citet{Cantat-Gaudin_2020}, mean PM components ($\mu_{\alpha}\cos\delta, \mu_{\delta}$) estimated in the study, distances ($d_{\rm MCMC}$) derived from isochrones fitting, and mean radial velocities ($V_{\rm R}$) with relevant uncertainties calculated from {\it Gaia} DR3 data. We assumed galacto\-cen\-tric distance and circular velocity for the Sun as $R_{\rm gc}=8.34$ kpc and $V_{\rm rot}=240$ km s$^{-1}$ \citep{Bobylev_2014}, respectively, and its distance from the Galactic plane as $27\pm 4$ pc \citep{Chen_2000}. Orbit integration was carried out with 1 Myr steps up to 3 Gyr to achieve a complete orbit, and relevant orbit parameters were obtained for 26 King OCs with available $V_{\rm R}$ measurements (except King 20). 

The kinematics and orbit analyses resulted in estimates for the parameters of apogalactic ($R_{\rm a}$) and perigalactic ($R_{\rm p}$) distances, guiding radii ($R_{\rm gui}$), orbit eccentricity ($e$), the maximum vertical distance from Galactic plane ($Z_{\rm max}$), space velocity components ($U$, $V$, $W$) and orbital period ($P_{\rm orb}$). The guiding radius is defined as the radius of a circular orbit with the same angular momentum as the actual orbit of the object, and it provides a measure of the mean orbital radius in the Galactic potential. The space velocities and the Galactic orbital parameters for the 26 King OCs are listed in Table~\ref{app:fundamental}. The results show that the guiding radii of the studied OCs lie within $6.54<R_{\rm gui}~{\rm (kpc)}<14.18$. Moreover, we introduced the traceback early orbital radius ($R_{\mathrm{teo}}$; birthplace), a parameter first defined by \citet{Akbaba_2024}. This parameter characterizes an object's orbital configuration at an earlier time by tracing its orbit backwards, assuming a static Galactic potential. In determining the birth regions of the King OCs, the ages derived in Section~\ref{sec:parameter} were taken into account, and the orbits were integrated backward in time up to their present ages to infer their possible birth locations. Unlike the commonly used birth radius, $R_{\mathrm{teo}}$ does not aim to identify a precise formation location but instead offers insight into the object's previous orbital state. The calculation of $R_{\mathrm{teo}}$ was conducted alongside other kinematic and dynamic orbital parameters.

For King 2 and King 12 the space velocity components ($U$, $V$, $W$) were derived as ($93.93 \pm 4.34$, $-72.66 \pm 8.85$, $-15.53 \pm 1.39$) and ($60.76 \pm 3.49$, $-16.18 \pm 1.89$, $-8.44\pm 0.70$) km s$^{-1}$, respectively. For the Local Standard of Rest (LSR) correction, we used the values of \citet{Coskunoglu_2011} as $(U, V, W)_{\rm LSR}$=$(8.83 \pm 0.24$, $14.19 \pm 0.34$, $6.57 \pm 0.21)$ km s$^{-1}$. These were applied to the 26 King OCs analyzed in the kinematics study. For King 2 and King 12, the LSR corrected space velocity components $(U, V, W)_{\rm LSR}$ were estimated to be ($102.76 \pm 4.35$, $-58.47 \pm 8.86$, $-8.96 \pm 1.40$) and ($69.59 \pm 3.49$, $-1.99 \pm 1.93$, $-1.87 \pm 0.72$) km s$^{-1}$, respectively. The total space velocities $S_{\rm LSR}$ for both OCs were derived as $118.56 \pm 9.97$ km s$^{-1}$ and $69.64 \pm 4.06$ km s$^{-1}$, respectively. \citet{Schuster_2012} mentioned that stars with space velocities within the $-50<V_{\rm LSR}$, $-180<V_{\rm LSR}\leq -50$ km s$^{-1}$ and $V_{\rm LSR} \leq -180$ km s$^{-1}$ belong to the thin disc, thick disc and halo populations of the Galaxy, respectively. According to the calculated $V_{\rm LSR}$ estimates for both OCs, we conclude that King 2 and King 12 belong to the thin-disc population. This finding has also been demonstrated for King 2 by \citet{Haroon_2025}. Also, the $V_{\rm LSR}$ estimates for the 26 King OCs in this kinematics study lie within the $-58.47 \pm 8.86$ km s$^{-1}$ (King 2) and $11.76 \pm 3.71$ km s$^{-1}$ (King 11). This indicates that all of the 26 King OCs belong to the thin-disc population of the Galaxy.

The resulting Galactic orbits for King 2 and King 12 are presented in Figure~\ref{fig:galactic_orbits}. This figure shows the $R_{\rm gc} \times Z$ plane with the side view of the orbits \citep[see, also][]{Koc_2022, Yucel_2024, Elsanhoury_2025}, while the right panels represent the $R_{\rm gc} \times t $ planes that define the change in the galacto-centric distance over time for King 2 (upper panel) and King 12 (lower panel). Yellow-filled circles and triangles in the panels denote the present-day and early orbital positions of the clusters, respectively. Pink and green-dashed lines with similar coloured triangles represent the orbits and early orbital positions of the clusters for upper and lower errors of input parameters. The uncertainties of the Galactic orbital parameters of the clusters were calculated by propagating the observational errors in the astrometric measurements and radial velocities. Figure \ref{fig:galactic_orbits} illustrates that King 2 and King 12 were formed outside the solar circle with  $R_{\rm teo}=12.83\pm 0.81$ and $R_{\rm teo}=10.10\pm 0.35$ kpc. Both OCs orbit entirely outside the solar circle as well. Moreover, analyses indicated that Kings 25, 26, and 27 were formed within the solar circle and remained confined there throughout their entire lifespans. However, results showed that remaining clusters formed outside the solar circle and orbited entirely outside the region (see also Table \ref{app:fundamental}).

\subsection{Comparison with UCC Parameters}
\label{sec:ucc_comparison}

To assess the novelty and reliability of the parameters derived in this study, we compared our results with the values listed in the Unified Cluster Catalogue \citep[UCC;][]{Perren_2023} for the 27 King OCs. The UCC provides a comprehensive compilation of published OC information and lists representative parameters collected from the literature whenever available. Therefore, it constitutes an appropriate reference against which the present homogeneous \textit{Gaia} DR3-based analysis can be evaluated. However, the UCC values necessarily combine results from different studies and may therefore reflect heterogeneous membership selections, photometric systems, extinction assumptions, isochrone sets, and radial-velocity samples. In contrast, the present work applies the same membership procedure, photometric filtering, MCMC isochrone-fitting framework, and kinematical analysis to the full King-cluster sample.

A statistical summary of the comparison between this study and the UCC is given in Table~\ref{tab:ucc_summary}, where the mean and median values of both datasets and their corresponding differences are listed. The astrometric parameters obtained in this study show very good agreement with the UCC values. The mean differences, defined as this study minus UCC, are $\Delta\mu_{\alpha}\cos\delta=-0.0009$ mas yr$^{-1}$, $\Delta\mu_{\delta}=-0.0048$ mas yr$^{-1}$, and $\Delta\varpi=+0.0096$ masas. The photometric distances are also generally consistent, with a median difference of $+0.10$ kpc and a median relative absolute difference of approximately 5.9\%. For the fundamental astrophysical parameters, we find median differences of $\Delta A_{\rm V}=-0.18$ mag, $\Delta\log t=0.00$ dex, and $\Delta{\rm [Fe/H]}=-0.04$ dex. The largest discrepancies occur mainly for young or highly reddened clusters, for which the inferred parameters are particularly sensitive to the adopted membership list, extinction treatment, and CMD morphology. These differences are therefore expected when comparing a homogeneous re-analysis with catalogue values compiled from multiple independent studies. The cluster-by-cluster comparisons from which these summary statistics were calculated are listed in Table~\ref{app:ucc_comparison}.

\begin{table*}
\centering
\caption{{\bf Statistical comparison between the parameters derived in this study and the corresponding UCC values for the King OCs. The differences are defined as this study minus UCC.}}
\label{tab:ucc_summary}
\resizebox{\textwidth}{!}{
\begin{tabular}{lccccccc}
\hline
Parameter 
& $N$ 
& \multicolumn{2}{c}{This study} 
& \multicolumn{2}{c}{UCC} 
& \multicolumn{2}{c}{This study $-$ UCC} \\
\cline{3-8}
& 
& Mean & Median 
& Mean & Median 
& Mean difference & Median difference \\
\hline
$\mu_{\alpha}\cos\delta$ (mas yr$^{-1}$) 
& 27 
& $-1.7726$ & $-2.3190$ 
& $-1.7717$ & $-2.3012$ 
& $-0.0009$ & $+0.0011$ \\

$\mu_{\delta}$ (mas yr$^{-1}$) 
& 27 
& $-1.4772$ & $-1.2610$ 
& $-1.4724$ & $-1.2542$ 
& $-0.0048$ & $-0.0021$ \\

$\varpi$ (mas) 
& 27 
& $0.3478$ & $0.3180$ 
& $0.3381$ & $0.3065$ 
& $+0.0096$ & $+0.0094$ \\

$d$ (kpc) 
& 27 
& $3.3309$ & $3.0200$ 
& $3.2213$ & $2.8700$ 
& $+0.1096$ & $+0.1000$ \\

$A_{\rm V}$ (mag) 
& 27 
& $2.1630$ & $2.0448$ 
& $2.2761$ & $2.2100$ 
& $-0.1131$ & $-0.1827$ \\

$\log t$ (dex) 
& 27 
& $8.5179$ & $8.3010$ 
& $8.5870$ & $8.6405$ 
& $-0.0691$ & $0.0000$ \\

${\rm [Fe/H]}$ (dex) 
& 27 
& $-0.0626$ & $-0.0200$ 
& $-0.0521$ & $-0.0350$ 
& $-0.0104$ & $-0.0400$ \\

$V_{\rm R}$ (km s$^{-1}$) 
& 24 
& $-25.5892$ & $-37.8250$ 
& $-29.5298$ & $-43.9057$ 
& $+3.9406$ & $+0.6331$ \\
\hline
\end{tabular}
}
\vspace{1mm}

\begin{minipage}{0.98\textwidth}
\footnotesize
{\bf \textit{Notes.} Mean and median differences were calculated from the individual cluster-by-cluster differences, i.e. $\Delta X = X_{\rm this~study}-X_{\rm UCC}$, rather than from the difference between the mean or median values of the two samples. The median relative absolute distance difference quoted in the text was calculated as ${\rm median}(|\Delta d|/d_{\rm UCC})\simeq5.9$\%. For $V_{\rm R}$, the comparison includes only the clusters for which radial velocities are available in both this study and the UCC.}
\end{minipage}
\end{table*}

The comparison of radial velocities further illustrates the importance of a homogeneous member-based reassessment. For some clusters, the UCC/literature radial velocities are unavailable or are based on very small numbers of stars. In this study, systemic radial velocities were derived after applying membership and quality criteria to \textit{Gaia} DR3 and supplementary spectroscopic data. This allowed us to determine radial velocities for King 12 and King 16, whereas King 20 was excluded from the kinematical analysis because no reliable member radial velocity was found under our adopted criteria. Moreover, for King 21, the available literature value is based on a star that is inconsistent with the cluster astrometry in our membership analysis, and was therefore not adopted as the systemic cluster radial velocity. These cases demonstrate that the present analysis is not a simple repetition of catalogue values, but provides an internally consistent reassessment of the input parameters required for kinematical and orbital analyses.

In addition to the fundamental astrophysical parameters, the present study provides kinematical and dynamical quantities that are not homogeneously available in the UCC for the King-cluster sample. In particular, we derive the Galactic space-velocity components $(U,V,W)$, present-day Galactocentric distances ($R_{\rm gc}$), guiding radii ($R_{\rm gui}$), traceback early orbital radii ($R_{\rm teo}$), perigalactic and apogalactic distances ($R_{\rm p}$ and $R_{\rm a}$), orbital eccentricities ($e$), maximum vertical distances from the Galactic plane ($Z_{\rm max}$), and orbital periods using the same input parameters and the same Galactic potential model. These quantities are essential for interpreting the metallicity distribution of the King clusters in a chemo-dynamical context, because the present-day Galactocentric radius alone may not represent the original formation radius or mean orbital radius of a cluster. Therefore, the novelty of the present work lies not only in the homogeneous re-determination of the fundamental parameters, but also in providing an internally consistent orbital framework for the full King-cluster sample.

\section{Radial Metallicity Gradient}
\label{sec:gradient}
\subsection{Metallicity Distribution}

Galactic OCs constitute powerful astrophysical laboratories for probing the chemical evolution of the Galactic disc, as they provide reliable estimates of both ages and chemical abundances. In this study, after determining the fundamental astrophysical parameters of the investigated King OCs, their Galactocentric distances were computed and projected onto the iron abundance-Galactocentric radius plane. This analysis provides a Galactic-context framework in which the homogeneously derived metallicities of the King clusters can be evaluated.
The King sample alone is not intended to redefine the global radial metallicity gradient of the Galactic disc. Instead, it is used to examine whether this historically heterogeneous OC subgroup follows the metallicity-radius behaviour traced by recent large OC samples. For this purpose, we compare the metallicity-radius relations derived for the King clusters with those obtained from the 164 spectroscopically analysed OCs of \citet{Otto_2026}. The two samples are first analysed separately, allowing the King-cluster gradient to be assessed against an external spectroscopic reference. In addition, physically motivated subsamples are examined to explore the effects of age and radial migration on the inferred gradients. For both the King and reference samples, the present-day Galactocentric distances ($R_{\rm gc}$), guiding radii ($R_{\rm gui}$), and model-based early orbital/birth-radius proxies ($R_{\rm teo}$) were used to quantify how the inferred radial metallicity gradients depend on the adopted definition of Galactocentric radius.

\begin{table*}[!t]
\renewcommand{\arraystretch}{0.9}
\centering
\caption{Radial metallicity gradients estimated from OCs over the past 25 years. Here, $d{\rm [Fe/H]}/dR_{\rm gc}$ represents the radial metallicity gradient, $R_{\rm gc}$ is the Galactocentric distance of the OC, $\tau$ denotes the cluster age, $N$ is the number of OCs used in the analysis, and the reference indicates the corresponding study.}
\label{tab:radial_gradient_literature}
\begin{tabular}{ccccc|ccccc}
\hline\hline
$d{\rm [Fe/H]}/dR_{\rm gc}$ & $R_{\rm gc}$ & $\tau$ & $N$ & Ref.& $d{\rm [Fe/H]}/dR_{\rm gc}$ & $R_{\rm gc}$ & $\tau$ & $N$ & Ref.\\
(dex\,kpc$^{-1}$) & (kpc) & (Gyr) & & & (dex\,kpc$^{-1}$) & (kpc) & (Gyr) & &\\
\hline\hline
$-0.060 \pm 0.010$ & 7--16 & $\tau > 0.7$ & 39 & 01 & $-0.066 \pm 0.005$ & $<$14 & -- & 157 & 14 \\
$-0.080 \pm 0.001$ & 6--17 & -- & 215 & 02 & $-0.076 \pm 0.009$ & 6--16.5 & -- & 134 & 15 \\
$-0.063 \pm 0.008$ & $<$17 & -- & 118 & 03 & $-0.073 \pm 0.002$ & 6--11.5 & -- & 94 & 16 \\
$-0.056 \pm 0.007$ & $<$17 & $\tau < 0.5$ & 488 & 04 & $-0.032 \pm 0.002$ & 11.5--16.0 & -- & 56 & 16\\
$-0.070 \pm 0.011$ & $<$13.5 & $\tau < 0.5$ & 485 & 04 & $-0.064 \pm 0.007$ & 5--24 & -- & 175 & 17\\
$-0.060 \pm 0.002$ & 5--19 & $\tau < 0.5$ & -- & 05 & $-0.054 \pm 0.008$ & 5--12 & -- & 503 & 18 \\
$-0.052 \pm 0.011$ & 6--12 & -- & 67 & 06 & $-0.062 \pm 0.011$ & 6--15 & $1.9 < \tau < 4$ & 136 & 19 \\
$-0.015 \pm 0.007$ & 12--24 & -- & 12 & 06 & $-0.058 \pm 0.004$ & $<$12 & -- & 136 & 19\\
$-0.066 \pm 0.007$ & $<$12 & -- & 82 & 07 & $-0.063 \pm 0.006$ & 6--21 & $1 < \tau < 3$ & 62 & 20 \\
$-0.061 \pm 0.004$ & 7--13 & $0.5 < \tau < 4.5$ & 19 & 08 & $-0.054 \pm 0.004$ & 6--21 & -- & 62 & 20\\
$-0.052 \pm 0.003$ & 6.5--13 & -- & 46 & 09 & $-0.070 \pm 0.002$ & 4.0--12.8 & -- & 1837 & 21 \\
$-0.077 \pm 0.007$ & 6.5--11 & -- & -- & 09 & $-0.005 \pm 0.018$ & 12.8--20.5 & -- & 35 & 21\\
$+0.018 \pm 0.009$ & $>$13 & -- & -- & 09 & $-0.048 \pm 0.009$ & 5--12 & $\tau > 0.5$ & -- & 22\\
$-0.050 \pm 0.001$ & 6--11 & $\tau < 2$ & 14 & 10 & $-0.045 \pm 0.006$ & 4--17 & -- & 1131 & 23\\
$-0.056 \pm 0.011$ & 6--11 & -- & 18 & 10 & $-0.069 \pm 0.008$ & 5--11.3 & -- & 71 & 24 \\
$-0.068 \pm 0.004$ & 6--13.9 & -- & 68 & 11 & $-0.025 \pm 0.011$ & $>$11.3 & -- & 28 & 24 \\
$-0.009 \pm 0.011$ & $>$13.9 & -- & 3 & 11 & $-0.048 \pm 0.008$ & 6--12 & -- & 203 & 25\\
$-0.053 \pm 0.004$ & 7--15 & $\tau < 3$ & 183 & 12 & $-0.075 \pm 0.006$ & 6--21 & -- & 164 & 26 \\
$-0.074 \pm 0.007$ & 7--11.5 & $\tau < 0.5$ & 225 & 13 & $-0.068 \pm 0.005$ & 6--18 & -- & 164 & 26\\
$-0.059 \pm 0.006$ & $<$14 & $\tau < 2$ & 133 & 14 & $-0.070 \pm 0.006$ & 5--14 & -- & 215 & 27\\
$-0.089 \pm 0.007$ & $<$14 & $2 < \tau < 4$ & 13 & 14 & -- & -- & -- & -- & --\\
\hline
\end{tabular}
\begin{minipage}{17cm}
\small
01) \citet{Friel_2002}, 02) \citet{Hou2002}, 03) \citet{Chen_2003}, 04) \citet{Wu_2009}, 05) \citet{Genovali_2014}, 06) \citet{Reddy_2016}, 07) \citet{Netopil_2016}, 08) \citet{Donor_2018}, 09) \citet{Carrera_2019}, 10) \citet{Casamiquela_2019}, 11) \citet{Donor_2020}, 12) \citet{Zhong_2020}, 13) \citet{Zhang_2021}, 14) \citet{Zhang_2021_b}, 15) \citet{Spina_2021}, 16) \citet{Myers_2022}, 17) \citet{Spina_2022}, 18) \citet{Gaia_DR3}, 19) \citet{Netopil_2022}, 20) \citet{Magrini_2023}, 21) \citet{Joshi_2024}, 22) \citet{Cavallo_2024}, 23) \citet{Zhang_2024}, 24) \citet{Carbajo_2024}, 25) \citet{Yang_2025}, 26) \citet{Otto_2026}, 27) \citet{Zhang_2026}
\end{minipage}
\end{table*}

Radial metallicity gradients in the Galactic disc constitute one of the fundamental observational constraints for understanding the chemical evolution of the Galaxy. The data on radial metallicity gradients obtained from various celestial objects have been compiled by \citet{Onal_Tas_2016}. Since this study focuses on metallicity gradients based on OCs, only metallicity gradients derived from OCs in the literature were considered to ensure homogeneity in the dataset. The findings from these studies are listed in Table~\ref{tab:radial_gradient_literature}. As shown in Table~\ref{tab:radial_gradient_literature}, radial metallicity gradients have been investigated over various Galactocentric distances ($R_{\rm gc}$) and age ($\tau$) intervals. In general, the inner disc exhibits a steeper metallicity gradient, while the outer disc tends to show a more uniform distribution. \citet{Netopil_2016} estimated the transition radius to be approximately $R_{\rm gc} \sim 12$ kpc, whereas \citet{Donor_2020} applied a two-line model and determined it to be around $R_{\rm gc} \sim 13.9$ kpc, although this value is strongly influenced by the adopted distance scale. Although the King OCs analysed in this work cover a relatively broad Galactocentric interval, $R_{\rm gc}=6.9$-$13.3$ kpc, the sample of 27 clusters alone is too limited to provide an independent revision of the global Galactic radial metallicity gradient. Instead, the King sample offers a homogeneous set of newly derived astrophysical, chemical, and kinematical parameters that can be examined within the broader metallicity structure of the Galactic disc. Therefore, we used the 164 OC sample analysed by \citet{Otto_2026} as an external spectroscopic reference. This reference sample, based on high-quality spectroscopic measurements from the SDSS-V/MWM DR19 dataset \citep{SDSS-2026}, provides a broader observational framework for evaluating the consistency of the King clusters with the metallicity pattern traced by recent OC studies. 
This selection is not used to define a new global gradient from the full combined catalogue, while selected physically motivated subsamples were later used to examine the effects of age and radial migration.
A comparison between the two datasets reveals that nine King OCs are common to both samples. For these overlapping clusters, \citet{Otto_2026} derived the mean metallicities using one star in six OCs (King 2, 5, 6, 8, 13, and 15), two stars in King 23, three stars in King 1, and six stars in King 7. The comparison between the metallicities obtained in this work and those reported by \citet{Otto_2026} yields a mean offset of 0.02 dex with a standard deviation of 0.13 dex. Given that several of the overlapping clusters in \citet{Otto_2026} are represented by only a small number of spectroscopic members, this level of agreement indicates that the two metallicity scales are broadly consistent within the expected observational and methodological uncertainties. Therefore, the \citet{Otto_2026} sample provides a useful external reference for assessing the placement of the King clusters in the Galactic metallicity-radius plane.

\begin{table*}
  \centering
  \small
  \caption{The radial metallicity relations estimated from two OC samples and the statistical results are presented.}
    \begin{tabular}{ccccccl}
    \hline
    Plane & Source & $N$ & Equation & $R$ & $\sigma$ & Remark \\
 \hline
    $\mathrm{[Fe/H]}-R_{\rm gc}$  & [1]     & 27  & $\mathrm{[Fe/H]}=-0.061\pm 0.168\times R_{\rm gc}+0.556\pm 0.016$  & 0.597 & 0.133 & All sample \\
    $\mathrm{[Fe/H]}-R_{\rm gc}$  & [2]     & 164 & $\mathrm{[Fe/H]}=-0.063\pm 0.040\times R_{\rm gc}+0.529\pm 0.004$  & 0.787 & 0.102 & All sample \\
    $\mathrm{[Fe/H]}-R_{\rm gui}$ & [1]     & 26  & $\mathrm{[Fe/H]}=-0.058\pm 0.164\times R_{\rm gui}+0.530\pm 0.016$ & 0.600 & 0.135 & All sample \\
    $\mathrm{[Fe/H]}-R_{\rm gui}$ & [2]     & 164 & $\mathrm{[Fe/H]}=-0.063\pm 0.039\times R_{\rm gui}+0.536\pm 0.004$ & 0.800 & 0.100 & All sample \\
    $\mathrm{[Fe/H]}-R_{\rm teo}$ & [1]     & 26  & $\mathrm{[Fe/H]}=-0.060\pm 0.166\times R_{\rm teo}+0.556\pm 0.016$ & 0.612 & 0.133 & All sample \\
    $\mathrm{[Fe/H]}-R_{\rm teo}$ & [2]     & 164 & $\mathrm{[Fe/H]}=-0.056\pm 0.038\times R_{\rm teo}+0.463\pm 0.004$ & 0.773 & 0.105 & All sample \\
    $\mathrm{[Fe/H]}-R_{\rm gc}$  & [1]+[2] & 63  & $\mathrm{[Fe/H]}=-0.054\pm 0.094\times R_{\rm gc} +0.461\pm 0.010$ & 0.571 & 0.126 & $\tau\leq 300$ Myr \\
    $\mathrm{[Fe/H]}-R_{\rm gui}$ & [1]+[2] & 63  & $\mathrm{[Fe/H]}=-0.051\pm 0.088\times R_{\rm gui}+0.436\pm 0.010$ & 0.583 & 0.126 & $\tau\leq 300$ Myr \\
    $\mathrm{[Fe/H]}-R_{\rm teo}$ & [1]+[2] & 63  & $\mathrm{[Fe/H]}=-0.047\pm 0.087\times R_{\rm teo}+0.394\pm 0.009$ & 0.552 & 0.129 & $\tau\leq 300$ Myr \\
    $\mathrm{[Fe/H]}-R_{\rm gui}$ & [1]+[2] & 115 & $\mathrm{[Fe/H]}=-0.061\pm 0.049\times R_{\rm gui}+0.520\pm 0.005$ & 0.758 & 0.108 & $\Delta R\leq 0.5$ \\
    $\mathrm{[Fe/H]}-R_{\rm gui}$ & [1]+[2] & 99  & $\mathrm{[Fe/H]}=-0.059\pm 0.053\times R_{\rm gui}+0.506\pm 0.005$ & 0.741 & 0.111 & $\Delta R\leq 0.5$ + $e_{\rm p}\leq 0.1$ \\
    \hline
        \end{tabular}%

\begin{minipage}{0.95\textwidth}
\footnotesize
\textit{Notes.} [1] denotes the King-cluster sample analysed in this study, and [2] denotes the \citet{Otto_2026} reference sample. The [1]+[2] rows do not represent a full combined catalogue fit; they correspond only to physically selected young or dynamically restricted subsamples.
\end{minipage}
         
  \label{tab:gradient-table}%
\end{table*}%

\begin{figure}[!t]
\centering
\includegraphics[width=0.5\textwidth]{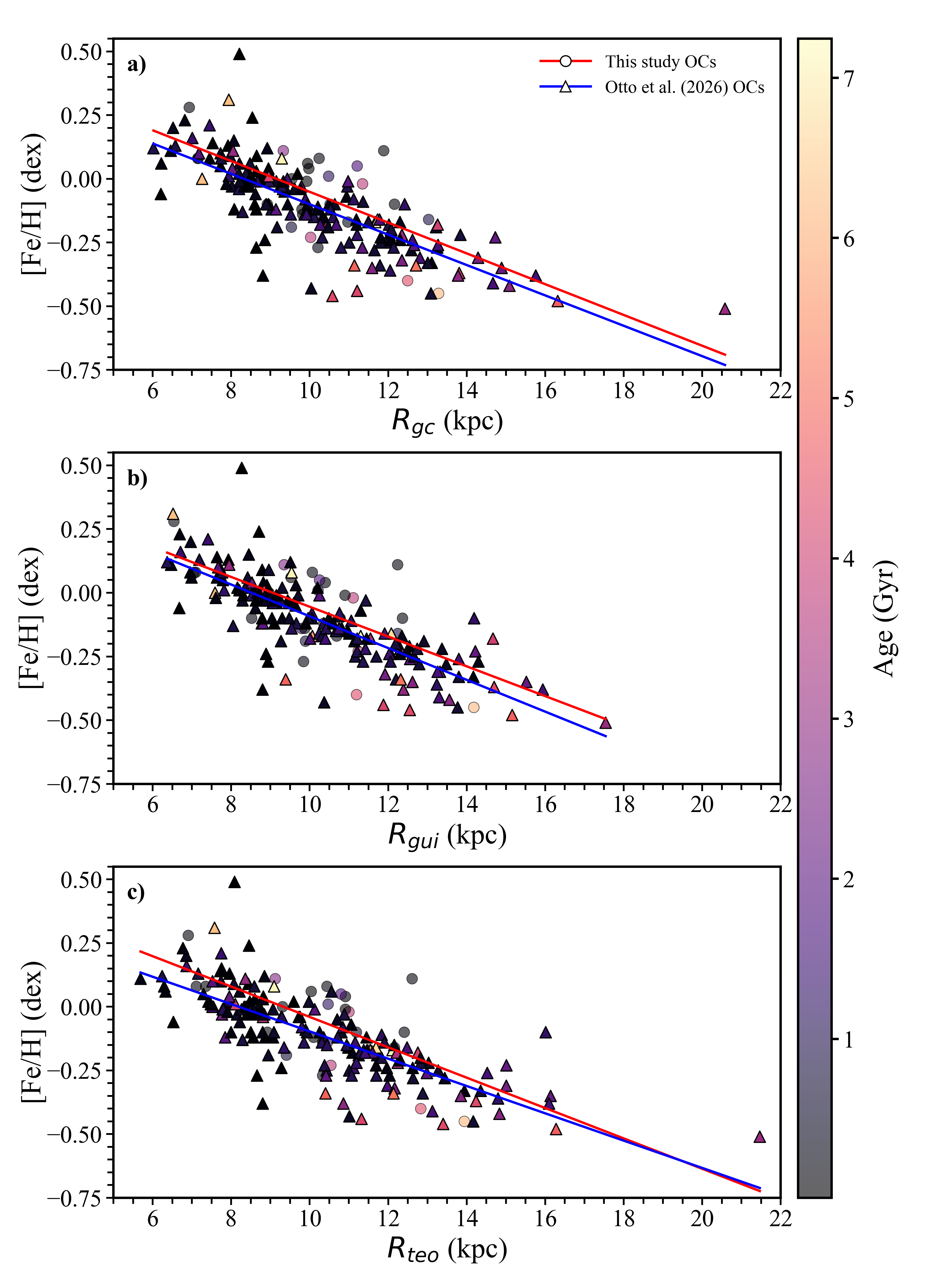}
\caption{\label{fig:gradient_1}
Distributions of the King clusters and the OCs compiled by \citet{Otto_2026} on the a) [Fe/H]-$R_{\rm gc}$, b) [Fe/H]-$R_{\rm gui}$, and c) [Fe/H]-$R_{\rm teo}$ planes. The solid lines represent the linear fits, and the color scale shown on the sides of the panels indicates the cluster ages.} 
\end{figure}

To evaluate the radial metallicity behaviour of the King clusters relative to the external reference sample, the distributions of clusters in the [Fe/H]-$R_{\rm gc}$, [Fe/H]-$R_{\rm gui}$, and [Fe/H]-$R_{\rm teo}$ planes were examined as a function of cluster age (Figure~\ref{fig:gradient_1}). In all three projections, both samples broadly follow the expected negative metallicity trend with increasing Galactocentric distance. Least-squares linear regressions were therefore applied separately to the King sample and to the \citet{Otto_2026} reference sample. The resulting fits are displayed in Figure~\ref{fig:gradient_1}, while the associated relations and slope values are listed in Table~\ref{tab:gradient-table}. An inspection of the fitted relations indicates that the radial metallicity gradients derived for the King clusters are consistent with those obtained from the \citet{Otto_2026} reference sample, within the limitations imposed by the smaller King sample.In the [Fe/H]-$R_{\rm gc}$ plane, the gradients were found to be $-0.061$ dex kpc$^{-1}$ for this study and $-0.063$ dex kpc$^{-1}$ for \citet{Otto_2026}, values that are fully consistent with the typical gradients reported in the literature for the Galactic thin disc (see Table~\ref{tab:radial_gradient_literature}). Similarly, the slopes derived in the [Fe/H]-$R_{\rm gui}$ plane amount to $-0.058$ dex kpc$^{-1}$ and $-0.063$ dex kpc$^{-1}$, while those obtained in the [Fe/H]-$R_{\rm teo}$ plane are $-0.060$ dex kpc$^{-1}$ and $-0.056$ dex kpc$^{-1}$, respectively. These values are broadly consistent with the typical gradients reported in the literature for the Galactic thin disc. These findings indicate that, within the adopted samples and radius definitions, the inferred slopes are similar and support the expected negative metallicity-radius trend in the Galactic thin disc.

In contrast, noticeable differences are observed in the zero points of the linear relations derived for all three planes. As widely discussed in the literature, such discrepancies are primarily attributed to systematic offsets arising from differences in spectroscopic analysis techniques, model atmosphere assumptions, adopted line lists, solar reference abundance scales, and sample selection criteria. In this context, \citet{Netopil_2022}, in their comprehensive compilation and homogenization of OC metallicities, demonstrated in detail that zero-point shifts of the order of $\sim 0.05-0.10$ dex frequently occur among different datasets, and that these offsets predominantly affect the intercepts rather than the slope values. Accordingly, the zero-point differences identified between this study and \citet{Otto_2026} are consistent with the level of systematic uncertainty commonly reported among independent OC metallicity datasets. These offsets should therefore be considered when interpreting the intercepts of the fitted relations, whereas the comparison of the slopes remains useful for assessing the relative placement of the King clusters within the Galactic metallicity distribution. The agreement between the separately derived slopes supports the conclusion that the King clusters are chemically consistent with the broader thin-disc OC population, without implying that the King sample alone provides a statistically dominant determination of the global gradient.

Nevertheless, as expected, a noticeable scatter is observed for some OCs in the metallicity–Galactocentric distance planes (see Figure~\ref{fig:gradient_1}). The literature suggests that such dispersion may arise from a combination of physical and observational effects, including, in particular, the derivation of cluster metallicities from a limited number of member stars, the broad age range spanned by the clusters, and radial migration processes within the Galactic disc \citep[e.g.,][]{Netopil_2016, Anders_2017, Casamiquela_2021}. In particular, the simultaneous consideration of young and old OC populations can blur the temporal signatures of chemical evolution, thereby enhancing the observed scatter in these planes.

\begin{figure}[!t]
\centering
\includegraphics[width=0.50\textwidth]{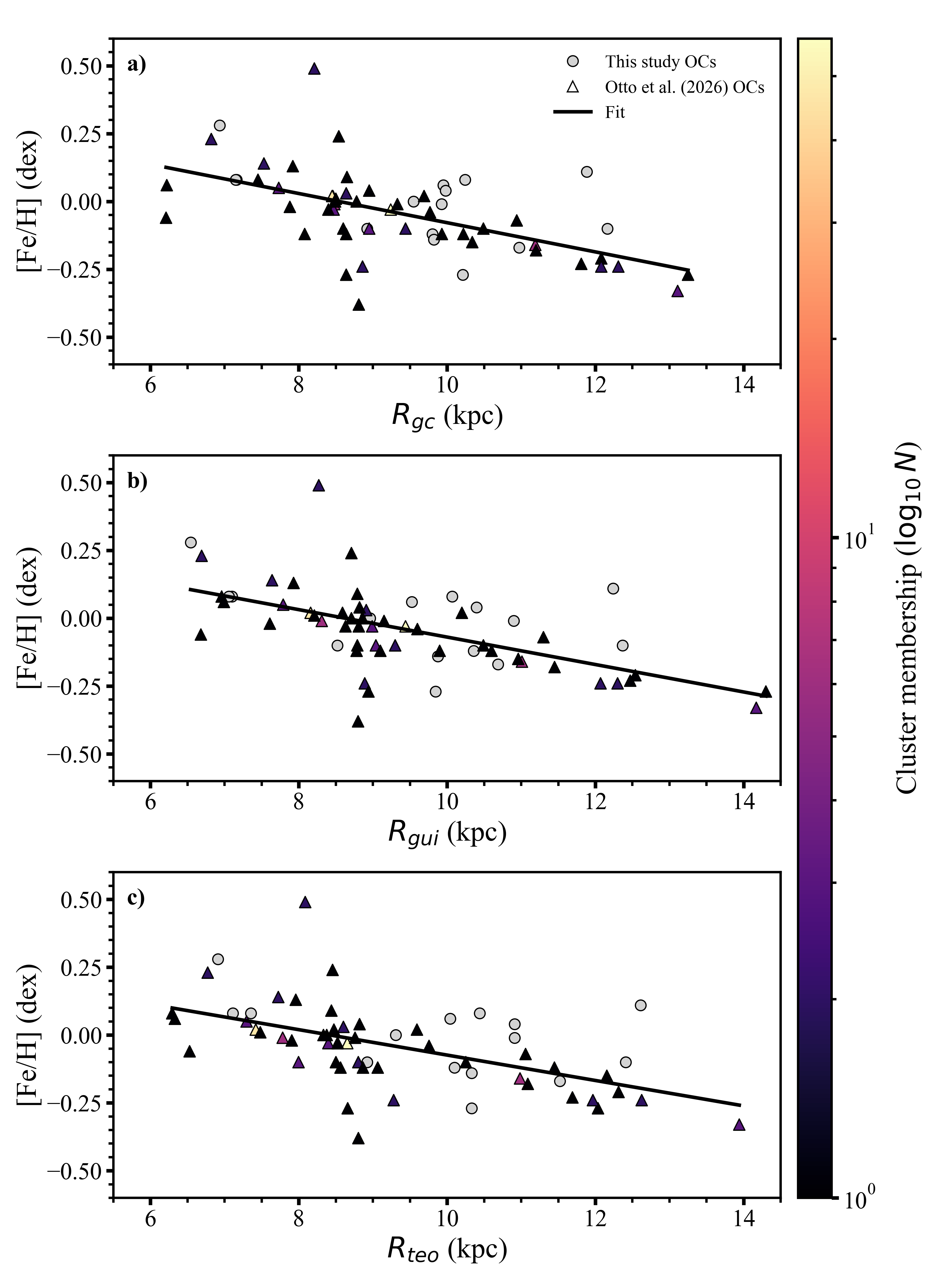}
\caption{\label{fig:gradient_2}
Distributions of the OCs younger than $\tau=300$ Myr in the King OCs and the sample compiled by \citet{Otto_2026} on the a) [Fe/H]-$R_{\rm gc}$, b) [Fe/H]-$R_{\rm gui}$, and c) [Fe/H]-$R_{\rm teo}$ planes are shown. The solid lines represent the linear fits, and the color scale indicates the logarithm of the number of member stars used to derive the mean cluster metallicity.} 
\end{figure}

To investigate the age dependence of the metallicity-radius relation, we constructed a physically selected young subsample consisting of OCs younger than $\tau=300$ Myr drawn from the King and \citet{Otto_2026} datasets. This selection is not used to define a new global gradient from the full combined catalogue, but to test whether restricting the analysis to a young population reduces the scatter and changes the inferred slope. The resulting subsample contains 63 OCs. In the distributions presented in Figure~\ref{fig:gradient_2}, the adopted color scale reflects the number of member stars used to derive the mean cluster metallicities. The linear regression relations obtained from fits to the OCs in each plane are summarized in Table~\ref{tab:gradient-table}, and the corresponding fits are displayed in Figure~\ref{fig:gradient_2}. Following the application of this age constraint, the scatter across the planes is significantly reduced, leading to radial metallicity gradients that converge toward more similar values and exhibit increased consistency in their zero points. For this young subsample, the slopes of the [Fe/H]-$R_{\rm gc}$, [Fe/H]-$R_{\rm gui}$, and [Fe/H]-$R_{\rm teo}$ relations were determined to be $-0.054$, $-0.051$, and $-0.047$ dex kpc$^{-1}$, respectively. These values are shallower than those derived from the full King and reference samples considered separately, and are consistent with the typical gradient range reported for young OCs in the literature \citep[e.g.,][]{Netopil_2016, Donor_2020, Spina_2021}. The relatively flatter gradients obtained for the young cluster population suggest that the chemical evolution of the Galactic disc over the past few hundred million years may have proceeded in a more homogeneous manner and/or that radial mixing processes have played a more prominent role in shaping the metallicity distribution of young populations. The changes in the fitted zero points among the $R_{\rm gc}$, $R_{\rm gui}$, and $R_{\rm teo}$ planes further illustrate that the adopted radial coordinate can affect the normalization of the metallicity relation. This behaviour is consistent with the expectation that present-day positions, guiding radii, and estimated birth radii do not necessarily encode identical chemical information, particularly for clusters that may have experienced dynamical evolution. Nevertheless, despite the imposed age constraint, the residual scatter in the planes does not vanish completely and remains particularly pronounced for clusters with metallicities derived from a small number of member stars. Indeed, a close inspection of Figure~\ref{fig:gradient_2} reveals that the majority of the clusters exhibiting the largest dispersion are those for which the metallicities are based on limited stellar samples. However, \citet{Otto_2026} prioritized high spectroscopic data quality in constructing their sample and therefore did not impose a strict lower limit on the number of member stars per cluster. While this approach effectively broadens the sample coverage, it may also contribute to the increased scatter observed for clusters represented by small stellar samples. Similar trends have been reported in previous studies \citep[e.g.,][]{Netopil_2016, Donor_2020, Spina_2021}, highlighting that this effect constitutes an important factor to consider when interpreting radial metallicity gradients derived from OCs.

A more complete interpretation of the metallicity distribution of Galactic OCs requires consideration of the dynamical processes experienced by stars and clusters over time. In this context, radial migration mechanisms within the disc, particularly the effects of churning and blurring, play a crucial role in reshaping the observed metallicity distributions \citep{Sellwood_2002, Schonrich2009, Minchev_2013}. Churning refers to the systematic change in the Galactocentric radii of stars due to long-term variations in their angular momentum, whereas blurring describes the transient radial displacements arising from epicyclic motions around the birth radii along eccentric orbits. Both processes, especially for long-lived objects such as OCs, can cause present-day locations to differ from their birth sites, thereby introducing additional scatter into radial metallicity relations.

\begin{figure}[!t]
\centering
\includegraphics[width=0.50\textwidth]{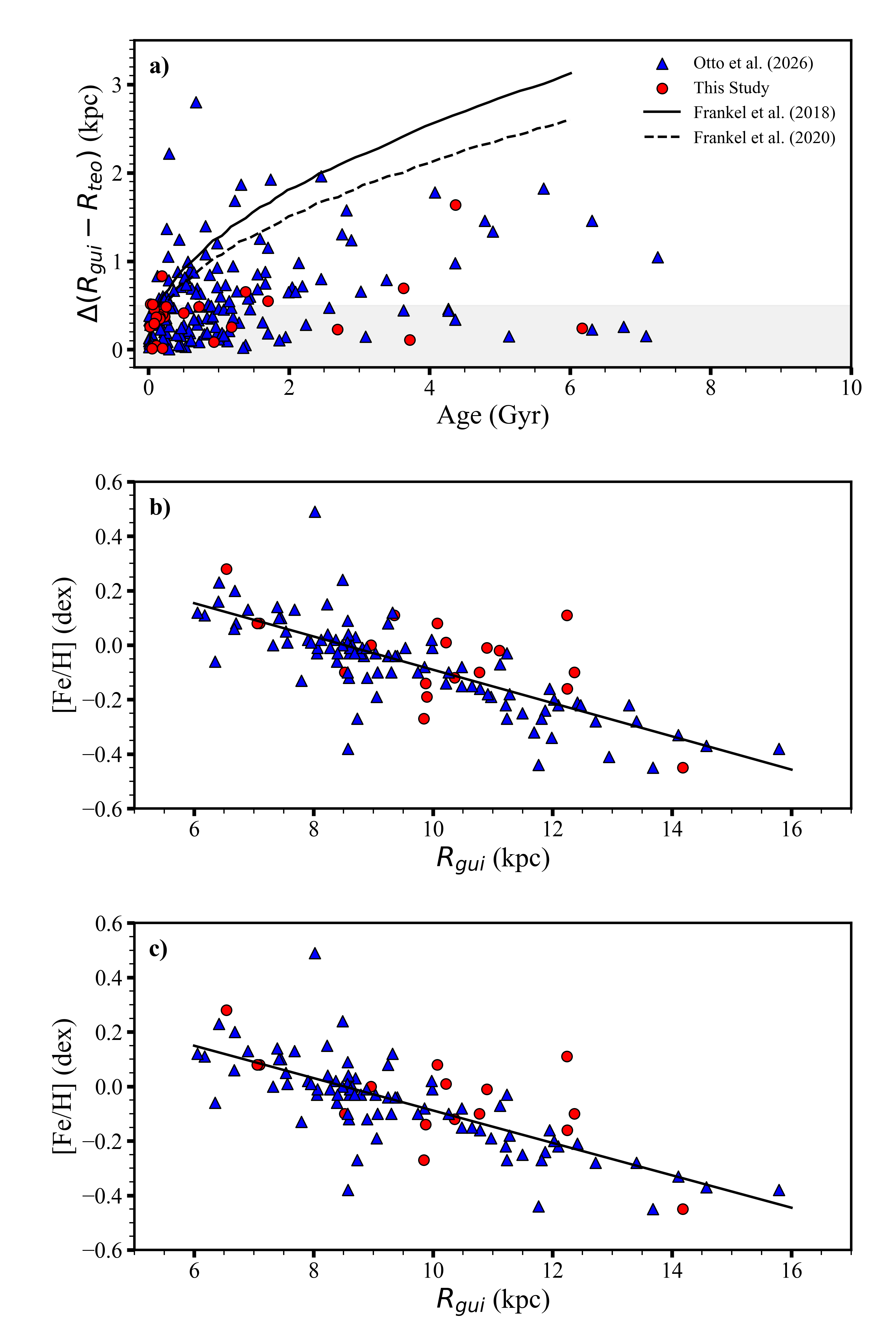}
\caption{\label{fig:gradient_3}
a) Distribution of the King (circles) and \citet{Otto_2026} (triangles) OCs on the distance–age diagram, including radial migration. The solid and dashed lines represent the radial migration models proposed by \citet{Frankel_2018} and \citet {Frankel_2020}, respectively. The shaded grey region in the upper panel indicates the clusters least affected by radial migration. (b) Radial metallicity distributions of the King and \citet{Otto_2026} OCs satisfying the condition $\Delta R \leq 0.5$ kpc. (c) Radial metallicity distributions of the King and \citet{Otto_2026} OCs satisfying the conditions $\Delta R \leq 0.5$ kpc and $e_{\rm p} \leq 0.1$. The solid lines in panels (b) and (c) indicate the linear fits.} 
\end{figure} 

As an additional dynamical assessment, we examined how the inferred metallicity-radius relation changes when the effects of churning and blurring are reduced through simple selection criteria. For this purpose, the King and \citet{Otto_2026} datasets were considered together only after applying physically motivated dynamical constraints. The differences between the guiding radii ($R_{\rm gui}$) and the model-based early orbital/birth-radius proxies ($R_{\rm teo}$) were computed as $\Delta R=|R_{\rm gui}-R_{\rm teo}|$ and examined as a function of cluster age (Figure~\ref{fig:gradient_3}a). In addition, the model predictions presented by \citet{Frankel_2018, Frankel_2020} were overlaid on the same diagram to facilitate a comparative evaluation of theoretical radial migration behaviors across different ages. Considering jointly the distributions of 164 OCs adopted from \citet{Otto_2026} and 26 King OCs kinematically analyzed in this study, we find that a non-negligible fraction of clusters are located beyond the theoretical radial migration limits, with several old clusters ($\tau > 2$ Gyr) exhibiting displacements of approximately 0.5-2 kpc (Figure~\ref{fig:gradient_3}a). This indicates that a substantial subset of OCs has experienced non-negligible radial displacement over time. To minimize the influence of radial migration on gradient measurements, we applied the criterion $\Delta R\leq 0.5$ kpc proposed by \citet{Frankel_2020}, selecting a total of 115 OCs from both samples that satisfy this condition. When projected onto the [Fe/H]-$R_{\rm gui}$ plane (Figure~\ref{fig:gradient_3}b), this subsample displays a broad Galactocentric radial coverage spanning 6-16 kpc, while the potential break region around $\sim 12$ kpc reported in the literature does not emerge prominently. Consequently, the OCs drawn from both catalogs were jointly analyzed using linear regression, yielding a radial metallicity gradient of -0.061 dex kpc$^{-1}$ (see Table~\ref{tab:gradient-table}).

The final dynamical constraint considered in this study is the blurring effect arising from the orbital eccentricities of OCs. To mitigate this effect, an additional criterion of $e_{\rm p} \leq 0.1$ \citep[e.g.,][]{Plevne_2015} was imposed on the sample of 115 OCs, thereby reducing the dataset to 99 OCs. The same analysis performed on this refined subsample (see Figure~\ref{fig:gradient_3}c) yield a radial metallicity gradient of $-0.059$ dex kpc$^{-1}$  (see Table~\ref{tab:gradient-table}). This result shows that applying simple cuts intended to reduce the effects of churning and blurring leads to a slope very similar to that obtained from the $\Delta R\leq0.5$ kpc subsample alone. Thus, within the limitations of the dynamically restricted subsamples, the negative metallicity-radius trend remains stable under these dynamical selections. This supports the use of the King clusters as a homogeneous reference subsample for examining the chemical and dynamical context of the Galactic thin disc, while avoiding the implication that the King sample alone provides a new independent determination of the global metallicity gradient.


\section{Summary and Conclusions}

We have presented a homogeneous \textit{Gaia} DR3–based structural, astrophysical, kinematical, and chemical analysis of 27 King OCs. By combining uniform membership determination with an MCMC isochrone-fitting framework and orbit integration within a consistent Galactic potential, we established an internally coherent parameter set for this historically heterogeneous cluster family. The primary objective of this work was to reassess the King clusters within a unified Gaia-era context and to evaluate their behaviour relative to the chemical and dynamical structure of the Galactic thin disc. King 2 and King 12, respectively the oldest and youngest members of the sample, were used throughout the analysis to illustrate key aspects of the methodology. The full parameter set is listed in Table~\ref{app:fundamental}.

1) Radial density profiles were constructed for all clusters and fitted with the empirical profile of \citet{King_1962}. The resulting limiting radii span 8-25 arcmin, and concentration parameters range from $C=0.21$ to 2.01, consistent with moderately to strongly centrally concentrated stellar distributions. The relatively higher $C$ values observed for OCs in the second and third Galactic quadrants can be attributed to their reduced exposure to disruptive gravitational effects within the Galactic disc, allowing them to retain their structural integrity and experience lower mass-loss rates \citep{Friel1995, Piskunov2007}. These results further support the view that the Galactic environment plays a crucial role in the dynamical and structural evolution of OCs.

2) Membership probabilities were determined using the {\sc upmask} algorithm applied to {\it Gaia} DR3 astrometry, followed by photometric refinement based on the cluster CMD loci. This approach yields clean member samples, with 474 and 141 probable members identified for King 2 and King 12, respectively.

3) Mean proper motions and trigonometric parallaxes were computed from the high-probability members. For King 2, we found $\mu_{\alpha}\cos\delta=-1.398\pm0.005$ and $\mu_{\delta}=-0.876\pm0.008$ mas yr$^{-1}$, while King 12 yields $\mu_{\alpha}\cos\delta=-3.437\pm0.003$ and $\mu_{\delta}=-1.471\pm0.003$ mas yr$^{-1}$. The corresponding parallaxes provide astrometric distances of $d_{\varpi}=6098\pm1190$ pc and $3058\pm234$ pc, respectively, consistent with expectations for clusters in the outer part of the thin disc.

4) Distances, ages, extinctions, and heavy-element mass fractions were derived self-consistently using an affine-invariant MCMC sampler applied to {\it Gaia} CMDs and a high-resolution grid of {\sc parsec} isochrones. This unified approach naturally accounts for parameter covariances and places all clusters on a common scale. The derived ranges span
$E(G_{\rm BP}-G_{\rm RP})=0.113$-1.933 mag,
[Fe/H] from $-0.40$ to $+0.28$ dex,
ages from 17 to 6166 Myr,
and distances from 739 to 6272 pc.
For King 2 and King 12, we obtained ages of $\sim4.4$ Gyr and $\sim17$ Myr, with colour excesses of 0.532 and 0.784 mag, respectively.
The conversion from heavy-element mass fraction $Z$ to [Fe/H] follows the {\sc parsec}-consistent scheme described in Section~\ref{sec:parameter}. While our findings align well with recent literature for King OCs \citep{Gokmen_2023, Haroon_2025, Nasser2025, Bisht2026}, systematic offsets may arise from the diverse analysis methods employed in these studies. In contrast, the present work utilises a homogeneous MCMC isochrone-fitting strategy combined with uniform membership criteria, thereby reducing methodological scatter and offering a consistent parameter set for the studied King OCs.

5) By transforming the MCMC-based distances into photometric parallaxes and comparing them with the corresponding {\it Gaia} DR3 mean cluster parallaxes, we derived cluster-by-cluster parallax differences defined as $\Delta\varpi = \varpi_{\rm Gaia} - \varpi_{\rm MCMC}$. The individual values span $-41 \leq \Delta\varpi~(\mu{\rm as}) \leq 29$. The ensemble yields a median offset of $\widetilde{\Delta\varpi} = -13~\mu$as, a mean value of $-14.1~\mu$as, and a standard deviation of $15.3~\mu$as. The 16th–84th percentile interval of $[-28.4,-5.2]~\mu$as corresponds to a half-width of $\sim 11.6~\mu$as, consistent with a scaled median absolute deviation of $\sim 12~\mu$as, indicating statistical robustness against outliers. No statistically significant correlation is found between $\Delta\varpi$ and Galactic latitude (Spearman $\rho=-0.07$, $p=0.73$) within the latitude range covered by the sample.

This comparison does not constitute an independent calibration of the {\it Gaia} DR3 parallax zero-point; rather, it provides an internal consistency assessment between the trigonometric and photometric distance scales within a homogeneous cluster sample. The inferred median offset of approximately $-13~\mu$as is fully consistent, within uncertainties, with recent zero-point determinations based on independent tracers, particularly those obtained for Galactic-plane sources.

6) Radial velocities were available for 26 of the 27 OCs. Weighted means were calculated for each system, resulting in $V_{\rm R}=-110.36\pm9.87$ km s$^{-1}$ for King 2 and $-41.26\pm0.17$ km s$^{-1}$ for King 12. Orbit integrations using {\sc galpy} and {\sc MWPotential2014} show that all clusters with full phase-space information belong kinematically to the thin disc. Most King clusters have birth radii and guiding radii outside the solar circle or remain confined within it, depending on age and orbital eccentricity; King 2 and King 12 follow orbits entirely exterior to the solar circle.

7) The radial metallicity behaviour of the King OCs was examined using three complementary Galactic distance definitions: the present-day Galactocentric radius ($R_{\rm gc}$), the guiding radius ($R_{\rm gui}$), and the model-based early orbital/birth-radius proxy ($R_{\rm teo}$). The gradients derived from the 27 King clusters are close to $-0.060$ dex kpc$^{-1}$ and are consistent, within the limitations of the sample size, with those obtained independently from the 164 spectroscopically analysed OCs of \citet{Otto_2026}. This agreement indicates that the homogeneously analysed King clusters follow the same broad metallicity-radius behaviour as the larger thin-disc OC population.

Physically selected subsamples were then used to examine the influence of age and radial migration. For OCs younger than $\tau=300$ Myr, the gradients become systematically flatter, suggesting a more homogeneous recent chemical enrichment and/or a stronger relative influence of radial mixing in young populations. After applying dynamical constraints intended to reduce the effects of churning and blurring, the gradient remains close to $-0.059$ dex kpc$^{-1}$, indicating that the negative metallicity-radius trend is stable under these selections.

The role of the King sample in this analysis is therefore not to provide an independent replacement for metallicity gradients derived from large OC catalogues, but to supply a newly derived, internally homogeneous set of astrophysical, chemical, and orbital parameters for a historically heterogeneous OC subgroup. When compared with the \citet{Otto_2026} reference sample and with dynamically selected OC subsamples, the King clusters provide an independent homogeneous check on the chemo-dynamical structure of the Galactic thin disc.

By providing a strictly homogeneous Gaia DR3-based analysis of the King OCs, this study reduces inter-study systematics and enables a more reliable interpretation of their structural, astrophysical, kinematical, and chemical properties. The comparison with UCC/literature values shows that our results are generally consistent with published parameters, while the remaining differences can be understood in terms of membership selection, extinction treatment, CMD morphology, metallicity scale, and radial-velocity quality control. The consistency between photometric and astrometric distance scales, together with the comparison of the King metallicity gradients with an external spectroscopic OC reference sample, demonstrates that King clusters are valuable tracers of the present-day thin disc and establishes them as a useful homogeneous benchmark for future studies of Galactic structure and evolution.


\section*{Acknowledgements}
We thank the anonymous referees for their insightful and constructive suggestions, which significantly improved the manuscript. This study has been supported in part by the Scientific and Technological Research Council (T\"UB\.ITAK) 122F109. It was funded by the Scientific Research Projects Coordination Unit of Istanbul University (project number: 39743). This research has made use of the Astrophysics Data System, funded by NASA under Cooperative Agreement 80NSSC21M0056, as well as the SIMBAD and VizieR databases operated at CDS, Strasbourg, France, and presents results from the European Space Agency (ESA) space mission Gaia, whose data are processed by the Gaia Data Processing and Analysis Consortium (DPAC), funded by national institutions participating in the Gaia MultiLateral Agreement (MLA); the Gaia mission and archive websites are https://www.cosmos.esa.int/gaia and https://archives.esac.esa.int/gaia, respectively.

\appendix
\section{Cluster-by-cluster comparison with UCC parameters}
\label{app:ucc_comparison}

The following tables provide the cluster-by-cluster comparison between the parameters derived in this study and the corresponding values listed in the Unified Cluster Catalogue (UCC). The differences are defined as this study minus UCC, i.e. $\Delta X=X_{\rm this\ study}-X_{\rm UCC}$. These individual differences were used to compute the mean and median offsets reported in Table~\ref{tab:ucc_summary}.

\begin{table*}
\centering
\caption{Cluster-by-cluster comparison of astrometric parameters derived in this study with the corresponding UCC values. Differences are defined as this study minus UCC.}
\label{tab:ucc_astrometric_comparison}
\resizebox{0.7\textwidth}{!}{%
\begin{tabular}{lrrrrrrrrr}
\hline
Cluster & $\mu_{\alpha}\cos\delta_{\rm this}$ & $\mu_{\alpha}\cos\delta_{\rm UCC}$ & $\Delta\mu_{\alpha}\cos\delta$ & $\mu_{\delta,{\rm this}}$ & $\mu_{\delta,{\rm UCC}}$ & $\Delta\mu_{\delta}$ & $\varpi_{\rm this}$ & $\varpi_{\rm UCC}$ & $\Delta\varpi$ \\
& \multicolumn{3}{c}{(mas yr$^{-1}$)} &
\multicolumn{3}{c}{(mas yr$^{-1}$)} &
\multicolumn{2}{c}{(mas)} & ($\mu$as) \\
\hline
King~1  & -4.935 & -4.9379 & +0.0029 & -1.302 & -1.2979 & -0.0041 & 0.550 & 0.5401 & +9.9 \\
King~2  & -1.398 & -1.3986 & +0.0006 & -0.876 & -0.8691 & -0.0069 & 0.164 & 0.1352 & +28.8 \\
King~3  & -0.700 & -0.7026 & +0.0026 & -0.131 & -0.1308 & -0.0002 & 0.170 & 0.1608 & +9.2 \\
King~4  & -0.598 & -0.5971 & -0.0009 & -0.174 & -0.1630 & -0.0110 & 0.395 & 0.3919 & +3.1 \\
King~5  & -0.303 & -0.3002 & -0.0028 & -1.261 & -1.2542 & -0.0068 & 0.413 & 0.4036 & +9.4 \\
King~6  & +3.830 & +3.8252 & +0.0048 & -1.891 & -1.9055 & +0.0145 & 1.382 & 1.3767 & +5.3 \\
King~7  & +1.127 & +1.1251 & +0.0019 & -1.191 & -1.1894 & -0.0016 & 0.318 & 0.3038 & +14.2 \\
King~8  & +0.452 & +0.4716 & -0.0196 & -1.700 & -1.7062 & +0.0062 & 0.184 & 0.1718 & +12.2 \\
King~9  & -2.294 & -2.2976 & +0.0036 & -1.013 & -1.0126 & -0.0004 & 0.147 & 0.1326 & +14.4 \\
King~10 & -2.697 & -2.7010 & +0.0040 & -2.118 & -2.1217 & +0.0037 & 0.283 & 0.2780 & +5.0 \\
King~11 & -3.388 & -3.3814 & -0.0066 & -0.666 & -0.6711 & +0.0051 & 0.306 & 0.3065 & -0.5 \\
King~12 & -3.437 & -3.4220 & -0.0150 & -1.471 & -1.4488 & -0.0222 & 0.327 & 0.3194 & +7.6 \\
King~13 & -2.728 & -2.7296 & +0.0016 & -0.872 & -0.8699 & -0.0021 & 0.256 & 0.2494 & +6.6 \\
King~14 & -3.237 & -3.2434 & +0.0064 & -1.073 & -1.0540 & -0.0190 & 0.393 & 0.3926 & +0.4 \\
King~15 & -2.319 & -2.3012 & -0.0178 & -0.895 & -0.8944 & -0.0006 & 0.307 & 0.2921 & +14.9 \\
King~16 & -3.009 & -3.0189 & +0.0099 & -0.537 & -0.5188 & -0.0182 & 0.339 & 0.3389 & +0.1 \\
King~17 & -0.195 & -0.1812 & -0.0138 & -1.405 & -1.3768 & -0.0282 & 0.256 & 0.2454 & +10.6 \\
King~18 & -2.634 & -2.6320 & -0.0020 & -2.085 & -2.0971 & +0.0121 & 0.338 & 0.3100 & +28.0 \\
King~19 & -4.794 & -4.7951 & +0.0011 & -2.712 & -2.7123 & +0.0003 & 0.364 & 0.3597 & +4.3 \\
King~20 & -2.668 & -2.6770 & +0.0090 & -2.656 & -2.6461 & -0.0099 & 0.519 & 0.5060 & +13.0 \\
King~21 & -3.243 & -3.2378 & -0.0052 & -1.722 & -1.7130 & -0.0090 & 0.337 & 0.3168 & +20.2 \\
King~22 & +0.778 & +0.7757 & +0.0023 & -0.092 & -0.0793 & -0.0127 & 0.185 & 0.1695 & +15.5 \\
King~23 & -0.477 & -0.4795 & +0.0025 & -0.876 & -0.8799 & +0.0039 & 0.279 & 0.2794 & -0.4 \\
King~24 & -2.960 & -2.9610 & +0.0010 & +2.483 & +2.4849 & -0.0019 & 0.235 & 0.2329 & +2.1 \\
King~25 & -1.388 & -1.3812 & -0.0068 & -3.964 & -3.9618 & -0.0022 & 0.329 & 0.3151 & +13.9 \\
King~26 & -2.388 & -2.3701 & -0.0179 & -4.982 & -4.9854 & +0.0034 & 0.423 & 0.4118 & +11.2 \\
King~27 & -2.257 & -2.2870 & +0.0300 & -4.702 & -4.6812 & -0.0208 & 0.191 & 0.1896 & +1.4 \\
\hline
\end{tabular}%
}
\begin{minipage}{0.98\textwidth}
\footnotesize
\textit{Notes.} The individual differences in this table were used to compute the astrometric mean and median offsets reported in Table~\ref{tab:ucc_summary}.
\end{minipage}
\end{table*}

\begin{table*}
\centering
\caption{Cluster-by-cluster comparison of distances, extinctions, and ages derived in this study with the corresponding UCC values. Differences are defined as this study minus UCC.}
\label{tab:ucc_cluster_comparison_1}
\resizebox{0.6\textwidth}{!}{%
\begin{tabular}{lrrrrrrrrr}
\hline
Cluster & $d_{\rm this}$ & $d_{\rm UCC}$ & $\Delta d$ & $A_{V,{\rm this}}$ & $A_{V,{\rm UCC}}$ & $\Delta A_V$ & $\log t_{\rm this}$ & $\log t_{\rm UCC}$ & $\Delta\log t$ \\
 & \multicolumn{3}{c}{(kpc)} & \multicolumn{3}{c}{(mag)} & \multicolumn{3}{c}{(dex)} \\
\hline
King~1  & 1.795 & 1.695 & +0.100 & 1.99 & 2.36 & -0.37 & 9.430 & 9.442 & -0.012 \\
King~2  & 5.947 & 5.500 & +0.447 & 1.18 & 0.98 & +0.20 & 9.640 & 9.670 & -0.030 \\
King~3  & 5.222 & 4.500 & +0.722 & 3.44 & 2.42 & +1.02 & 8.041 & 8.840 & -0.799 \\
King~4  & 2.517 & 2.690 & -0.173 & 2.08 & 2.56 & -0.48 & 8.369 & 8.100 & +0.269 \\
King~5  & 2.208 & 2.270 & -0.062 & 2.08 & 2.29 & -0.21 & 9.140 & 9.009 & +0.131 \\
King~6  & 0.739 & 0.730 & +0.009 & 1.55 & 1.58 & -0.03 & 8.290 & 8.340 & -0.050 \\
King~7  & 3.020 & 2.870 & +0.150 & 3.68 & 3.94 & -0.26 & 8.270 & 8.800 & -0.531 \\
King~8  & 4.828 & 5.365 & -0.537 & 1.66 & 1.83 & -0.17 & 8.970 & 8.915 & +0.055 \\
King~9  & 6.272 & 6.310 & -0.038 & 1.12 & 1.33 & -0.21 & 9.570 & 9.490 & +0.080 \\
King~10 & 3.452 & 3.250 & +0.202 & 3.37 & 3.55 & -0.18 & 7.477 & 7.447 & +0.030 \\
King~11 & 3.023 & 2.930 & +0.093 & 2.63 & 3.02 & -0.39 & 9.560 & 9.320 & +0.240 \\
King~12 & 2.719 & 2.490 & +0.229 & 1.75 & 1.86 & -0.11 & 7.230 & 7.146 & +0.084 \\
King~13 & 3.678 & 3.230 & +0.448 & 1.70 & 2.06 & -0.36 & 8.700 & 8.690 & +0.010 \\
King~14 & 2.515 & 2.400 & +0.115 & 1.23 & 1.56 & -0.33 & 7.919 & 7.699 & +0.220 \\
King~15 & 3.080 & 3.150 & -0.070 & 1.69 & 2.06 & -0.37 & 8.400 & 8.400 & +0.000 \\
King~16 & 2.685 & 2.440 & +0.245 & 2.22 & 2.43 & -0.21 & 7.748 & 7.778 & -0.030 \\
King~17 & 3.713 & 3.320 & +0.393 & 1.21 & 1.74 & -0.53 & 8.199 & 8.450 & -0.252 \\
King~18 & 2.811 & 2.750 & +0.061 & 1.97 & 1.98 & -0.00 & 8.190 & 8.468 & -0.278 \\
King~19 & 2.599 & 2.555 & +0.044 & 2.04 & 1.94 & +0.10 & 8.860 & 8.652 & +0.207 \\
King~20 & 1.973 & 1.850 & +0.123 & 2.51 & 2.21 & +0.30 & 7.959 & 8.320 & -0.361 \\
King~21 & 2.919 & 2.810 & +0.109 & 2.33 & 2.52 & -0.19 & 7.672 & 7.699 & -0.027 \\
King~22 & 5.222 & 5.620 & -0.398 & 1.58 & 1.45 & +0.13 & 9.790 & 9.630 & +0.160 \\
King~23 & 3.487 & 3.210 & +0.277 & 0.25 & 0.41 & -0.16 & 9.230 & 9.072 & +0.158 \\
King~24 & 3.876 & 3.970 & -0.094 & 2.16 & 2.64 & -0.48 & 9.070 & 9.000 & +0.070 \\
King~25 & 2.772 & 2.830 & -0.058 & 4.30 & 4.20 & +0.10 & 8.049 & 8.378 & -0.329 \\
King~26 & 2.321 & 2.280 & +0.041 & 3.04 & 3.22 & -0.18 & 8.301 & 8.640 & -0.339 \\
King~27 & 4.542 & 3.960 & +0.582 & 3.61 & 3.33 & +0.29 & 7.908 & 8.450 & -0.542 \\
\hline
\end{tabular}%
}
\begin{minipage}{0.98\textwidth}
\footnotesize
\textit{Notes.} $A_{{\rm V},{\rm this}}$ was obtained from the MCMC-derived $A_{\rm G}$ values using $A_{\rm V}=A_{\rm G}/0.83627$. The age values were converted to $\log t$ as $\log_{10}({\rm Age~[Myr]}\times10^6)$. Mean and median offsets quoted in the text were calculated from the individual cluster-by-cluster differences.
\end{minipage}
\end{table*}

\begin{table*}
\centering
\caption{Cluster-by-cluster comparison of metallicities and radial velocities derived in this study with the corresponding UCC values. Differences are defined as this study minus UCC.}
\label{tab:ucc_cluster_comparison_2}
\resizebox{0.6\textwidth}{!}{%
\begin{tabular}{lrrrrrr}
\hline
Cluster & ${\rm [Fe/H]}_{\rm this}$ & ${\rm [Fe/H]}_{\rm UCC}$ & $\Delta{\rm [Fe/H]}$ & $V_{R,{\rm this}}$ & $V_{R,{\rm UCC}}$ & $\Delta V_R$ \\
 & \multicolumn{3}{c}{(dex)} & \multicolumn{3}{c}{(km s$^{-1}$)} \\
\hline
King~1 & +0.11 & -0.05 & +0.16 & -53.23 & -53.11 & -0.12 \\
King~2 & -0.40 & -0.36 & -0.04 & -110.36 & -137.65 & +27.29 \\
King~3 & -0.10 & +0.03 & -0.13 & -44.18 & -44.80 & +0.62 \\
King~4 & +0.08 & -0.04 & +0.12 & -45.23 & -46.10 & +0.87 \\
King~5 & -0.14 & -0.15 & +0.01 & -43.50 & -43.01 & -0.49 \\
King~6 & -0.10 & +0.28 & -0.38 & -21.56 & -21.76 & +0.20 \\
King~7 & -0.17 & -0.11 & -0.06 & -10.77 & -9.83 & -0.94 \\
King~8 & -0.16 & -0.32 & +0.16 & -1.02 & -1.83 & +0.81 \\
King~9 & -0.02 & -0.38 & +0.36 & -59.94 & -60.59 & +0.65 \\
King~10 & +0.06 & +0.09 & -0.03 & -62.72 & -76.08 & +13.36 \\
King~11 & -0.23 & -0.27 & +0.04 & -27.94 & -24.48 & -3.46 \\
King~12 & -0.12 & +0.06 & -0.18 & -41.26 & -- & -- \\
King~13 & -0.10 & -0.16 & +0.06 & -59.15 & -59.85 & +0.70 \\
King~14 & -0.14 & -0.03 & -0.11 & -54.82 & -74.18 & +19.36 \\
King~15 & -0.27 & +0.00 & -0.27 & -64.73 & -67.25 & +2.52 \\
King~16 & +0.04 & +0.04 & +0.00 & -46.49 & -- & -- \\
King~17 & +0.11 & -0.19 & +0.30 & -2.56 & +2.55 & -5.11 \\
King~18 & +0.00 & -0.04 & +0.04 & -56.48 & -62.04 & +5.56 \\
King~19 & -0.19 & -0.01 & -0.18 & -51.61 & -60.53 & +8.92 \\
King~20 & +0.01 & +0.14 & -0.12 & -- & -31.59 & -- \\
King~21 & -0.01 & +0.04 & -0.05 & -32.15 & -78.27 & +46.12 \\
King~22 & -0.45 & -0.37 & -0.08 & -3.74 & -2.73 & -1.01 \\
King~23 & +0.05 & -0.12 & +0.17 & +53.87 & +53.09 & +0.78 \\
King~24 & +0.01 & -0.34 & +0.35 & +73.58 & +73.60 & -0.02 \\
King~25 & +0.28 & +0.42 & -0.14 & +13.24 & +21.39 & -8.15 \\
King~26 & +0.08 & +0.22 & -0.14 & +24.80 & +24.80 & +0.00 \\
King~27 & +0.08 & +0.22 & -0.14 & +26.06 & +39.96 & -13.90 \\
\hline
\end{tabular}%
}
\end{table*}

\clearpage
\nolinenumbers

\renewcommand{\thefigure}{B\arabic{figure}}
\setcounter{figure}{0}
\renewcommand{\thetable}{C\arabic{table}}
\setcounter{table}{0}

\onecolumn

\section{CMDs of King OCs.}
\label{app:all_cmds}
\renewcommand{\thefigure}{B\arabic{figure}}
\setcounter{figure}{0}

\begin{figure}[H]
\centering
\includegraphics[width=\textwidth,height=0.89\textheight,keepaspectratio]{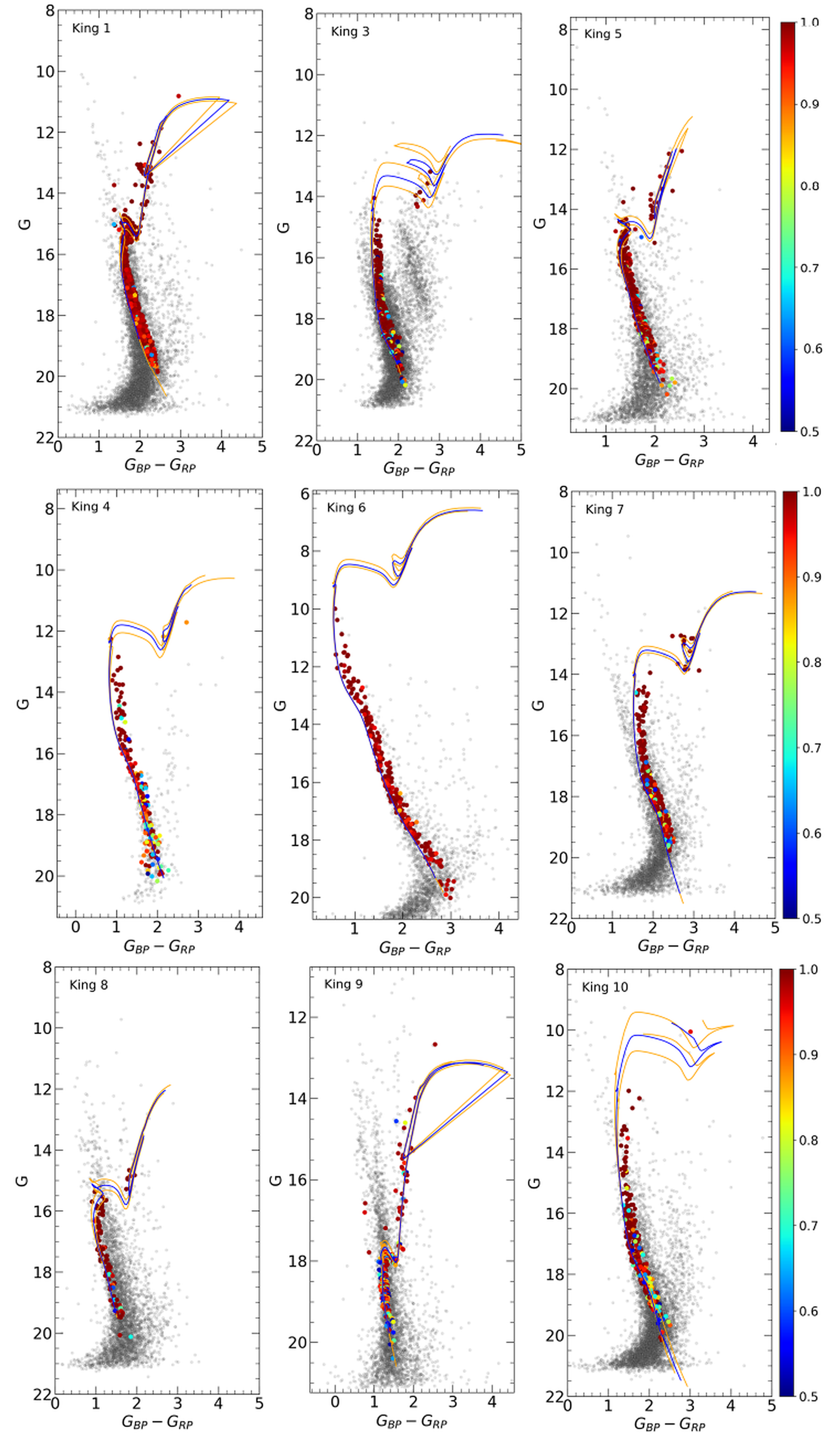}
\caption{CMDs and MCMC-derived best-fitting PARSEC isochrones for the remaining King OCs, excluding King~2 and King~12. Stars are colour-coded according to their membership probabilities, while grey points represent field stars or low-probability sources. The blue curves show the best-fitting isochrones, and the orange curves indicate the corresponding $1\sigma$ uncertainty envelopes.}
\label{fig:appendix_cmds_01}
\end{figure}

\clearpage

\begin{figure}[H]
\centering
\includegraphics[width=\textwidth,height=0.9\textheight,keepaspectratio]{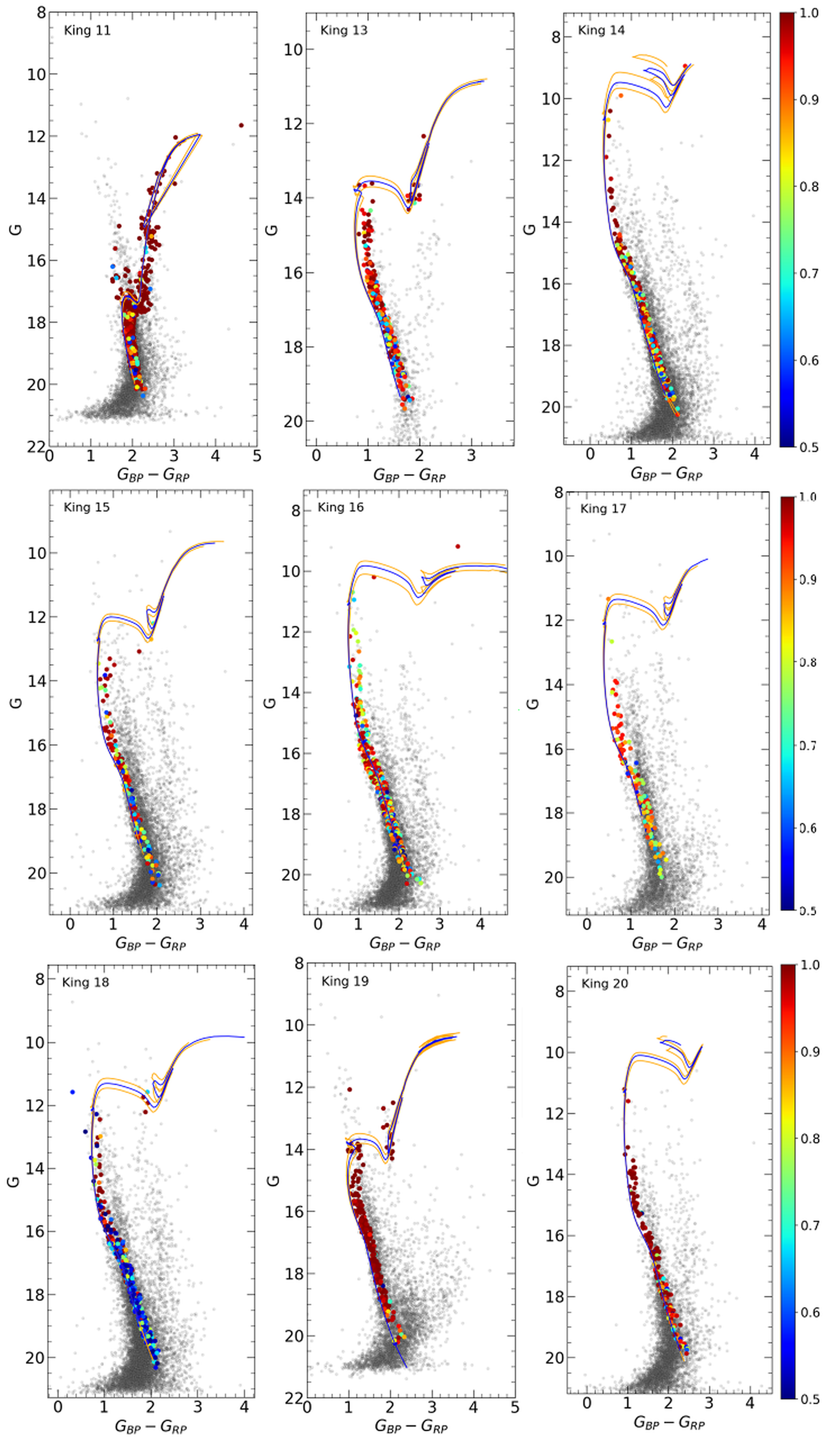}
\caption{Continued.}
\label{fig:appendix_cmds_02}
\end{figure}

\clearpage

\begin{figure}[H]
\centering
\includegraphics[width=0.70\textwidth,height=0.9\textheight,keepaspectratio]{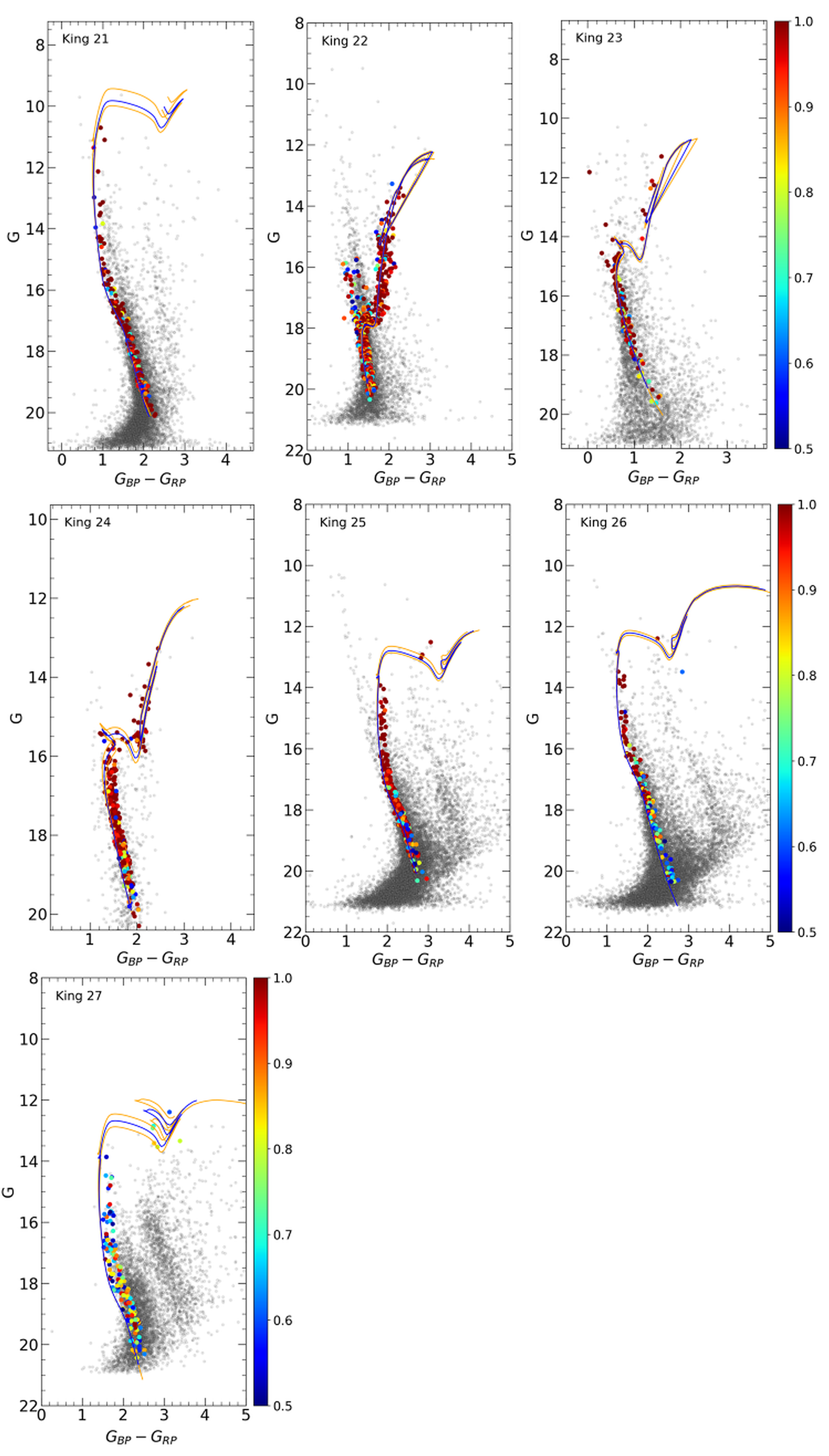}
\caption{Continued.}
\label{fig:appendix_cmds_03}
\end{figure}

\clearpage

\clearpage
\onecolumn

\section{Fundamental Parameters of the 27 King OCs}
\label{app:fundamental}

\renewcommand{\thetable}{C\arabic{table}}
\setcounter{table}{0}

\newsavebox\mybox
\newlength\mylength
\newcommand\boxup[2]{%
  \savebox\mybox{#1}%
  \setlength\mylength{\wd\mybox}%
  \parbox{\mylength}{#1 \\[1pt] #2}%
}

\begingroup
\setlength{\tabcolsep}{5pt}
\renewcommand{\arraystretch}{1.18}
\footnotesize

\begin{center}
\captionof{table}{Fundamental parameters of 27 King OCs.}

\adjustbox{width=\textwidth,totalheight=0.82\textheight,keepaspectratio,center}{%
\begin{tabular}{lccccc}

\hline
Cluster Name	                                   &	King 1	                 &	King 2	                   &	King 3	                 &	King 4	                 &	King 5	\\[2pt]
\hline
$(\alpha, \delta)_{\rm J2000}$ 	  &	   \boxup{00:22:01.21}{+64:22:58.8}	     &	\boxup{00:50:57.84}{+58:11:16.8}	   &	\boxup{01:36:24.48}{+64:32:24}	 &	\boxup{02:36:06.72}{+59:01:26.4}	 &	\boxup{03:14:43.68}{+52:41:42}	\\
$(l, b)_{\rm J2000}$ ($\deg$)	                  &	119.758, \,1.692	         &	122.869,\,$-4.684$	       &	127.744,\,2.089	         &	136.054,\,$-1.155$	     &	143.768,\,$-4.281$   \\
$f_0$ (stars arcmin$^{-2}$)	                      &	6.636$\pm$0.428	             &	87.067$\pm$2.531	       &	157.097$\pm$4.282	     &	10.770$\pm$0.253	     &	13.478$\pm$0.645	\\
$f_{\rm bg}$ (stars arcmin$^{-2}$)	              &	14.786$\pm$0.324	         &	7.496$\pm$0.408	           &	13.092$\pm$1.496	     &	5.426$\pm$0.253	         &	5.001$\pm$0.306	\\
$r_{\rm c}$ (arcmin)	                          &	4.686$\pm$0.655	             &	0.490$\pm$0.005	           &	0.819$\pm$0.045	         &	1.541$\pm$0.074	         &	2.021$\pm$0.240	\\
 $r_{\rm lim}^{\rm obs}$ (arcmin) 	                         & 20	                        &  15	                      &	   8	                    &  12	                    &  20	\\
$r_{\rm lim}^{\rm obs}$ (pc)	                              &	10.07	                     &	25.88	                   &    11.94	                 &	8.5	                     &	12.75	\\
$C$                                               & 0.63                         &  1.49                      &    0.99                     &  0.89                     &  1.00 \\
Cluster members ($P \geq 0.5$)	                  &	574	                         &	474	                       &	241	                     &	202	                     &	363	\\
$\mu_{\alpha} \cos \delta$ (mas yr$^{-1}$)	      &	$-4.935\pm$0.003	         &	$-1.398\pm$0.005	       &	$-0.700\pm$0.004	     &	$-0.598\pm$0.005	     &	$-0.303\pm$0.004	\\
$\mu_{\delta}$ (mas yr$^{-1}$)	                  &	$-1.302\pm$0.003	         &	$-0.876\pm$0.008	       &	$-0.131\pm$0.006	     &	$-0.174\pm$0.005	     &	$-1.261\pm$0.004	\\
$\varpi$ (mas)	                                  &	0.550$\pm$0.048	             &	0.164$\pm$0.032	           &	0.170$\pm$0.019	         &	0.395$\pm$0.051	         &	0.413$\pm$0.039	\\
$d_{\varpi}$ (pc)	                              &	1816$\pm$157	             &	6098$\pm$1190	           &	5894$\pm$647	         &	2529$\pm$327	         &	2421$\pm$229	\\
$E(B-V)$ (mag)	                                  & $0.632^{+0.042}_{-0.038}$    & $0.377^{+0.099}_{-0.103}$   & $1.096^{+0.068}_{-0.065}$   & $0.662^{+0.065}_{-0.061}$ & $0.662^{+0.110}_{-0.114}$ \\
$E(G_{\rm BP} - G_{\rm RP})$ (mag)	              &	$0.891^{+0.059}_{-0.054}$    & $0.532^{+0.139}_{-0.145}$   & $1.546^{+0.096}_{-0.091}$   & $0.934^{+0.091}_{-0.086}$ & $0.934^{+0.155}_{-0.161}$ \\
$A_{\rm G}$  (mag)	                              &	$1.66^{+0.11}_{-0.10}$       & $0.99^{+0.26}_{-0.27}$      & $2.88^{+0.18}_{-0.17}$      & $1.74^{+0.17}_{-0.16}$    & $1.74^{+0.29}_{-0.30}$ \\
$Z$                                               & $0.0190^{+0.0014}_{-0.0019}$ & $0.0063^{+0.0009}_{-0.0007}$& $0.0122^{+0.0011}_{-0.0013}$& $0.0181^{+0.0024}_{-0.0028}$& $0.0110^{+0.0017}_{-0.0012}$ \\
${\rm [Fe/H]}$ (dex)	                          &	$0.11\pm0.04$                & $-0.40\pm 0.06$                    & $-0.10\pm 0.04$                      & $0.08\pm 0.07$                    &  $-0.14\pm 0.06$ \\ 
 Age (Myr)	                                     &	$2692^{+282}_{-273}$        & $4365^{+6350}_{-1795}$     & $110^{+25}_{-24}$             & $234^{+32}_{-28}$        & $1380^{+125}_{-132}$ \\
Isochrone distance (pc)	                          &	 $1795^{+160}_{-188}$        & $5947^{+202}_{-189}$       & $5222^{+177}_{-201}$         & $2517^{+258}_{-270}$      & $2208^{+280}_{-291}$ \\
$X,\,Y,\,Z$ (pc)             & $(-891,1558,53)$  & $(-3217,4978,486)$& $(-3194,4127,190)$& $(-1812,1746,51)$ & $(-1776,1301,165)$ \\
$R_{\rm gc}$ (kpc)           & $9.329$           & $12.496$         & $12.161$         & $10.241$         & $10.140$ \\
$V_{\rm R}$ (km\,s$^{-1}$)   & $-53.23\pm0.13$   & $-110.36\pm9.87$ & $-44.18\pm1.27$  & $-45.23\pm0.14$  & $-43.50\pm0.06$ \\
$N$                          & $67$              & $11$             & $8$              & $1$              & $27$ \\
$U_{\rm LSR}$ (km\,s$^{-1}$) & $72.46\pm3.53$    & $102.76\pm4.35$  & $48.77\pm0.44$   & $45.45\pm0.39$   & $42.12\pm0.37$ \\
$V_{\rm LSR}$ (km\,s$^{-1}$) & $-10.51\pm2.21$   & $-58.47\pm8.86$  & $-10.44\pm1.37$  & $-13.12\pm0.601$ & $-15.40\pm0.59$ \\
$W_{\rm LSR}$ (km\,s$^{-1}$) & $-1.22\pm0.61$    & $-8.96\pm1.40$   & $-1.24\pm0.21$   & $2.77\pm0.47$    & $-3.06\pm1.63$ \\
$S_{\rm LSR}$ (km\,s$^{-1}$) & $73.23\pm4.21$    & $118.56\pm9.97$  & $49.89\pm1.45$   & $47.38\pm0.86$   & $44.95\pm1.77$ \\
$Z_{\rm max}$ (kpc)          & $0.083\pm0.008$   & $0.546\pm0.020$  & $0.240\pm0.012$  & $0.063\pm0.003$  & $0.149\pm0.029$ \\
$R_{\rm a}$ (kpc)            & $10.529\pm0.294$  & $12.976\pm0.001$ & $13.653\pm0.404$ & $10.447\pm0.163$ & $10.459\pm0.174$ \\
$R_{\rm p}$ (kpc)            & $8.389\pm0.246$   & $9.873\pm0.922$  & $11.293\pm0.115$ & $9.715\pm0.373$  & $9.214\pm0.323$ \\
$e$                          & $0.113\pm0.001$   & $0.136\pm0.046$  & $0.095\pm0.009$  & $0.036\pm0.012$  & $0.063\pm0.009$ \\
$R_{\rm teo}$ (kpc)          & $9.123\pm0.410$   & $12.830\pm0.813$ & $12.410\pm1.078$ & $10.442\pm0.341$ & $10.459\pm0.313$ \\
$R_{\rm gui}$ (kpc)      & $9.350\pm0.268$   & $11.194\pm0.573$ & $12.365\pm0.237$ & $10.068\pm0.276$ & $9.796\pm0.256$ \\
$P_{\rm orb}$ (Myr)          & $246\pm8$         & $303\pm13$       & $335\pm8$        & $264\pm8$        & $256\pm8$ \\

\hline
\end{tabular}%
}
\end{center}
\endgroup

\clearpage
\begingroup
\setlength{\tabcolsep}{5pt}
\renewcommand{\arraystretch}{1.18}
\footnotesize

\begin{center}
\addtocounter{table}{-1}
\captionof{table}{Fundamental parameters of 27 King OCs (continued).}

\adjustbox{width=\textwidth,totalheight=0.86\textheight,keepaspectratio,center}{%
\begin{tabular}{lccccc}
\hline
Cluster Name	                                 &	King 6	                 &	King 7	                 &	King 8	                 &	King 9	                 &	King 10	\\[2pt]
\hline

$(\alpha, \delta)_{\rm J2000}$ 	 &	\boxup{03:27:55.67}{+56:26:38.4}	 &	\boxup{03:59:07.92}{+51:47:06}	 &	\boxup{05:49:17.76}{+33:37:58.8}	  &	\boxup{22:15:30.48}{+54:24:18}	 &	\boxup{22:54:59.53}{+59:10:12}	\\
$(l, b)_{\rm J2000}$~($\deg$)	                 &	143.345,\,$-0.095$	     &	149.800,\,$-1.017$       &	176.384,\, 3.101	      &	101.441,\,$-1.828$	     &	108.493,\,$-0.398$	\\
$f_0$ (stars arcmin$^{-2}$)	                     &	2.006$\pm$0.168	         &	41.644$\pm$0.836	     &	68.247$\pm$1.776	      &	238.349$\pm$3.789	     &	66.781$\pm$1.629	\\
$f_{\rm bg}$ (stars arcmin$^{-2}$)	             &	2.129$\pm$0.121	         &	7.279$\pm$0.340	         &	10.489$\pm$0.364	      &	41.479$\pm$0.815	     &	14.216$\pm$0.497	\\
$r_{\rm c}$ (arcmin)	                         &	4.880$\pm$0.890	         &	1.688$\pm$0.071	         &	0.536$\pm$0.024	          &	0.643$\pm$0.018	         &	1.116$\pm$0.053	\\
$r_{\rm lim}^{\rm obs}$ (arcmin)	                         &	25	                     &	20	                     &	10	                      &	12	                     &	15	\\
$r_{\rm lim}^{\rm obs}$ (pc)	                             &	5.26	                 &	16.47	                 &	14.27	                  &	21.78	                 &	14.53	\\
$C$                                              & 0.71                      & 1.07                      & 1.27                    & 1.27                    &  1.13 \\
Cluster members ($P \geq 0.5$)	                 &	265	                     &	366                      &	104	                      &	208	                     &	333	\\
$\mu_{\alpha} \cos \delta$ (mas yr$^{-1}$)	     &	3.830$\pm$0.003	         &	1.127$\pm$0.007	         &	0.452$\pm$0.010	          &	$-2.294\pm$0.011	     &	$-2.697\pm$0.005	\\
$\mu_{\delta}$ (mas yr$^{-1}$)	                 &	$-1.891\pm$0.003	     &	$-1.191\pm$0.005	     &	$-1.700\pm$0.007	      &	$-1.013\pm$0.009	     &	$-2.118\pm$0.004	\\
$\varpi$ (mas)	                                 &	1.382$\pm$0.044	         &	0.318$\pm$0.044	         &	0.184$\pm$0.035	          &	0.147$\pm$0.035	         &	0.283$\pm$0.032	\\
$d_{\varpi}$ (pc)	                             &	723$\pm$23	             &	3146$\pm$436	         &	5425$\pm$1016	          &	6801$\pm$1626	         &	3529$\pm$399	\\
$E(B-V)$ (mag)	                                 &	$0.495^{+0.023}_{-0.023}$  & $1.173^{+0.008}_{-0.015}$ & $0.529^{+0.011}_{-0.011}$ & $0.358^{+0.099}_{-0.103}$ & $1.074^{+0.042}_{-0.042}$ \\
$E(G_{\rm BP} - G_{\rm RP})$ (mag)	             &	$0.698^{+0.032}_{-0.032}$  & $1.654^{+0.011}_{-0.021}$ & $0.746^{+0.016}_{-0.016}$ & $0.505^{+0.139}_{-0.145}$ & $1.514^{+0.059}_{-0.059}$ \\
$A_{\rm G}$  (mag)	                             &	$1.30^{+0.06}_{-0.06}$     & $3.08^{+0.02}_{-0.04}$    & $1.39^{+0.03}_{-0.03}$    & $0.94^{+0.26}_{-0.27}$ & $2.82^{+0.11}_{-0.11}$ \\
$Z$                                              & $0.0121^{+0.0008}_{-0.0009}$& $0.0105^{+0.0015}_{-0.0011}$ & $0.0108^{+0.0019}_{-0.0011}$ & $0.0146^{+0.0013}_{-0.0017}$ & $0.0171^{+0.0023}_{-0.0017}$ \\
${\rm [Fe/H]}$ (dex)	                         & $-0.10\pm 0.04$           & $-0.17\pm 0.06$          & $-0.16\pm 0.06$            & $-0.02\pm 0.04$          & $0.06\pm 0.05 $ \\
 Age (Myr)	                                    & $195^{+19}_{-17}$         & $186^{+10}_{-11}$        & $933^{+97}_{-95}$          & $3715^{+320}_{-325}$     & $30^{+8}_{-9}$ \\
Isochrone distance (pc)	                         & $739^{+11}_{-11}$         & $3020^{+103}_{-91}$      & $4828^{+269}_{-289}$       & $6272^{+133}_{-147}$     & $3452^{+115}_{-79}$ \\
$X,\,Y,\,Z$ (pc)                                 & $(-593,441,1)$            & $(-2610,1519,54)$       & $(-4811,304,261)$          & $(-1243,6144,200)$       & $(-1095,3274,24)$ \\
$R_{\rm gc}$ (kpc)                               & $8.921$                   & $10.974$                & $13.030$                   & $11.347$                 & $9.949$ \\
$V_{\rm R}$ (km\,s$^{-1}$)                       & $-21.56\pm1.45$           & $-10.77\pm0.48$         & $-1.02\pm2.43$             & $-59.94\pm0.51$          & $-62.72\pm0.20$ \\
$N$                                              & $49$                      & $27$                    & $5$                        & $9$                      & $4$ \\
$U_{\rm LSR}$ (km\,s$^{-1}$)                     & $17.28\pm1.32$            & $6.43\pm0.84$           & $6.81\pm2.59$              &  $92.54\pm1.12$          & $80.71\pm1.32$ \\
$V_{\rm LSR}$ (km\,s$^{-1}$)                     & $-10.57\pm0.77$           & $-11.42\pm0.55$         & $-24.47\pm2.09$            & $-29.56\pm0.82$          & $-27.99\pm0.73$ \\
$W_{\rm LSR}$ (km\,s$^{-1}$)                     & $8.66\pm0.21$             & $4.30\pm0.21$           & $-4.52\pm0.31$             & $21.73\pm0.38$           & $-5.13\pm0.38$ \\
$S_{\rm LSR}$ (km\,s$^{-1}$)                     & $22.03\pm1.54$    & $13.80\pm1.03$    & $25.80\pm3.34$    & $99.55\pm1.44$    & $85.58\pm1.56$ \\
$Z_{\rm max}$ (kpc)                              & $0.136\pm0.001$   & $0.101\pm0.003$   & $0.312\pm0.018$   & $0.612\pm0.026$   & $0.076\pm0.006$ \\
$R_{\rm a}$ (kpc)                                & $9.006\pm0.008$   & $11.608\pm0.136$  & $13.176\pm0.252$  & $12.488\pm0.234$  & $10.081\pm0.057$ \\
$R_{\rm p}$ (kpc)                                & $8.091\pm0.072$   & $9.893\pm0.012$   & $11.453\pm0.085$  & $10.074\pm0.111$  & $9.024\pm0.166$ \\
$e$                                              & $0.054\pm0.005$   & $0.080\pm0.005$   & $0.070\pm0.006$   & $0.107\pm0.004$   & $0.055\pm0.007$ \\
$R_{\rm teo}$ (kpc)                              & $8.922\pm0.011$   & $11.528\pm0.776$  & $12.161\pm0.373$  & $11.008\pm0.743$  & $10.046\pm0.498$ \\
$R_{\rm gui}$ (kpc)                          & $8.522\pm0.035$   & $10.688\pm0.065$  & $12.245\pm0.158$  & $11.109\pm0.160$  & $9.525\pm0.117$ \\
$P_{\rm orb}$ (Myr)                              & $220\pm1$         & $283\pm2$         & $330\pm6$         & $300\pm5$         & $248\pm3$ \\
\hline
\end{tabular}%
}
\end{center}
\endgroup

\clearpage
\begingroup
\setlength{\tabcolsep}{5pt}
\renewcommand{\arraystretch}{1.18}
\footnotesize

\begin{center}
\addtocounter{table}{-1}
\captionof{table}{Fundamental parameters of 27 King OCs (continued).}

\adjustbox{width=\textwidth,totalheight=0.86\textheight,keepaspectratio,center}{%
\begin{tabular}{lccccc}
\hline
Cluster Name	                                     &	King 11	                  &	King 12	                 &	King 13	                &	King 14	                   &	King 15	\\[2pt]
\hline
$(\alpha, \delta)_{\rm J2000}$ 	                     &	\boxup{23:47:38.89}{+68:38:09.6}	  &	\boxup{23:53:03.61}{+61:57:10.8}	 &	\boxup{00:10:12.73}{+61:10:30} &	\boxup{00:31:56.89}{+63:09:46.8}	   &	\boxup{00:33:02.89}{+61:51:14.4}	\\
$(l, b)_{\rm J2000}$~($\deg$)	                     &	117.151,\,6.484	          &	116.128,\,$-0.145$	     &	117.982,\,$-1.300$	     &	120.734,\,0.375	           &	120.765,\,$-0.941$	\\
$f_0$ (stars arcmin$^{-2}$)	                         &	56.338$\pm$1.241	      &	15.933$\pm$3.642	     &	17.847$\pm$0.315	     &	9.936$\pm$0.663	           &	19.033$\pm$1.011	\\
$f_{\rm bg}$ (stars arcmin$^{-2}$)	                 &	9.069$\pm$0.468	          &	2.358$\pm$0.093          &	8.453$\pm$0.131	         &	23.182$\pm$0.510	       &	18.188$\pm$0.223	\\
$r_{\rm c}$ (arcmin)	                             &	1.526$\pm$0.069	          &	0.097$\pm$0.054	         &	2.067$\pm$0.076	         &	3.950$\pm$0.587	           &	0.595$\pm$0.058	\\
$r_{\rm lim}^{\rm obs}$ (arcmin)	                             &	20	                      &	10	                     &	20	                     &	20	                       &	11	\\
$r_{\rm lim}^{\rm obs}$ (pc)	                                 &	18.52	                  &	8.27	                 &	20.71	                 &	14.44	                   &	10.37	\\
$C$                                                  &  1.12                     & 2.01                  &  0.99                     &  0.7                       &    1.27 \\
Cluster members ($P \geq 0.5$)	                     &	705	                      &	141	                     &	295	                     &	313	                       &	174	\\
$\mu_{\alpha} \cos \delta$ (mas yr$^{-1}$)	         &	$-3.388\pm$0.005	      &	$-3.437\pm$0.003	     &	$-2.728\pm$0.003	     &	$-3.237\pm$0.004	       &	$-2.319\pm$0.005	\\
$\mu_{\delta}$ (mas yr$^{-1}$)	                     &	$-0.666\pm$0.005	      &	$-1.471\pm$0.003	     &	$-0.872\pm$0.003	     &	$-1.073\pm$0.004	       &	$-0.895\pm$0.005	\\
$\varpi$ (mas)	                                     &	0.306$\pm$0.035	          &	0.327$\pm$0.025	         &	0.256$\pm$0.026	         &	0.393$\pm$0.058	           &	0.307$\pm$0.049	\\
$d_{\varpi}$ (pc)	                                 &	3263$\pm$375	          &	3058$\pm$234	         &	3908$\pm$390	         &	2545$\pm$377	           &	3261$\pm$516	\\
$E(B-V)$ (mag)	                                     &	$0.838^{+0.106}_{-0.106}$ & $0.556^{+0.110}_{-0.118}$ & $0.540^{+0.080}_{-0.087}$ & $0.392^{+0.099}_{-0.099}$  & $0.537^{+0.072}_{-0.068}$ \\
$E(G_{\rm BP} - G_{\rm RP})$ (mag)	                 &	$1.181^{+0.150}_{-0.150}$ & $0.784^{+0.155}_{-0.166}$ & $0.762^{+0.113}_{-0.123}$ & $0.553^{+0.139}_{-0.139}$  & $0.757^{+0.102}_{-0.096}$ \\
$A_{\rm G}$  (mag)	                                 &	$2.20^{+0.28}_{-0.28}$    & $1.46^{+0.29}_{-0.31}$    & $1.42^{+0.21}_{-0.23}$    & $1.03^{+0.26}_{-0.26}$     & $1.41^{+0.19}_{-0.18}$ \\
$Z$                                                  & $0.0092^{+0.0036}_{-0.0031}$ & $0.0116^{+0.0045}_{-0.0057}$ & $0.0123^{+0.0021}_{-0.0014}$ & $0.0111^{+0.0028}_{-0.0024}$ & $0.0083^{+0.0008}_{-0.0010}$ \\
${\rm [Fe/H]}$ (dex)	                             &	$-0.23\pm 0.16$                   & $-0.12\pm 0.22$                   & $-0.10\pm 0.06$                   & $-0.14\pm 0.10$                  & $-0.27\pm 0.05$ \\ 
 Age (Myr)	                                        & $3631^{+315}_{-322}$       & $17^{+6}_{-7}$            & $501^{+48}_{-45}$         & $83^{+11}_{-12}$        & $251^{+18}_{-17}$ \\
Isochrone distance (pc)	                             &	$3023^{+278}_{-311}$      & $2719^{+237}_{-223}$      & $3678^{+244}_{-251}$      & $2515^{+356}_{-347}$     & $3080^{+51}_{-59}$ \\
$X,\,Y,\,Z$ (pc)             & $(-1371,2673,341)$ & $(-1197,2441,7)$ & $(-1725,3247,83)$ & $(-1285,2162,16)$ & $(-1575,2646,51)$ \\
$R_{\rm gc}$ (kpc)           & $10.025$         & $9.803$          & $10.520$         & $9.821$          & $10.210$ \\
$V_{\rm R}$  (km\,s$^{-1}$)  & $-27.94\pm1.77$  & $-41.26\pm0.17$  & $-59.15\pm0.37$  & $-54.82\pm0.17$  & $-64.73\pm0.19$ \\
$N$                          & $69$             & $2$              & $15$             & $1$              & $5$ \\
$U_{\rm LSR}$ (km\,s$^{-1}$) & $65.60\pm3.43$   & $69.59\pm3.49$     & $80.26\pm2.72$ & $70.72\pm4.61$     & $71.85\pm0.44$ \\
$V_{\rm LSR}$ (km\,s$^{-1}$) & $11.76\pm3.71$   & $-1.99\pm1.93$     & $-15.01\pm1.88$& $-12.71\pm2.96$    & $-23.82\pm0.56$ \\
$W_{\rm LSR}$ (km\,s$^{-1}$) & $6.10\pm0.55$    & $-1.87\pm0.72$     & $0.48\pm0.51$  & $-3.63\pm1.35$     & $-2.99\pm0.24$ \\
$S_{\rm LSR}$ (km\,s$^{-1}$) & $66.92\pm5.08$   & $69.64\pm4.06$   & $81.65\pm3.35$   & $71.95\pm5.64$   & $75.75\pm0.75$ \\
$Z_{\rm max}$ (kpc)          & $0.469\pm0.065$  & $0.029\pm0.010$  & $0.063\pm0.004$  & $0.066\pm0.021$  & $0.050\pm0.003$ \\
$R_{\rm a}$ (kpc)            & $12.744\pm0.879$ & $11.047\pm0.461$ & $11.197\pm0.391$ & $10.609\pm0.458$ & $10.538\pm0.037$ \\
$R_{\rm p}$ (kpc)            & $10.076\pm0.197$ & $9.741\pm0.201$  & $10.385\pm0.309$ & $9.228\pm0.502$  & $9.229\pm0.098$ \\
$e$                          & $0.117\pm0.025$  & $0.063\pm0.011$  & $0.038\pm0.003$  & $0.070\pm0.006$  & $0.066\pm0.004$ \\
$R_{\rm teo}$ (kpc)          & $10.547\pm0.776$ & $10.104\pm0.354$ & $11.196\pm1.082$ & $10.337\pm1.021$ & $10.335\pm0.947$ \\
$R_{\rm gui}$ (kpc)      & $11.236\pm0.470$ & $10.358\pm0.317$ & $10.776\pm0.348$ & $9.875\pm0.485$  & $9.844\pm0.071$ \\
$P_{\rm orb}$ (Myr)          & $302\pm16$       & $273\pm9$        & $284\pm11$       & $259\pm14$       & $258\pm2$ \\
\hline
\end{tabular}%
}
\end{center}
\endgroup

\clearpage
\begingroup
\setlength{\tabcolsep}{5pt}
\renewcommand{\arraystretch}{1.18}
\footnotesize

\begin{center}
\addtocounter{table}{-1}
\captionof{table}{Fundamental parameters of 27 King OCs (continued).}

\adjustbox{width=\textwidth,totalheight=0.86\textheight,keepaspectratio,center}{%
\begin{tabular}{lccccc}
\hline
Cluster Name	                                 &	King 16	                  &	King 17	                 &	King 18	                 &	King 19	                  &	King 20	\\[2pt]
\hline
$(\alpha, \delta)_{\rm J2000}$	 &	\boxup{00:43:39.61}{+64:10:19.2}	  &	\boxup{05:08:21.60}{+39:04:08.4}	 &	\boxup{22:52:13.21}{+58:17:20.4}	 &	\boxup{23:08:12.73}{+60:31:22.8}	  &	\boxup{23:33:13.21}{+58:28:08.4}	\\
$(l, b)_{\rm J2000}$~($\deg$)	                 &	122.084,\,1.313	          &	167.297,\,$-0.745$	     &	107.785,\,$-1.032$	     &	110.563,\,0.155	          &	112.839,\,$-2.861$	\\
$f_0$ (stars arcmin$^{-2}$)	                     & 	35.124$\pm$1.774	      &	13.891$\pm$0.688	     &	5.813$\pm$0.349	         &	31.879$\pm$0.510	      &	7.893$\pm$0.248	\\
$f_{\rm bg}$ (stars arcmin$^{-2}$)	             &	17.025$\pm$0.533	      &	11.530$\pm$0.207	     &	5.335$\pm$0.180	         &	6.337$\pm$0.097	          &	5.624$\pm$0.089	\\
$r_{\rm c}$ (arcmin)	                         &	0.959$\pm$0.094	          &	1.093$\pm$0.107	         &	13.021$\pm$0.602	     &	0.422$\pm$0.012	          &	1.694$\pm$0.106	\\
$r_{\rm lim}^{\rm obs}$ (arcmin)	                         &	15	                      &	15	                     &	25	                     &	10	                      &	15	\\
$r_{\rm lim}^{\rm obs}$ (pc)	                             &	12.53	                  &	16.82	                 &	20.82	                 &	7.57	                  &	8.03	\\
$C$                                              &  1.19                     & 1.14                    &  0.28                     & 1.37                      & 0.95 \\
Cluster members ($P \geq 0.5$)	                 &	355	                      &	142	                     &	349	                     &	212	                      &	169	\\
$\mu_{\alpha} \cos \delta$ (mas yr$^{-1}$)	     &	$-3.009\pm$0.003	      &	$-0.195\pm$0.009	     &	$-2.634\pm$0.005	     &	$-4.794\pm$0.004	      &	$-2.668\pm$0.005	\\
$\mu_{\delta}$ (mas yr$^{-1}$)	                 &	$-0.537\pm$0.004	      &	$-1.405\pm$0.006	     &	$-2.085\pm$0.004	     &	$-2.712\pm$0.004	      &	$-2.656\pm$0.005	\\
$\varpi$ (mas)	                                 &	0.339$\pm$0.042	          &	0.256$\pm$0.029	         &	0.338$\pm$0.041	         &	0.364$\pm$0.043	          &	0.519$\pm$0.041	\\
$d_{\varpi}$ (pc)	                             &	2953$\pm$368	          &	3910$\pm$443	         &	2961$\pm$358	         &	2744$\pm$322	          &	1926$\pm$151	\\
$E(B-V)$ (mag)	                                 &	$0.709^{+0.034}_{-0.034}$ & $0.384^{+0.065}_{-0.061}$& $0.628^{+0.084}_{-0.084}$ & $0.651^{+0.106}_{-0.103}$  & $0.799^{+0.106}_{-0.106}$ \\
$E(G_{\rm BP} - G_{\rm RP})$ (mag)	             &	$0.999^{+0.048}_{-0.048}$ & $0.542^{+0.091}_{-0.086}$& $0.886^{+0.118}_{-0.118}$ & $0.918^{+0.150}_{-0.145}$  & $1.127^{+0.150}_{-0.150}$ \\
$A_{\rm G}$ (mag)	                             &	$1.86^{+0.09}_{-0.09}$    & $1.01^{+0.17}_{-0.16}$   & $1.65^{+0.22}_{-0.22}$    & $1.71^{+0.28}_{-0.27}$     & $2.10^{+0.28}_{-0.28}$ \\
$Z$                                              & $0.0166^{+0.0022}_{-0.0018}$ & $0.0193^{+0.0012}_{-0.0017}$ & $0.0153^{+0.0007}_{-0.0019}$ & $0.0099^{+0.0025}_{-0.0022}$ & $0.0156^{+0.0011}_{-0.0009}$ \\
${\rm [Fe/H]}$ (dex)	                         &	$0.04\pm 0.06$             & $0.11\pm 0.04$           & $0.00\pm 0.04$           & $-0.19\pm 0.11$            & $0.01\pm 0.03$ \\
 Age (Myr)	                                    &  $56^{+7}_{-9}$ & $158^{+16}_{-13}$ & $155^{+15}_{-13}$ & $724^{+68}_{-74}$ & $91^{+8}_{-8}$ \\
Isochrone distance (pc)	                         &	$2685^{+226}_{-243}$ & $3713^{+464}_{-471}$ & $2811^{+209}_{-217}$ & $2599^{+252}_{-261}$ & $1973^{+133}_{-124}$ \\
$X,\,Y,\,Z$ (pc)             & $(-1426,2274,62)$& $(-3622,816,48)$ & $(-858,2676,51)$ & $(-913,2433,7)$  & $(-765,1816,98)$ \\
$R_{\rm gc}$ (kpc)           & $9.979$          & $11.886$         & $9.549$          & $9.535$          & $9.256$ \\
$V_{\rm R}$  (km\,s$^{-1}$)  & $-46.49\pm0.30$  & $-2.56\pm3.48$   & $-56.48\pm3.39$  & $-51.61\pm2.05$  & -- \\
$N$                          & $1$              & $1$              & $5$              & $12$             & -- \\
$U_{\rm LSR}$ (km\,s$^{-1}$) & $66.06\pm2.66$   & $7.64\pm3.87$    & $67.81\pm2.06$   & $90.05\pm5.45$   & -- \\
$V_{\rm LSR}$ (km\,s$^{-1}$) & $-4.64\pm2.06$   & $-3.76\pm1.47$   & $-26.38\pm4.22$  & $-10.44\pm4.25$  & -- \\
$W_{\rm LSR}$ (km\,s$^{-1}$) & $-0.17\pm0.49$   & $-10.93\pm2.08$  & $-1.80\pm0.78$   & $-1.49\pm0.78$   & -- \\
$S_{\rm LSR}$ (km\,s$^{-1}$) & $66.22\pm3.40$   & $13.86\pm4.63$   & $72.78\pm4.76$   & $90.67\pm6.96$   & -- \\
$Z_{\rm max}$ (kpc)          & $0.092\pm0.007$  & $0.230\pm0.065$  & $0.034\pm0.010$  & $0.037\pm0.007$  & -- \\
$R_{\rm a}$ (kpc)            & $11.025\pm0.414$ & $12.624\pm0.565$ & $9.629\pm0.098$  & $11.049\pm0.597$ & -- \\
$R_{\rm p}$ (kpc)            & $9.833\pm0.263$  & $11.897\pm0.326$ & $8.373\pm0.526$  & $8.938\pm0.353$  & -- \\
$e$                          & $0.057\pm0.006$  & $0.030\pm0.009$  & $0.070\pm0.026$  & $0.106\pm0.007$  & -- \\
$R_{\rm teo}$ (kpc)          & $10.917\pm0.672$ & $12.618\pm0.703$ & $9.310\pm0.213$  & $9.408\pm2.444$  & -- \\
$R_{\rm gui}$ (kpc)      & $10.397\pm0.331$ & $12.241\pm0.435$ & $8.962\pm0.340$  & $9.894\pm0.456$  & -- \\
$P_{\rm orb}$ (Myr)          & $274\pm10$       & $328\pm14$       & $232\pm8$        & $261\pm14$       & -- \\
\hline
\end{tabular}%
}
\end{center}
\endgroup


\clearpage
\begingroup
\setlength{\tabcolsep}{5pt}
\renewcommand{\arraystretch}{1.18}
\footnotesize

\begin{center}
\addtocounter{table}{-1}
\captionof{table}{Fundamental parameters of 27 King OCs (continued).}

\adjustbox{width=\textwidth,totalheight=0.86\textheight,keepaspectratio,center}{%
\begin{tabular}{lccccc}
\hline
Cluster Name	                                 &	King 21	                  &	King 22	                  &	King 23	                  &	King 24	                   &	King 25	\\[2pt]
\hline
$(\alpha, \delta)_{\rm J2000}$ 	 &	\boxup{23:49:54.97}{+62:42:18}	  &	\boxup{05:22:07.44}{+45:26:31.2}	  &	\boxup{07:21:50.16}{$-$00:59:16.8}	  &	\boxup{07:50.28.8}{$-$29:51:10.8}	   &	\boxup{19:24:29.28}{+13:41:45.7}	\\
$(l, b)_{\rm J2000}$~($\deg$)	                 &	115.945,\,0.671	          &	163.589,\,5.030	          &	217.296,\,6.305	          &	245.890,\,$-1.737$	       &	48.860,\,$-0.934$	\\
$f_0$ (stars arcmin$^{-2}$)	                     &	28.017$\pm$0.700	      &	5.093$\pm$0.328	          &	13.140$\pm$0.371	      &	26.486$\pm$1.081	       &	10.327$\pm$0.674	\\
$f_{\rm bg}$ (stars arcmin$^{-2}$)	             &	9.111$\pm$0.157	          &	2.833$\pm$0.195	          &	10.573$\pm$0.107	      &	4.252$\pm$0.236	           &	9.178$\pm$0.505	\\
$r_{\rm c}$ (arcmin)	                         &	6.231$\pm$0.406	          &	3.398$\pm$0.471	          &	1.016$\pm$0.055	          &	0.595$\pm$0.043	           &	3.385$\pm$0.478	\\
$r_{\rm lim}^{\rm obs}$ (arcmin)	                         &	10	                      &	25	                      &	10	                      &	10	                       &	20	\\
$r_{\rm lim}^{\rm obs}$ (pc)	                             &	8.41	                  &	36.31	                  &	9.83	                  &	11.07	                   &	15.10	\\
$C$                                              &  0.21                      & 0.87                      & 0.99                     & 1.23                     &    0.77 \\
Cluster members ($P \geq 0.5$)	                 &	209	                      &	556	                      &	150	                      &	320	                       &	238	\\
$\mu_{\alpha} \cos \delta$ (mas yr$^{-1}$)	     &	$-3.243\pm$0.006	      &	0.778$\pm$0.007	          &	$-0.477\pm$0.004	      &	$-2.960\pm$0.005	       &	$-1.388\pm$0.007	\\
$\mu_{\delta}$ (mas yr$^{-1}$)	                 &	$-1.722\pm$0.006	      &	$-0.092\pm$0.005	      &	$-0.876\pm$0.004	      &	2.483$\pm$0.005	           &	$-3.964\pm$0.006	\\
$\varpi$ (mas)	                                 &	0.337$\pm$0.047	          &	0.185$\pm$0.020	          &	0.279$\pm$0.033	          &	0.235$\pm$0.031	           &	0.329$\pm$0.044	\\
$d_{\varpi}$ (pc)	                             &	2966$\pm$409	          &	5412$\pm$589	          &	3582$\pm$427	          &	4248$\pm$560	           &	3036$\pm$404	\\
$E(B-V)$ (mag)	                                 &	$0.743^{+0.095}_{-0.095}$ & $0.343^{+0.072}_{-0.087}$ & $0.08^{+0.023}_{-0.027}$  & $0.689^{+0.095}_{-0.099}$  & $1.371^{+0.118}_{-0.121}$ \\
$E(G_{\rm BP} - G_{\rm RP})$ (mag)	             &	$1.047^{+0.134}_{-0.134}$ & $0.709^{+0.102}_{-0.123}$ & $0.113^{+0.032}_{-0.038}$ & $0.972^{+0.134}_{-0.139}$  & $1.933^{+0.166}_{-0.171}$ \\
$A_{\rm G}$  (mag)	                             &	$1.95^{+0.25}_{-0.25}$    & $1.32^{+0.19}_{-0.23}$    & $0.21^{+0.06}_{-0.07}$    & $1.81^{+0.25}_{-0.26}$     & $3.60^{+0.31}_{-0.32}$ \\
$Z$                                              & $0.0149^{+0.0023}_{-0.0026}$& $0.0056^{+0.0003}_{-0.0005}$ & $0.0168^{+0.0010}_{-0.0009}$ & $0.0155^{+0.0030}_{-0.0026}$ & $0.0278^{+0.0018}_{-0.0019}$ \\
${\rm [Fe/H]}$ (dex)	                         &	$-0.01\pm 0.08$            & $-0.45\pm 0.04$      & $0.05\pm 0.02$        & $0.01\pm 0.08$             & $0.28\pm 0.03$ \\
 Age (Myr)	                                    & $47^{+5}_{-6}$    & $6166^{+441}_{-457}$ & $1698^{+80}_{-85}$ & $1175^{+105}_{-103}$ & $112^{+8}_{-7}$ \\
Isochrone distance (pc)	                         &	$2919^{+198}_{-206}$ & $5222^{+510}_{-480}$ & $3487^{+144}_{-160}$ & $3876^{+89}_{-139}$ & $2772^{+202}_{-240}$ \\
$X,\,Y,\,Z$ (pc)             & $(-1277,2625,34)$ & $(-4990,1470,458)$& $(-2757,-2100,383)$& $(-1583,-3536,117)$ & $(1823,2087,45)$ \\
$R_{\rm gc}$ (kpc)           & $9.925$           & $13.284$          & $11.211$          & $10.483$          & $6.933$ \\
$V_{\rm R}$ (km\,s$^{-1}$)   & $-32.15\pm0.94$   & $-3.74\pm1.03$    & $53.87\pm0.40$    & $73.58\pm0.47$    & $13.24\pm1.30$ \\
$N$                          & $3$               & $39$              & $14$              & $16$              & $5$ \\
$U_{\rm LSR}$ (km\,s$^{-1}$) & $67.10\pm2.57$    & $10.04\pm1.21$    & $-29.37\pm0.28$   & $-82.02\pm2.03$   & $58.50\pm4.04$ \\
$V_{\rm LSR}$ (km\,s$^{-1}$) & $6.95\pm2.33$     & $0.57\pm0.96$     & $-26.50\pm0.66$   & $-24.94\pm0.53$   & $-11.81\pm1.85$ \\
$W_{\rm LSR}$ (km\,s$^{-1}$) & $-6.37\pm0.82$    & $20.82\pm1.70$    & $-1.10\pm0.51$    & $-19.28\pm0.62$   & $-2.24\pm0.78$ \\
$S_{\rm LSR}$ (km\,s$^{-1}$) & $67.76\pm3.56$    & $23.12\pm2.30$    & $39.57\pm0.88$    & $87.87\pm2.18$    & $59.73\pm4.51$ \\
$Z_{\rm max}$ (kpc)          & $0.127\pm0.020$   & $0.979\pm0.140$   & $0.413\pm0.018$   & $0.372\pm0.018$   & $0.034\pm0.007$ \\
$R_{\rm a}$ (kpc)            & $12.007\pm0.538$  & $15.508\pm0.669$  & $11.407\pm0.153$  & $10.563\pm0.084$  & $6.922\pm0.063$ \\
$R_{\rm p}$ (kpc)            & $9.966\pm0.139$   & $13.208\pm0.410$  & $9.341\pm0.074$   & $9.943\pm0.196$   & $6.197\pm0.156$ \\
$e$                          & $0.093\pm0.015$   & $0.080\pm0.006$   & $0.100\pm0.003$   & $0.030\pm0.006$   & $0.055\pm0.008$ \\
$R_{\rm teo}$ (kpc)          & $10.916\pm0.599$  & $13.945\pm1.594$  & $10.801\pm2.529$  & $10.470\pm0.873$  & $6.908\pm0.215$ \\
$R_{\rm gui}$ (kpc)      & $10.900\pm0.308$  & $14.182\pm0.508$  & $10.250\pm0.106$  & $10.214\pm0.142$  & $6.541\pm0.114$ \\
$P_{\rm orb}$ (Myr)          & $291\pm11$        & $391\pm17$        & $272\pm4$         & $269\pm4$         & $165\pm3$ \\
\hline
\end{tabular}%
}
\end{center}
\endgroup

\clearpage
\begingroup
\setlength{\tabcolsep}{5pt}
\renewcommand{\arraystretch}{1.18}
\footnotesize

\begin{center}
\addtocounter{table}{-1}
\captionof{table}{Fundamental parameters of 27 King OCs (continued).}

\adjustbox{width=\textwidth,totalheight=0.86\textheight,keepaspectratio,center}{%
\begin{tabular}{lcc}
\hline
Cluster Name	                                 &	King 26	                      &	King 27	\\[2pt]
\hline
$(\alpha, \delta)_{\rm J2000}$ 	 &	\boxup{19:28:56.88}{$+14$:52:55.3}	  &	\boxup{19:42:38.64}{$+21$:08:42.1}	\\
$(l, b)_{\rm J2000}$~($\deg$)	                 &	50.417,\,$-1.321$	          &	57.477,\,$-1.114$	\\
$f_0$ (stars arcmin$^{-2}$)	                     &	8.709$\pm$0.464	              &	41.951$\pm$1.679	\\
$f_{\rm bg}$ (stars arcmin$^{-2}$)	             &	5.296$\pm$0.165	              &	46.148$\pm$1.270	\\
$r_{\rm c}$ (arcmin)	                         &	1.368$\pm$0.149	              &	3.421$\pm$0.301	\\
$r_{\rm lim}^{\rm obs}$ (arcmin)	                         &	15	                          &	19	\\
$r_{\rm lim}^{\rm obs}$ (pc)	                             &	9.43	                      & 25.03	\\
$C$                                              &  1.04                         & 0.74 \\ 
Cluster members ($P \geq 0.5$)	                 &	150	                          &	156\\
$\mu_{\alpha} \cos \delta$ (mas yr$^{-1}$)	     &	$-2.388\pm$0.006	          &	$-2.257\pm$0.010	\\
$\mu_{\delta}$ (mas yr$^{-1}$)	                 &	$-4.982\pm$0.007	          &	$-4.702\pm$0.012	\\
$\varpi$ (mas)	                                 &	0.423$\pm$0.039	              &	0.191$\pm$0.021	\\
$d_{\varpi}$ (pc)	                             &	2365$\pm$220	              &	5243$\pm$566	\\
$E(B-V)$ (mag)	                                 &	$0.967^{+0.061}_{-0.061}$     & $1.15^{+0.103}_{-0.099}$ \\
$E(G_{\rm BP} - G_{\rm RP})$ (mag)	             &	$1.364^{+0.086}_{-0.086}$     & $1.621^{+0.145}_{-0.139}$ \\
$A_{\rm G}$  (mag)	                             &	$2.54^{+0.16}_{-0.16}$        & $3.02^{+0.27}_{-0.26}$ \\
$Z$                                              &  $0.0182^{+0.0017}_{-0.0018}$  & $0.0180^{+0.0011}_{-0.0012}$ \\
${\rm [Fe/H]}$ (dex)	                         &	$0.08\pm 0.04$                & $0.08\pm 0.03$	\\
Age (Myr)	                                     &	$200^{+11}_{-12}$             & $81^{+11}_{-9}$ \\\href{}{}
Isochrone distance (pc)	                         &	$2321^{+154}_{-163}$          & $4542^{+253}_{-264}$ \\
$X,\,Y,\,Z$ (pc)             & $(1479,1788,54)$    & $(2442,3829,88)$ \\
$R_{\rm gc}$ (kpc)           & $7.161$             & $7.149$ \\
$V_{\rm R}$(km\,s$^{-1}$)    & $24.80\pm0.53$      & $26.06\pm1.13$ \\
$N$                          & $1$                 & $6$ \\
$U_{\rm LSR}$ (km\,s$^{-1}$) & $71.37\pm3.46$      & $117.19\pm5.70$ \\
$V_{\rm LSR}$ (km\,s$^{-1}$) & $-5.44\pm2.20$      & $-24.20\pm2.33$ \\
$W_{\rm LSR}$ (km\,s$^{-1}$) & $2.87\pm0.32$       & $-2.02\pm0.58$ \\
$S_{\rm LSR}$ (km\,s$^{-1}$) & $71.63\pm4.11$      & $119.68\pm6.19$ \\
$Z_{\rm max}$ (kpc)          & $0.052\pm0.001$     & $0.078\pm0.007$ \\
$R_{\rm a}$ (kpc)            & $7.377\pm0.133$     & $7.422\pm0.091$ \\
$R_{\rm p}$ (kpc)            & $6.836\pm0.077$     & $6.726\pm0.040$ \\
$e$                          & $0.038\pm0.004$     & $0.049\pm0.009$ \\
$R_{\rm teo}$ (kpc)          & $7.112\pm0.003$     & $7.352\pm0.087$ \\
$R_{\rm gui}$ (kpc)      & $7.097\pm0.103$     & $7.057\pm0.020$ \\
$P_{\rm orb}$ (Myr)          & $179\pm3$           & $179\pm1$ \\
\hline
\end{tabular}%
}
\end{center}
\endgroup


\clearpage

\nolinenumbers
\label{lastpage}

\end{document}